\documentclass[%
 reprint,
amsmath,amssymb,aps,
]{revtex4-1}

\usepackage{graphicx}
\usepackage{dcolumn}
\usepackage{bm}

\usepackage{bm}

\newcommand{\Lumi}{\mathcal{L}}

\begin{document}

\title{Modern and Future Colliders}

\author{V.~Shiltsev}
\affiliation{Fermilab, PO Box 500, MS339, Batavia, IL 60510,USA}
\author{F.~Zimmermann} 
\affiliation{European Organization for Nuclear Research, CERN, 1211 Geneve, Switzerland }
\date{\today{}}

\begin{abstract}
Since the initial development of charged particle colliders in the middle of the 20th century, these advanced scientific instruments have been at the forefront of scientific discoveries in high energy physics. Collider accelerator technology and beam physics have progressed immensely and modern facilities now
operate at energies and luminosities many orders of magnitude greater than the pioneering colliders of the early 1960s. In addition, the field of colliders 
remains extremely dynamic and continues to develop many innovative approaches. 
Indeed, several novel concepts are currently being considered for designing and constructing even more powerful future colliders. In this paper, we first review the colliding beam method and the history of colliders, and then present the major achievements of operational machines and the key 
features  of near-term collider projects that are currently under development. We conclude with an 
analysis of numerous proposals and studies for far-future colliders.
The evaluation of their respective potentials reveals tantalizing prospects for further significant breakthroughs in the collider field. 
\end{abstract}


\maketitle

\tableofcontents{}

\section{Introduction}
\label{intro}
Particle accelerators are unique scientific instruments which offer access to unprecedented energy per constituent, using well-focused high density beams of electrons ($e^-$), positrons ($e^+$), protons ($p$), antiprotons ($\bar {p}$), ions, muons ($\mu^+$, $\mu^-$), mesons, photons and gamma quanta ($\gamma$), among others \cite{sessler, scharf, livingston}. They have been widely used for physics research since the early 20th century and have greatly progressed both scientifically and technologically since. Analysis of all Nobel-Prize winning research in physics since 1939
\cite{nobelprize} 
--- the year the Nobel Prize was awarded to Ernest O.~Lawrence for invention of the first modern accelerator, the cyclotron \cite{lawrence1932} --- reveals that accelerators have played an integral role in influencing more than a quarter of physics-prize recipients by either inspiring them or 
facilitating their research. 
On average, accelerators have contributed to one 
Nobel Prize for Physics every three years \cite{HC}. Four Nobel prizes have directly honored breakthroughs in accelerator science and technology; aside from E.O.~Lawrence, John Cockcroft and Ernest Walton received the prize in 1951 for their invention of
the eponymous linear accelerator \cite{cockcroft}, and Simon van der Meer in 1984 for conceiving and developing the novel method of stochastic cooling \cite{vandermeer1985}.  To gain  an insight into the physics of  elementary particles, one accelerates them to very high kinetic energy, lets them strike other particles, and detects products of the ensuing 
reactions that transform the particles into new particles, such as the Higgs boson, which was discovered in the debris of proton-proton collisions at the Large Hadron Collider (LHC) \cite{bruning2007} and celebrated with the 2013 Nobel Prize in Physics \cite{englert, higgs}.  Recently, accelerator-based synchrotron radiation sources were instrumental for a number of Nobel-Prize winning research achievements in chemistry and biology, recognized 
in 1997, 2003, 2006, 2009, and 2012. At present, about 140 accelerators of all types worldwide are devoted to fundamental research \cite{ELSA}.  
In the United States alone, the Department of Energy (DOE) Office of Science is supporting 16 large accelerator-based user facilities open for basic research --- such as colliders, light sources and neutron sources --- with a total annual budget for operation and construction exceeding \$2B \cite{DOE}. These facilities enable scientific research to be carried out by about 20,000 users from academia, industry, and government laboratories. Europe's leading particle physics laboratory,  CERN, with an annual budget of about 1.15 BCHF \cite{CERN}, operates the world's largest accelerator complex and brings together 17,000 physicists, engineers, and technicians from more than 110 different countries. 

Colliders are the most sophisticated of all accelerator types and employ the most advanced technologies and beam physics techniques to push the envelope of their performance. What makes them the instruments of choice for particle physics is their kinematic advantage of a high center-of-mass energy resulting in larger momentum transfers. 
Indeed, the center of mass energy (c.m.e.) 
$E_{cme}$ 
(also often cited as $\sqrt{s}$, the square root of one of the Lorentz-invariant \textit{Mandelstam variables} in the kinematics of reactions --- see, e.g., \cite{perkins})  for the head-on collision of two particles of masses $m_1$ and $m_2$ with energies $E_1$ and $E_2$ colliding at a crossing angle $\theta_c$ is   
\begin{eqnarray}
\lefteqn{ 
E_{cme}  =  
\biggl(
2E_1 E_2 + (m_1^2+m_2^2)c^4  + 
}
\nonumber \\
& & 
+ 2\cos{\theta_c} \sqrt{E_1^2-m_1^2c^4}\sqrt{E_2^2-m_2^2c^4} 
\biggr)
^{1/2} \; ,
\label{E1}
\end{eqnarray}
where $c$ denotes the speed of light.

\begin{table}
\begin{tabular}{|c|c|c|c|c|c|}
\hline \hline & Species & $E_{b}$, GeV & $C$, m & $\Lumi_{peak}^{max}$ & Years \\
\hline
AdA & $e^+e^-$ & 0.25 & 4.1 & $10^{25}$  & 1964 \\
VEP-1 & $e^-e^-$ & 0.16 & 2.7 & $5\times10^{27}$  & 1964-68 \\
CBX & $e^-e^-$ & 0.5 & 11.8 &$ 2\times10^{28}$  & 1965-68 \\
VEPP-2 & $e^+e^-$ & 0.67 & 11.5 &$ 4 \times10^{28}$  & 1966-70 \\
ACO & $e^+e^-$ & 0.54 & 22 & $10^{29}$  & 1967-72 \\
ADONE & $e^+e^-$ & 1.5 & 105 & $6\times10^{29}$  & 1969-93 \\
CEA & $e^+e^-$ & 3.0 & 226 & $0.8\times10^{28}$  & 1971-73 \\
ISR & $pp$ & 31.4 & 943 & $1.4\times10^{32}$  & 1971-80 \\
SPEAR & $e^+e^-$ & 4.2 & 234 & $1.2\times 10^{31}$  & 1972-90 \\
DORIS & $e^+e^-$ & 5.6 & 289 & $3.3\times10^{31}$  & 1973-93 \\
VEPP-2M & $e^+e^-$ & 0.7 & 18 & $5\times10^{30}$  & 1974-2000 \\
VEPP-3 & $e^+e^-$ & 1.55 & 74 & $2\times10^{27}$  & 1974-75 \\
DCI & $e^+e^-$ & 1.8 & 94.6 & $2\times10^{30}$  & 1977-84 \\
PETRA & $e^+e^-$ & 23.4 & 2304 & $2.4\times10^{31}$  & 1978-86 \\
CESR & $e^+e^-$ & 6 & 768 & $1.3\times10^{33}$  & 1979-2008 \\
PEP & $e^+e^-$ & 15 & 2200 & $6\times10^{31}$  & 1980-90 \\
S$p\bar{p}$S & $p \bar{p}$ & 455 & 6911 & $6\times10^{30}$  & 1981-90 \\
TRISTAN & $e^+e^-$ & 32 & 3018 & $4\times10^{31}$  & 1987-95 \\
Tevatron & $p \bar{p}$ & 980 & 6283 & $4.3\times 10^{32}$  & 1987-2011 \\
SLC & $e^+e^-$ & 50 & 2920 & $2.5\times10^{30}$  & 1989-98 \\
LEP & $e^+e^-$ & 104.6 & 26660 & $10^{32}$  & 1989-2000 \\
HERA & $ep$ & 30+920& 6336 & $7.5\times10^{31}$  & 1992-2007 \\
PEP-II & $e^+e^-$ & 3.1+9 & 2200 & $1.2\times10^{34}$  & 1999-2008 \\
KEKB & $e^+e^-$ & 3.5+8.0 & 3016 & $2.1\times10^{34}$  & 1999-2010 \\
\hline
VEPP-4M & $e^+e^-$ & 6 & 366 & $2\times10^{31}$  & 1979- \\
BEPC-I/II & $e^+e^-$ & 2.3 & 238 & $10^{33}$  & 1989- \\
DA$\Phi$NE & $e^+e^-$ & 0.51 & 98 & $4.5\times10^{32}$  & 1997- \\
RHIC & $p,i$ & 255 & 3834 & $2.5\times10^{32}$  & 2000- \\
LHC & $p,i$ & 6500 & 2669 & $2.1\times10^{34}$  & 2009- \\
VEPP2000 & $e^+e^-$ & 1.0 & 24 & $4\times10^{31}$  & 2010- \\
S-KEKB & $e^+e^-$ & 7+4 & 3016 & $8\times10^{35}$ $^*$ & 2018- \\
\hline \hline
\end{tabular}
\caption{Past and present particle colliders: their particle species, maximum beam energy $E_b$, circumference or length $C$, maximum luminosity  $\Lumi$, and years of luminosity operation ($i$ is for ions; $^*$ design; luminosity is in units of cm$^{-2}$s$^{-1}$, it is defined in Eq.(\ref{eq:lumi}) and discussed below.)} \label{T1}
\end{table}

For many decades throughout the first half of the 20th century, the only arrangement for accelerator experiments involved a fixed-target setup, where a beam of charged 
particles accelerated with a particle accelerator hit a stationary target set into  the path of the beam. In this case, as follows from Eq.~(\ref{E1}), for high energy accelerators $E \gg { mc^2}$, $E_{cme} \approx \sqrt{2 E \times mc^2}$. For example, the collision of $E_b$=7000 GeV protons with stationary protons $mc^2\approx$1 GeV can produce reactions with $E_{cme}$ of about 120 GeV. A more effective colliding beam set-up, in which  two beams of particles are accelerated and directed against each other, offers a much higher center of mass energy of $E_{cme}\approx 2 \sqrt{E_1E_2}$, assuming a typically small or zero crossing angle $\theta_c\approx 0$. In the case of two equal masses of colliding particles (e.g., ~protons and protons, or protons and antiprotons) with the same energy of 7000 GeV, one obtains $E_{cme}=2E_b$ or 14 000 GeV. Several machines operate with beams of unequal energies, either because the colliding particles have different masses (electron-proton collisions at HERA) or because of the need to generate new short-lived particles, such as $B$ mesons, with a Lorentz boost so as to more easily detect and analyze their decays (asymmetric $B$-factories KEKB, PEP-II, and SuperKEKB).

In total, 31 colliders have so far reached the operational stage  (some in several successive configurations) and seven of these are operational now (2019) --- see Table \ref{T1}. These facilities essentially shaped modern particle physics \cite{hoddeson, ellis, barger}. The idea of exploring collisions in the center of mass system to fully exploit the energy of accelerated particles was first given serious consideration by the Norwegian engineer and inventor Rolf Wider\"{o}e, who in 1943 had filed a patent for the collider concept (and received the patent in 1953) \cite{Wideroe, waloschek2013}. This idea was further developed by Donald Kerst \cite{kerst} and Gerry O\textsc{\char13}Neill \cite{o'neill}, and in the late 1950s three teams started working on colliding beams: (i) a Princeton-Stanford group in the US that included William Barber, Bernard Gittelman, Gerry O\textsc{\char13}Neill, and Burton Richter, who in 1959 proposed building a couple of tangent rings to study M\o ller scattering $e^-e^- \rightarrow e^-e^-$ (Stanford colliding-beam experiment CBX \cite{barber}); (ii) a somewhat similar project initiated by Gersh Budker in the Soviet Union, where an electron-electron collider VEP-1 was under construction in 1958 \cite{budker1962}; and (iii) an Italian group at the Laboratori Nazionali di Frascati, led by Bruno Touschek, which began the design of the first electron-positron collider AdA \cite{bernardini1960}. In the early 1960s, almost concurrently, these first colliders went  into operation in the Soviet Union \cite{budker1967, levichev2018}, France (to where the AdA had been moved) \cite{bernardini1964, bernardini2004}, and the USA \cite{gittelman1965, rees1986}. 

\begin{figure}[htbp]
\centering
\includegraphics[width=0.99\linewidth]{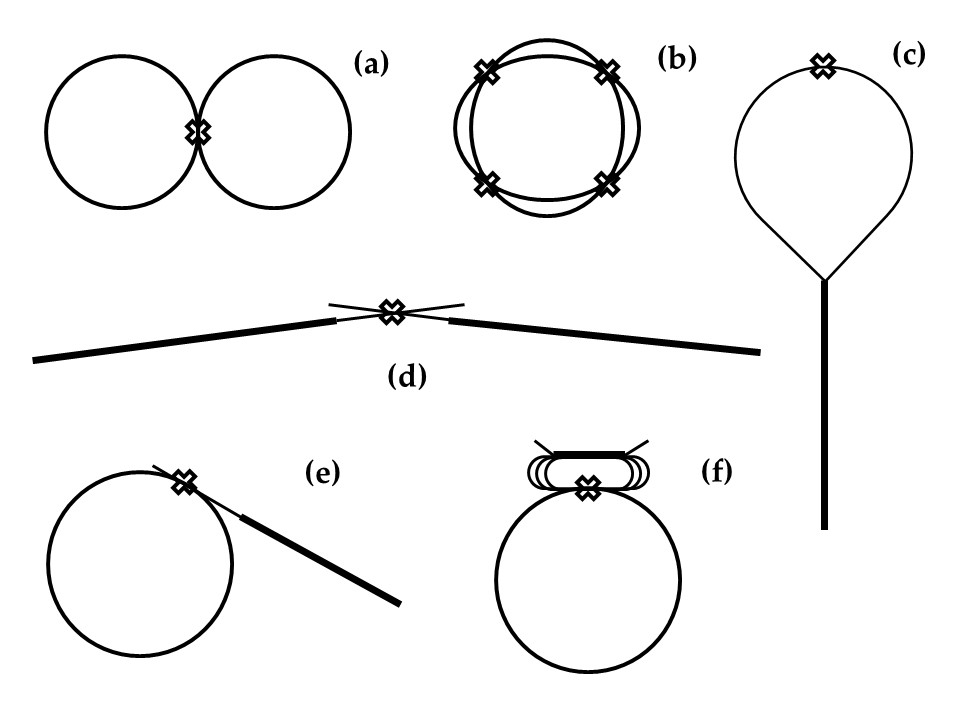}
\caption{Schematics of particle collider types.} \label{fig:types}
\end{figure}

Figure~\ref{fig:types} presents the most common arrangements of colliding beams. In storage ring configurations - Fig.~\ref{fig:types}a and \ref{fig:types}b $--$ particles of each beam circulate and repeatedly collide. Historically, a single ring was often used for colliding particle and antiparticle beams of equal energy. Modern and future storage-ring colliders (e.g.~LHC, DA$\Phi$NE, BEPC-II, FCC, CEPC, SppC, etc.) utilize double rings to achieve
extremely high luminosity by colliding a large number of bunches. 
The two rings may store particles of the same type, or particles and their antiparticles, or two different particle types, 
like electrons and hadrons.  
In linear colliders, first proposed in Ref.~\cite{tigner1965} and then 
further developed for higher energy, 
e.g.~in Refs.~\cite{Amaldi:1975hi,vlepp79}, the two colliding beams are accelerated in linear accelerators (linacs) and transported to a collision point, either with use of the same linac and two arcs as in Fig.~\ref{fig:types}c, or in a simple two-linac configuration as depicted in Fig.~\ref{fig:types}d.   
Other configurations are possible and were considered: e.g., linac-ring schemes as in Fig.~\ref{fig:types}e or collision of beams circulating in a ring and a few-pass energy recovery linac (ERL) (Fig.~\ref{fig:types}f). 

\begin{figure}[htbp]
\centering
\includegraphics[width=0.99\linewidth]{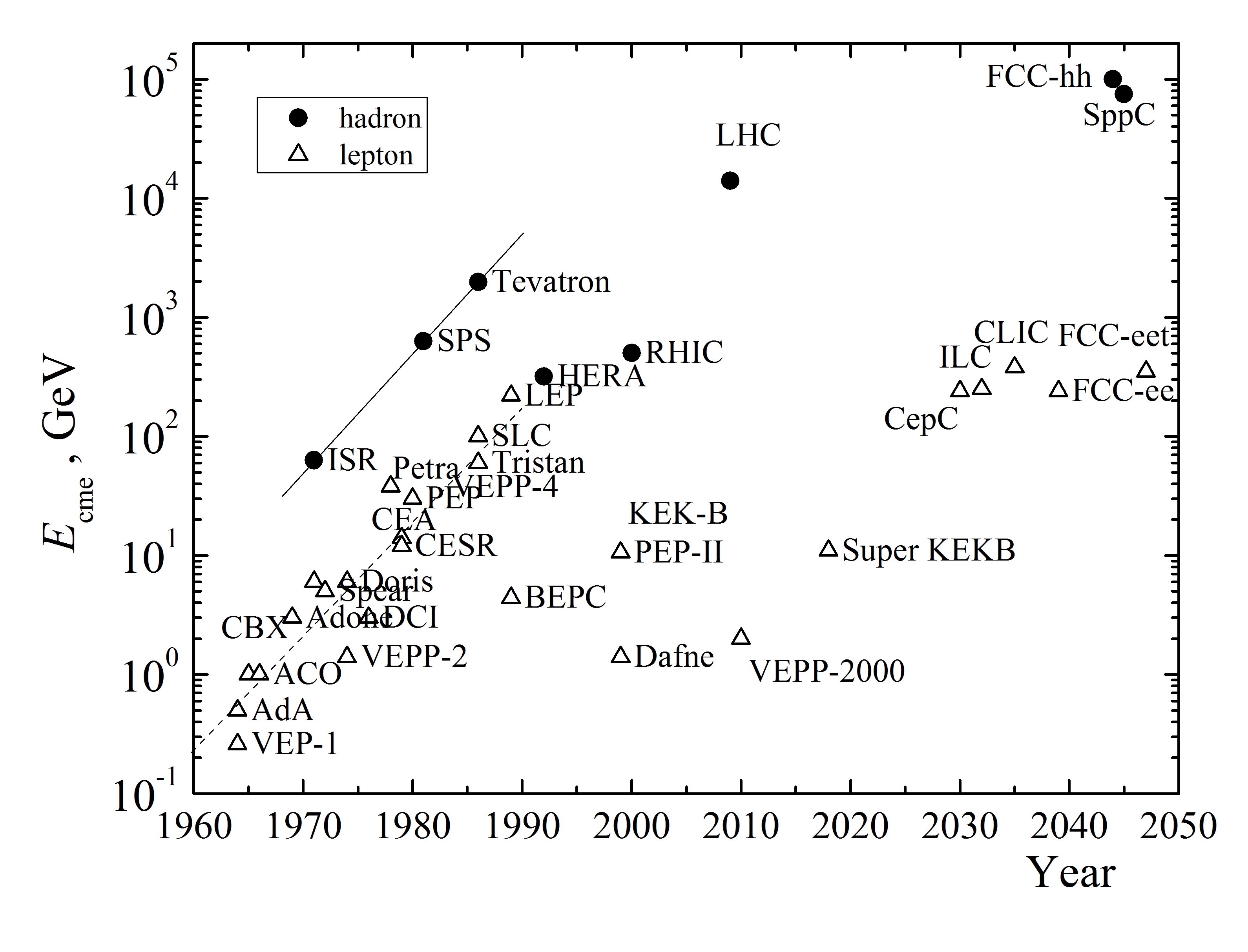}
\caption{Center of mass energy reach of particle colliders vs their start of operation. Solid and dashed lines indicate a ten-fold increase per decade for hadron (circles) and lepton (triangles) colliders (adapted from \cite{shiltsev2012}).}
\label{fig:colliders_E}
\end{figure}

In contrast to other types of accelerators, which have many diverse applications, colliders have exclusively served the needs of frontier particle physics research (or what is 
nowadays called high-energy physics (HEP) and nuclear physics).
The ever-growing demands of particle physics research drove the increase in energy of colliders, as is demonstrated in Fig.~\ref{fig:colliders_E}. In this figure, the triangles represent maximum c.m.e.~and the start of operation for lepton colliders (mostly $e^+e^-$), while full circles represent hadron (protons, antiprotons, ions, proton-electron) colliders. One can see that until the early 1990s, the c.m.e.~increased on average by a factor of ten every decade. Notably, hadron colliders were 10--20 times more energetic (though hadrons are not elementary particles and only a fraction of their energy is available to produce new particles during collisions). Since then, the paths of different colliders diverged: hadron colliders continued the quest for record high energies in particle reactions and the LHC was built at CERN, while in parallel, highly productive $e^+e^-$ colliders called \textit{particle factories} focused on precise exploration of rare phenomena at much lower energies.

\begin{figure}[htbp]
\centering
\includegraphics[width=0.99\linewidth]{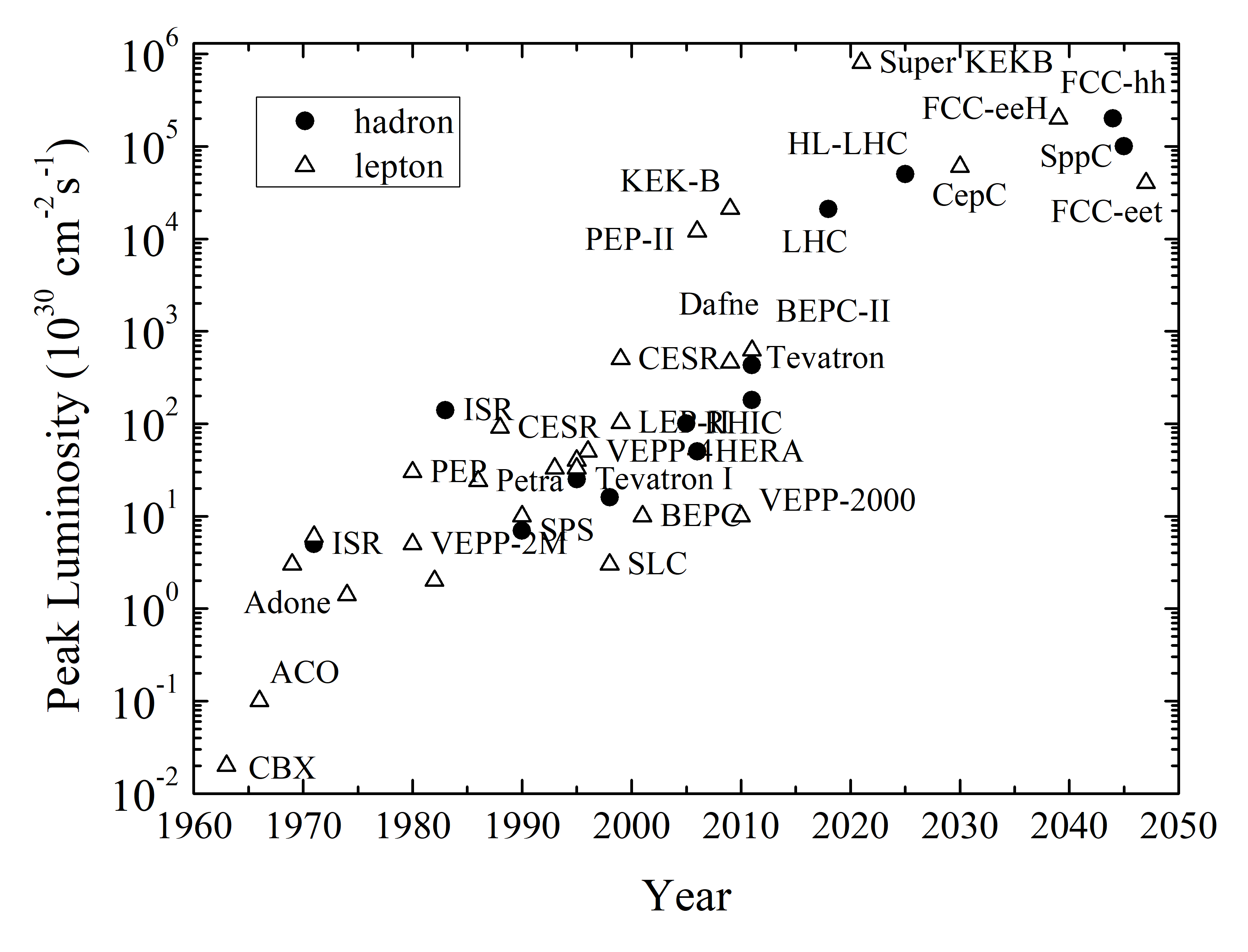}
\caption{Luminosities of particle colliders (triangles are lepton colliders and full circles are hadron colliders, adapted from \cite{shiltsev2012}).}
\label{fig:colliders_L}
\end{figure}

The exploration of rare particle physics phenomena requires not only an appropriately high energy, but also a sufficiently large number of detectable reactions. The number of events of interest $N_{exp}$ is given by 
the product of the cross section of the reaction under study, $\sigma_{\textup{exp}}$, and the time integral over the instantaneous {\it luminosity}, $\Lumi$:
\begin{equation}
N_{\textup{exp}} =  \sigma_{\textup{exp}} \cdot \int \Lumi (t) dt.
\label{eq:intlumi}
\end{equation}
The luminosity dimension is [length]$^{-2}$[time]$^{-1}$. The integral on the right is referred to as integrated luminosity $\Lumi_{\rm int}$, and, reflecting the smallness of typical particle-interaction cross-sections, or the correspondingly high integrated luminosity 
required, 
is often reported in  units of inverse pico-, femto- or attobarn, where 1 barn=10$^{-24}$~cm$^2$, e.g., 1 fb=10$^{-39}$~cm$^{2}$. In fixed target mode,  luminosity is a product of the extracted high-energy particle flux rate times target density and length. 
Collider luminosity is defined only by the beam parameters and is critically dependent on beam densities, which are typically orders of magnitude lower than those of liquid or solid targets. Colliders usually employ bunched beams of particles with approximately Gaussian distributions, and for two bunches containing $N_1$ and $N_2$ particles colliding head-on with frequency $f_{\rm coll}$, a basic expression for the 
luminosity is 
\begin{equation}
    \Lumi = f_{\rm coll} { N_1  N_2\over 4\pi\sigma_x^{\ast} \sigma_y^{\ast}}\; ,
\label{eq:lumi}
\end{equation}
where $\sigma_x^{\ast}$ and $\sigma_y^{\ast}$ characterize the rms transverse beam sizes in the horizontal and vertical directions at the point of interaction \cite[Ch.~6.4]{myers2013accelerators}. To achieve high luminosity, one therefore has to maximize the  population and number of bunches, either producing these narrowly or focusing them tightly and colliding them at high frequencies at  dedicated locations where products of their reactions can be registered by detectors. Figure \ref{fig:colliders_L} demonstrates the impressive progress in luminosity of  colliding beam facilities since the invention of the method --- over the last 50 years, the performance of colliders has improved by more than six orders of magnitude and reached record high values of over $10^{34}$~cm$^{-2}$s$^{-1}$. At such a luminosity, one can expect to produce 5 million events over one year of operation (effectively, about $10^7$ s) for a reaction cross section is 50 picobarn (pb)=$10^{-36}$~cm$^2$. An example process with this magnitude is Higgs particle production $pp\rightarrow H+X$ at $14$ TeV c.m.e.~in the LHC  \cite{lhchiggs}. 

Luminosity considerations differ significantly for lepton and hadron colliders. For point-like colliding particles such as leptons, the reach of the collider, defined as the highest mass $M$ of a particle that can be produced there, just equals $E_{cme}/c^2$. 
Due to the ``1/$s$'' scaling of the Feynman propagator for hard-scattering processes
(here, $s = (p_1+p_2)^2$ again 
is the Mandelstam variable, with $p_1$ and $p_2$ denoting the four-momenta of the two incoming particles),
 the production cross-section of $M$ is  proportional to $E_{cme}^{-2}$ \cite{henley}. To detect new particles of increasing mass,  integrated luminosity should increase as $E_{cme}^2$.  As hadrons are quark$-$gluon composite objects, the probability of creating a new mass $M$ depends on the Quantum Chromodynamics (QCD) parton distribution functions in the nucleon; the corresponding cross sections scale as $\sigma_{\textup{exp}} \propto E_{cme}^{-2}f(Mc^2/E_{cme})$, where $f(x)$ is a sharply falling function \cite{eichten1984partonluminosities, quigg2011}. In consequence, the collider mass reach is a strong function of $E_{cme}$ and a rather 
weak function of $\Lumi_{\rm int}$ \cite{salam2017}. For example, with a rough approximation of  $f(x)\propto x^{-6}$, the mass discovery reach of a hadron collider scales as $M \propto E_{cme}^{2/3} \cdot \Lumi_{\rm int}^{1/6}$ \cite{teng2001}. This peculiar characteristics of very-high-energy hadron colliders has been proven again 
and again in the past (see the next section), 
and it is often invoked to qualify them as ``discovery machines''. 
In general, the key components of the experimental program toward understanding of the structure of matter are particle detectors and 
accelerators. Remarkable advances in the detector technology and key challenges of the detectors for future hadron and lepton colliders are outside the scope of this review and can be found in review publications and textbooks such as \cite[Ch.~34]{pdg2018},  \cite{green2000,grupen2008}, \cite[Ch.~11]{eppsu2020granada}.  

Over the past five decades, the quest for higher energy particles led to some five orders of magnitude boost of collider energies and an even greater increase in their luminosities --- see Figs.~\ref{fig:colliders_E} and \ref{fig:colliders_L}. Simultaneously, the size, complexity and cost of colliding beam facilities have also increased. 
Modern colliders employ numerous technologies for tunneling, geodesy and alignment, power converters and power supplies, ultra-high vacuum systems, particle sources, injection and extraction systems, cooling water and cryogenic cooling, beam diagnostics,  accelerator control, personnel safety and machine protection, among others. While at the dawn of accelerator and collider technology, most of these required dedicated 
and often pioneering developments, 
nowadays many such technologies are widely available from specialized industries. 
Still left almost solely to the pursuit of 
accelerator engineers and scientists are the ``core technologies'' required for 
accelerating particles to high energies ---
normal- and superconducting radio-frequency (RF) acceleration systems, and normal- and superconducting accelerator magnets --- and ``beam physics techniques'' to attain the necessary beam qualities such as intensity, brightness and sometimes polarization, including  beam cooling, manipulation and collimation, the production of exotic particles like antiprotons or muons, mitigation of beam instabilities, and countermeasures against beam-size blow up caused by space-charge (SC) and beam-beam effects, or intra-beam scattering (IBS), among others. 
The energy reach of a collider is mostly defined by core accelerator technologies, while its luminosity is grossly dependent on the sophistication of beam physics techniques. 

Being arguably the biggest and the most technologically advanced scientific instruments, colliders were and remain costly, often at the brink of financial and political affordability. That makes them prone to various risks and, in the past, several have been terminated, even after the start of construction. Most notable in this respect 
are energy frontier hadron colliders. In 1983, the construction of the 400 GeV c.m.e.~ISABELLE $pp$ collider (briefly renamed CBA) at the Brookhaven National Laboratory in the USA was stopped  \cite{month2003, crease2005a, crease2005b}, and in the early 1990s two other flagship projects were terminated: the 6 TeV c.m.e.~proton-proton complex UNK \cite{yarba1990, kuiper1994} in Protvino, Russia, and the 40 TeV c.m.e.~proton-proton Superconducting Super Collider (SSC) in Texas, USA, in 1993 \cite{wojcicki2009, riordan2015}.
Notwithstanding the above, advances in core accelerator technologies --- including the SC magnet developments for ISABELLE/CBA, UNK and SSC  --- have led to substantial reductions in collider cost per GeV \cite{shiltsev2014}. This progress, together with the growing strength of the high-energy particle physics community, enabled development of frontier machines, such as the currently operational multi-billion dollar LHC.
Even larger \$10B-scale future collider projects are generally considered feasible (see the following sections). 

On average, the colliders listed in Table \ref{T1} operated for 13 years, with many remarkable facilities operating for even  twice that time (Adone, VEPP-2, CESR, Tevatron, VEPP-4M, BEPC-II). Contrary to other research accelerators, such as light sources, where user groups and experiments are numerous and each might take as little beam time 
as weeks or a few days, over their lifetime most of these colliders served just one, two, or four permanently installed 
particle detector experiments surrounding the beam collision points. For example, PETRA, TRISTAN, LEP, RHIC and the LHC each had (or have) 
four main collision points and detectors \cite{fernow1989, hauptman2011particle}. The colliding-beam facilities usually consist of several machines needed to prepare and accelerate the beams and are generally quite complex, featuring several layers of structural hierarchy --- numerous primary elements, combined in technical subsystems, composed in individual accelerators, highly interconnected and working synchronously as one complex. The largest of these require hundreds of highly skilled personnel for operation, including a 
sizable number of PhD physicists. The complexity and scale of the colliders result in substantial lengths of time, usually many years, being required for full commissioning and for attaining the ultimate luminosities \cite{shiltsev2011a}. It is characteristic for colliders to continuously proceed through a series of minor operational improvements, interleaved with a few major upgrades, and to see their performance increase all through their lifetimes. 

Particle physics has not yet fully exploited the potential of the colliding-beam  technique and is largely betting its future on it \cite{ellis2018future}. 
The current consensus is that ``...no other instrument or research programme can replace high-energy colliders in the search for the fundamental laws governing the universe'' \cite{giudice2019}.

In Section \ref{History} below, we briefly outline the development of colliders and the corresponding core accelerator technologies and beam physics techniques. 
Seven currently operational collider facilities will be described and discussed in Section \ref{moderncoll}. The next generation of colliders, believed to be technically feasible and affordable, and which could be constructed over the next two or three decades, is the subject of Section \ref{futurecolliders}. Finally, in Section \ref{advancedcolliders}, we assess opportunities offered by emerging accelerator technologies and attempt to look beyond the current horizon and outline possible changes in the collider paradigm which might enable far-future, ultimate colliders.

\section{Development of colliders}
\label{History}

Modern and future colliders are extensively based on the accelerator technology and beam physics techniques developed and appraised by their predecessors.  In this section, we introduce and elaborate on 
major collider issues from a historical perspective. 
More detailed considerations and comprehensive lists of references can be found, e.g., in \cite{chao2013handbook, myers2013accelerators, bruning2016challenges}.

In an accelerator, charged particles gain energy from an electric 
field, which usually varies in time 
at a  high frequency ranging from 10s of MHz to 10s of GHz. 
The accelerating field gradients in RF cavities are 
usually orders of magnitude higher than in direct-current (DC) 
systems; RF cavities are, therefore, commonly used in colliders. 
At present, the highest beam accelerating gradients ever achieved in operational machines or beam test facilities are some $G\approx 100$~MV/m in 12 GHz normal-conducting (NC) RF cavities  \cite{senes2018} 
and 31.5 MV/m in 1.3 GHz superconducting RF (SRF) ones \cite{broemmelsiek2018}. 
In a linear-collider arrangement, illustrated in Figs.~\ref{fig:types}c, d and e, the beam energy $E_b$ is the product of the average accelerating gradient $G$ and the length of the linac $L$:
\begin{equation}
    E_b = e G \cdot L
    \; ,
\label{eq:energy_l}
\end{equation}
where $e$ denotes the elementary (electron) charge, assuming the acceleration
of singly charged particles like electrons or protons. 
For example, reaching 1 TeV energy requires either 
10 km of NC RF accelerator or 
$\sim$30 km of SRF linac, if the RF cavities occupied all available space --- which they do not.  
Cost considerations (see below) often call for minimization of RF acceleration, e.g., through repeated use of the same RF system which in that case would boost the energy in small portions $\Delta E_b=eV_{RF}$ per turn every time a particle passes through the total cavity voltage $V_{RF}$. Such an arrangement 
can be realized both in the form of circular colliders (Fig.~\ref{fig:types}a, b), which have proven extremely  successful, and also through novel schemes based on ERLs  (Fig.~\ref{fig:types}f). 
Circular colliders are most common; here, the momentum and energy of ultra-relativistic particles are determined by the bending radius inside the dipole magnets, $\rho$, and by the average magnetic field $B$ of these magnets: 
\begin{equation}
    p=e B \cdot \rho \quad \textup{or} \quad E_b\; \textup{[GeV]} = 0.3 (B\rho)\; \textup{[Tm]} \; . 
\label{eq:energy_c}
\end{equation}
In such a scheme, the field $B$ needs to be synchronously increased to track particle energy gains after each passage through the accelerating RF cavities. Such synchrotron conditions allow the beam orbit to remain inside the rather limited space provided by the accelerator beam pipe passing   
through the magnet apertures.  
The maximum field of NC magnets is about 2 Tesla (T), 
due to the saturation of ferromagnetic materials, and while this 
is sufficient for lower energy colliders, such as most  $e^+e^-$ storage rings, it is not adequate for frontier-energy hadron (or muon) beams, because of the implied 
need for excessively long accelerator tunnels and prohibitively high total magnet power consumption. The development of superconducting (SC) magnets that employ high electric current carrying Nb-Ti wires cooled by liquid helium below 5~K, opened up the way towards higher fields and to hadron colliders at record energies \cite{tollestrup2008}. The latest of these, the 14 TeV c.m.e.~LHC at CERN, uses double bore magnets with a maximum field of 8.3 T at a temperature of 1.9 K, 
in a tunnel of $C=26.7$ km circumference (dipole-magnet bending radius $\rho=2,800$~m).

\subsection{Basic technologies and beam physics principles}
\label{basics}

\subsubsection{Magnets and RF structures}
\label{technology} 
Magnets form the core of all types of colliders. Besides bending magnets, several other field shapes are required in order to focus and control the beams and manipulate 
beam polarization. Accelerator magnets typically are long (up to a few m), and feature transversely small apertures (few cm), which accommodate the beam vacuum pipes. The magnetic field components are normally oriented in the $(x,y)$ plane of the magnet cross section. In such a 2D configuration, the most common representation of the field is given by a complex multipole expansion:  
\begin{equation}
B_y+iB_x = \sum_{n=1}^{\infty} (B_n +i A_n) (x+iy)^{n-1} \; ,
\label{Bmult}
\end{equation}
 where $B_n$ and $A_n$ represent the normal and skew multipole components 
 of the field, and $2n$ signifies the number of poles. For example, in an ideal horizontally deflecting, normal dipole magnet ($n=1$), we have $B_y=B_1$ and $B_x=0$. 
 
 For an ideal quadrupole magnet ($n=2$), the fields are $B_y=B_2x$, $B_x=B_2y$. So this type of magnet can be used as a focusing element as it deflects a particle proportionally to its transverse offset $x$ (or $y$)  from the magnet axis. 
 Namely, to first approximation, we have $\Delta x'=Kx$, where 
 $x'\equiv p_{x}/p_{s}$ is the slope of the particle trajectory (horizontal momentum $p_{x}$ 
 divided by the longitudinal momentum $p_{s}$), 
 $\Delta x'$ the change in slope after passing through the
 quadrupole, and $K$
 the normalized strength of a quadrupole of length $l$, 
 which is given by 
 $K=B_2l/(B\rho)$, with the magnetic rigidity $(B\rho) = p_{s}/e$.  
 
 Higher order multipole magnets, such as sextupoles ($n=3$) and octupoles ($n=4$), are also widely used, e.g., to control an accelerator's chromaticity --- the dependence of its focusing property on particle momentum --- and for beam stabilization, respectively. 
 Other commonly employed magnets are wigglers and undulators, 
 sequences of short dipole magnets with alternating polarity,  
 which yield a periodic field variation along the beam trajectory, causing the beam to wiggle and to lose energy, emitting electromagnetic radiation \cite{clarke2004science}. High-field, few T solenoids are commonly deployed in collider detectors \cite{yamamoto2003detectors}; solenoid magnets are also used for spin rotation and beam polarization control \cite{barber1984solenoid}; and for focusing of mostly lower-energy 
 beams, e.g., in electron coolers \cite{parkhomchuk2000electron} 
 and injectors \cite{Carlsten:1995wn}. 
 
 \begin{figure}[htbp]
\centering
\includegraphics[width=0.99\linewidth]{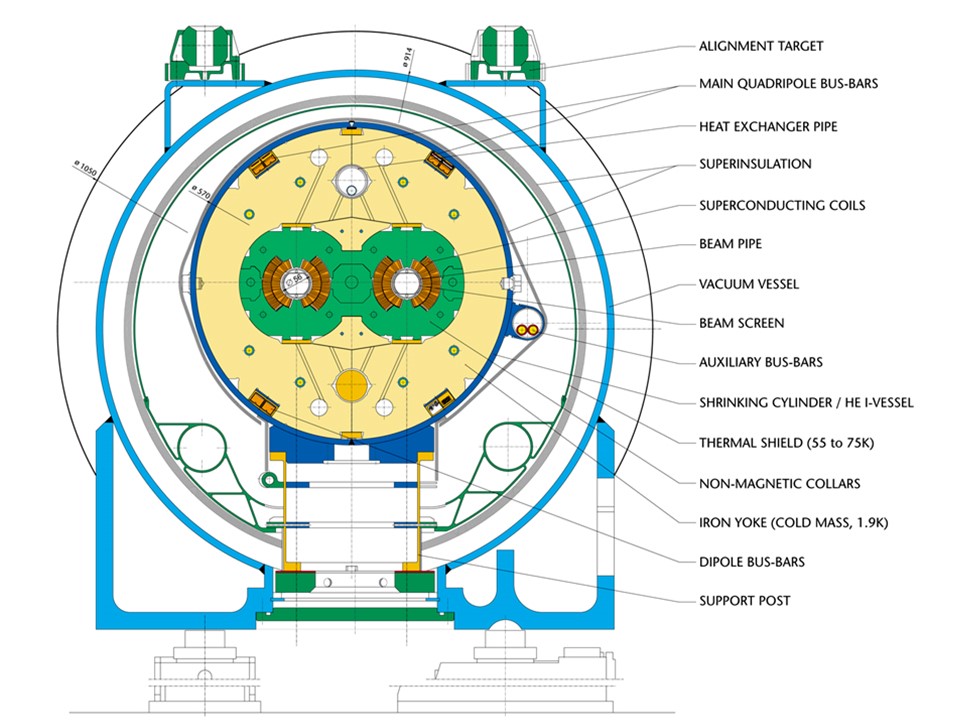}
\caption{Cross section of the 14.3 m long superconducting magnet of the Large Hadron Collider \cite{evans2016lhcdipolecrosssection}. 
The design field of 8.33 T is vertical and opposite in the 
two 56 mm diameter bores for the two counter-rotating beams,
with a horizontal beam-to-beam separation of 194 mm. The LHC comprises
1232 main dipoles, each weighing about 35 tons.}
\label{fig:LHC_dipole}
\end{figure}
 
Collider beam dynamics is highly sensitive to magnet field quality, understood as the relative deviation of the actual field from its ideal design value, 
and requires the unwanted components $(B_n, A_n)$ to be of the level of
a few $10^{-4}$ of the main field, e.g., of the corresponding primary dipole or quadrupole field, 
and to be even smaller for a few special, strong magnets used to ultimately focus or transversely compress the beams at the collider's interaction points (IPs). 
In normal-conducting (NC) magnets, 
a steel or iron yoke is employed to direct and shape the magnetic field inside 
the magnet aperture, so that the field quality is usually assured by the proper shaping of the magnet poles \cite{tanabe2005iron}. 
For field levels above 1.7--2.0 T, as are typical for NC magnets,
such an approach no longer works.  
However, significantly higher fields can be obtained with superconducting (SC) magnets. 
In SC magnets, the iron of the yoke does not play a major role in field formation. Instead, the achievement of the target field quality 
requires a conductor-coil placement accuracy and position stability of less than a few dozen micron, which is to be obtained 
while the coil is subjected to enormous magnetic forces, 
sometimes exceeding a million Newton per meter \cite{mess1996scmag}  \cite[Ch.8.1]{myers2013accelerators}. 

In addition, ramping the SC magnets induces so-called {\it persistent currents} inside the superconducting cables, which can result in dangerous systematic sextupole field components of order $B_3/(B_1 R_0^{-2}) \sim 20 \times 10^{-4}$ (where $R_0$ denotes a reference radius for the good field region, typically chosen as 
about 2/3 of the magnet's aperture) \cite{tollestrup2008}. These and other time-dependent effects require sophisticated systems of weak, but numerous, corrector magnets, adding to the complexity of collider operation, 
while assuring its efficiency. There are many other difficulties related to the operation of SC magnets, such as the need for cooling by liquid helium, 
quench detection and protection, alternating-current (AC) losses, and 
the careful control of the mega- to giga-joules of stored magnetic field energy; 
all of these have been generally resolved and do not outweigh the major advantages of superconductive systems, namely, a few orders of magnitude lower electric wall-plug power consumption and the ability to generate much higher magnetic fields in Nb-Ti based magnets, of up to 9 T, as demonstrated in the Tevatron (4.5 T), HERA (4.7 T), RHIC (3.5 T) and LHC (8.3 T). Even higher fields, of up to 12--16 T, can be achieved with Nb$_3$Sn conductor \cite{zlobin2019nb3sn}, and over 20 T are expected with certain high temperature superconductors (HTS) \cite{rossibottura2012scmag}. 
It should be noted that the field values cited above are lower than the critical fields of these materials, since the operation of systems of many (hundreds to thousands) accelerator magnets demands significant margins in temperature and critical current to achieve an acceptable stability in an environment characterized by powerful heat sources, due to beams circulating just a few cm away from SC coils (irradiation by local beam loss, vacuum pipe heating due to electron cloud effects and image currents, synchrotron radiation, etc.) --- see Fig.~\ref{fig:LHC_dipole}. 
\begin{figure}[htbp]
\centering
\includegraphics[width=0.99\linewidth]{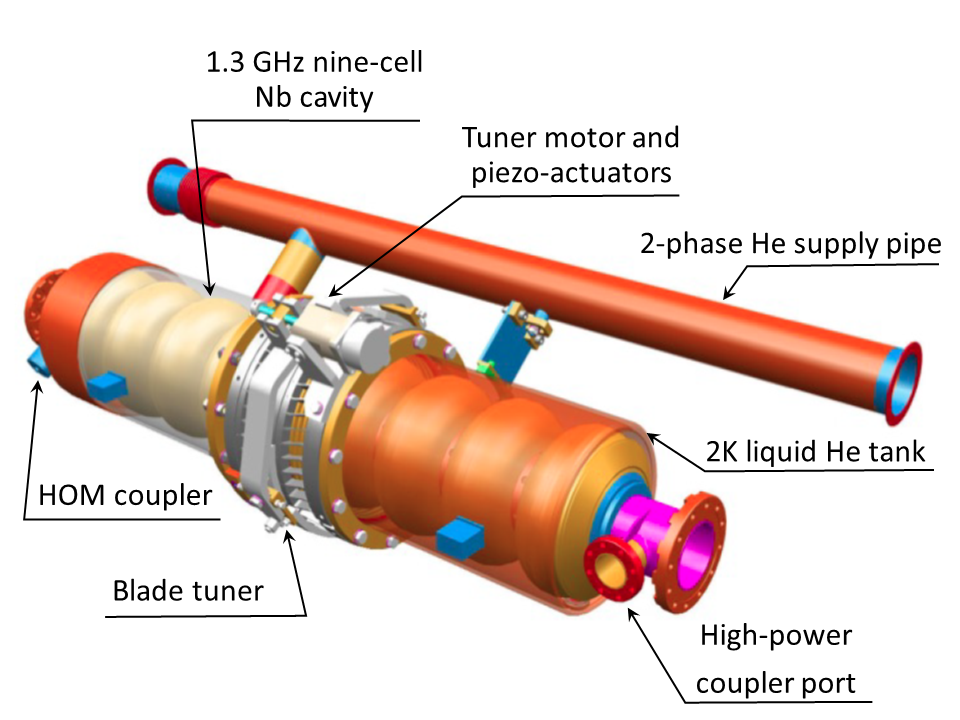}
\caption{The baseline superconducting cavity package (“dressed cavity”) of the International Linear Collider; the titanium helium tank is shown as transparent for a view of the 1 m long 9-cell niobium RF cavity inside (adapted from \cite{phinney2007ilc}).}
\label{fig:ILC_cavity}
\end{figure}

The RF systems of colliders are mostly needed to increase or maintain the particles' energy using time-varying longitudinal electric fields. They typically operate at carefully preselected frequencies $f_{RF}$ in the range of several tens of MHz to tens of GHz and consist of three main elements --- power converter, RF amplifier, and RF cavity --- together with control loops (low level RF subsystem, including the master oscillator) and ancillary systems: water cooling, vacuum, and cryogenics in the case of superconducting RF (SRF) cavities. The RF system is essentially a device that transforms electrical energy taken from the grid into energy transferred to a beam of particles in three major steps, each with its own technology and efficiency: 1) the transformation of AC power from the electric grid (alternate, low voltage, high current) to DC power (continuous, high voltage (HV), low current) that takes place in a power converter with some 90\% efficiency; 2) the transformation of the DC power into RF power (high-frequency) that takes place in an RF power source: RF tube, klystron, transistor, etc., with efficiency in the range of 50--70\% or more, depending on the specific device and mode of operation (continuous wave (CW) or pulsed); 3) transformation of RF power into the particle beam power gain that takes place in the gap of an accelerating cavity, with efficiencies that may reach 30\% or $\ge$50\% 
in a pulsed NC or SC linac \cite{bailey2010casRF} or be approximately 100\% in case of a continuous wave (CW) SRF system for a storage ring.  
Thanks to cost saving considerations, all these efficiencies have been constantly improving with increasing RF power demands, which for large modern and future colliders can be as high as dozens or even several hundreds of MW \cite{jpdmuon}.  

The energy gain of a particle traversing an RF cavity is  
\begin{equation}
\Delta E_b =  e \int \bm{E} \bm{v} dt = eV_{acc} \cos{(\omega_{RF}t+\phi)} \, , 
\label{Eacc}
\end{equation}
where $\bm{E}$, $V_{acc}$, $\omega_{RF}=2\pi f_{RF}$, and $\phi$ denote the cavity's electric field, accelerating voltage, frequency and phase, respectively, and $\bm{v}$ is the particle's velocity, usually parallel to the accelerating field, at the time of passage $t$. 
For synchrotrons and storage rings, 
the condition of synchronicity over subsequent acts of acceleration calls for the RF frequency $f_{rev}$ to be a harmonic of the revolution frequency $f_{RF}= h f_{rev}$, with the integer $h$ known as the harmonic number. 
The RF power supplied to the cavity from the source goes into the increase of beam power, $P_{b}$, and into sustaining an  
accelerating field that otherwise would decay, $P_{loss}$, 
due to the finite cavity surface conductivity: 
\begin{equation}
P_{RF} = P_{b}+P_{loss}=I_b\Delta E_b + \frac {V_{acc}^2}{2 R_s} \; . 
\label{Prf}
\end{equation}
Here, $I_b$ is the beam current and $R_s$ is the so-called {\it shunt impedance} (the resonant resistance of an equivalent RCL circuit) $R_s=Q \cdot (R/Q)$ is the product of the RF cavity's quality factor $Q$, related to the power dissipation on the cavity surface, 
and a factor $(R/Q)$, depending only on the cavity geometry. 
Typical $Q$ values for NC cavities are in the range of some 10$^{4}$, while they can reach a few 10$^\textup{10}$ in SRF cavities \cite{martinello2018field}. 
The factor $(R/Q)$, which is independent of the cavity size and of the surface resistance, is commonly used as a figure of merit --- see, e.g. \cite{padamsee2008srf}. It typically varies between 196 Ohm per  resonant cell, obtained for a TM$_{\textup{010}}$ mode pillbox cavity with minimal opening for beam passage, and some 100 Ohm per cell for large aperture elliptical cavities, such as used in SRF systems. 

The largest linac built to date was a 3 km long linac at the Stanford Linear Accelerator Center (SLAC), which operated NC copper structures at a frequency of 2.856 GHz (corresponding to RF wavelength of $\lambda_{RF}$=10.5 cm). It provided a total of 50 GeV acceleration in one pass, with average gradient $G=\Delta E_b/eL$ of about 21 MV/m \cite{erickson1984slac}. Only 80 kW of 10 MW of total RF power went into the power of the two colliding beams $2P_{b}$ \cite{phinney2000slc, lavine1992review}. 

Circular colliders are much more efficient due to repetitive energy transfer from RF cavities to beams over many turns, but at highest energies they face a serious impediment in the form of synchrotron radiation. The latter causes an energy loss per turn of 
\cite{Sands:1970ye} 
\begin{equation}
\Delta E_{SR} = \frac{1}{3 
\epsilon_{0}} \frac {e^2 
\beta^{3} \gamma^4}{\rho} \, , 
\label{SR}
\end{equation}
which increases with the fourth power of energy and scales with the inverse of the bending radius.
Inserting numerical values for electrons,  the energy loss of a particle during one revolution becomes $\Delta E_{SR}=$ 0.089 [MeV/turn] $E_b^4$[GeV]/$\rho$[m]. Even for the largest circumference $e^+e^-$ collider LEP with an average bending radius $\rho$ of about 3.1 km, maintaining maximum beam energy at 104.5 GeV required continuous wave (CW) operation of the 353.2 MHz RF system with a voltage of  $eV_{acc} \cos{(\phi_s)}=\Delta E_{SR}=$3.4 GeV per turn, to compensate for about 23 MW total synchrotron radiation beam power loss \cite{myers2000lep}. Adequately, the final LEP RF system consisted of 
272 superconducting Nb sputter-coated Cu cavities and 16 solid Nb cavities,    
with an average gradient of $G=$5--6 MV/m, and 56 lower voltage normal-conducting Cu cavities,
with a gradient of 1.5 MV/m, together provided a maximum total RF voltage of $V_{acc}=3.63$ GV. In the last year of LEP operation the 288 SRF cavities were powered by 36 klystrons with an average power of 0.6 MW \cite{assmannmyers2001,brown2001leprf}. The pure RF-to-beam-power efficiency was $\eta\approx 100\%$. An effective ``RF-to-beam-power efficiency'' of about 75\% was computed by taking into account the additional cryogenic power needs resulting from both RF-related heating and static heat load of the cryostat, in conjunction with the low cooling efficiency at cryogenic temperature \cite{Weingarten:308015}). 
Also adding AC-to-DC conversion and LEP
klystron efficiencies 
\cite{Butterworth:1097407}, plus waveguide losses, the total wall-plug-to-beam-power efficiency, including cryogenics, was close to 50\%.

Besides beam acceleration, RF systems are also 
employed for various other 
beam manipulations, such as longitudinal bunching, bunch compression, splitting, coalescing and flattening \cite{minty2013measurement} and, in some cases, to provide a time-varying  transverse deflection to particle bunches in {\it crab cavities} for linear and circular colliders \cite{palmer1988crabs, oide1989crabs, calaga2009crabs}, as streaking devices for time-varying diagnostics 
\cite{crabstreak},  
 and for bunch separation or bunch combination, e.g.~as RF deflectors in the drive-beam complex of the proposed CLIC linear collider and its past CLIC Test Facility 3 \cite{ctf3defl}.

Synchrotron-radiation power of protons is smaller than for electrons by a significant factor $(\gamma_p / \gamma_e)^4 \propto (m_e/m_p)^4 \approx 1.3 \times 10 ^\textup{13}$ at the same energy --- see Eq.~(\ref{SR}) --- but it can still become a significant concern in  highest-energy, high-current SC accelerators like the LHC. The reason is that synchrotron radiation leads to heating and outgassing of the beam vacuum pipe. The former poses problems for the cryogenic system of the SC magnets, while the latter may impede attainment of vacuum gas pressures of 1--10 nTorr or better, that are needed to guarantee sufficiently long lifetimes of the continually circulating beams. Fortunately, these technological challenges have been successfully resolved in modern colliding beam facilities \cite{barron1985cryogenic, lafferty1998foundations}, \cite[Ch.8.3, Ch.8.5]{myers2013accelerators}. 

For many modern colliders, especially for hadron colliders and linear colliders, the costs of core accelerator components, magnets and RF structures, dominate construction costs, followed by the costs for tunnels, electric power infrastructure and auxiliary systems for ultra-high vacuum, cryogenics, beam control and stabilization, among others. The growing demands for higher-energy beams have motivated a large segment of the accelerator community to search for, and to develop, 
new cost-effective technological concepts and advances. 

\subsubsection{Beam dynamics}
\label{beamdynamics}

Given the enormous and highly concentrated power carried by high energy particle beams, the main concern of beam dynamics in colliders is stability. Below, we briefly introduce major physics phenomena affecting the dynamics of individual particles in accelerators, both single 
high-intensity beams of many particles moving together, and 
also colliding beams. Comprehensive definitions and explanations of these subjects can be found in textbooks \cite{edwards2008introduction, peggs2017introduction,sylee2018accelerator,Chao:1490001}.  

While a reference particle proceeds along the design trajectory (reference orbit) mostly determined by transverse dipole fields, other particles in the bunch are kept close by through the focusing effect of quadrupole fields.  Generally following \cite{edwards2008introduction}, let us  assume that the reference particle carries a right-handed Cartesian coordinate system, with the co-moving $z$-coordinate pointed in the direction of motion along the reference trajectory, $z=s-vt$ (with $v$ the reference particle velocity, and $t$ time).  
The independent variable is the distance $s$ of the reference particle along this trajectory,  rather than time $t$, and for simplicity this reference path is taken to be planar.  The transverse coordinates are $x$ (horizontal) and $y$ (vertical), 
where $\{x,z\}$ defines the plane of the reference trajectory.

Several time scales are involved, and this is reflected in the approximations used in formulating the 
equations of motion.  All of today's high-energy colliders are alternating gradient synchrotrons \cite{chao2013handbook} and their shortest time scale is set by so-called {\it betatron oscillations}.  The linearized equations of motion of a particle displaced from the reference trajectory are:
\begin{eqnarray}
    x'' + K_x x &= & 0 \; \; {\rm with} \; \;  K_x\equiv\frac{e}{p}\frac{\partial B_y}{\partial x}+\frac{1}{\rho^2}    \,  
\nonumber 
\\ 
      y'' + K_y y & = &  0 \; \; {\rm with} \; \; K_y\equiv -\frac{e}{p}\frac{\partial B_y}{\partial x}    
      \nonumber
      \\
      z' &= &  -x/\rho   \, 
\label{betamo}
\end{eqnarray}
 where the magnetic field $B_y(s)$ is only in the $y$ direction, contains only dipole and quadrupole terms, and is here treated as static in time, but $s$-dependent. We take into account the Maxwell equation in vacuum $\bm{\nabla} \times  \bm{B}=\bm{0}$ to eliminate $B_x(s)$, using the relation $\partial B_x / \partial y =\partial B_y / \partial x$. The radius of curvature due to the field on the reference orbit is $\rho$ ($\rho=e/pB_{y}$); $p$ and $e$ are the particle's total momentum and charge, respectively.  The prime denotes $d/ds$.

 The equations for $x$ and $y$ are those of harmonic oscillators but with a restoring force periodic in $s$, that is, they are instances of Hill's equation \cite{magnus2013hill}. The solutions are:  
\begin{eqnarray}
\label{betatron}
  x(s)&= & \sqrt{2 J_x \beta_x}\;\cos\psi_x \; , \\ 
  x'(s) &= & -\sqrt{\frac{2 J_x}{\beta_x}}\left[\alpha\cos\psi_x+\sin\psi_x\right]  \; , 
\end{eqnarray}
where the {\it action} $J_x$ is a constant of integration,  $\alpha_x=\alpha_x (s)\equiv -(1/2) d\beta_x(s)/ds$, and the envelope of oscillations is modulated by the {\it amplitude function} {$\beta_x$}, commonly called the $beta$-{\it function}. 
A solution of the same form describes the motion in $y$.  The betatron oscillation phase advances according to $d\psi_x /ds=1/\beta_x$;  that is, $2\pi \beta_x$ also plays the  role of a local wavelength of oscillations along the orbit. An extremely important parameter is the {\it tune}, $Q_x$, which is the number of such oscillations per turn about the closed path: 
\begin{equation}
\label{tune}
Q_x=\frac{1}{2\pi}\oint d\psi_x=\frac{1}{2\pi}\oint \frac{ds} {\beta_x(s)}. 
\end{equation}
While the integer part of the tune [$Q_{x,y}$] generally characterizes the extent of the focusing lattice, it is the fractional part of the tune {$Q_x$} that needs to be well defined and controlled by the machine operators in order to stay away from potentially detrimental resonances, which may occur under conditions of $kQ_x+lQ_y=m$, where $k, l,$ and $m$ are integers. For example, for the LHC a combination of horizontal and vertical tunes --- also called the {\it working point} --- equal to ($Q_x, Q_y$)=(64.31, 59.32) has been selected, such that resonances up to the order of $|k|+|l|=10$ or $12$ are avoided  \cite{Gareyte:488276,persson2017lhc}. 
These resonances are driven by high order multipole components $B_n, A_n$ of the fields in the magnets if $k+l=n$, by self-fields of the beam, or by the electromagnetic fields of the opposite bunch. Normally, the nonlinear components are very weak compared to linear ones $B_1, B_2, A_2$. However, when the nonlinear resonance condition is encountered, the amplitudes of particle oscillations $A_{x,y}$ could grow over the beam lifetime, resulting in the escape of the particles to the machine aperture, in the increase of the average beam size, or in both; either of these are highly undesirable phenomena. Careful analysis of nonlinear beam dynamics is instrumental in determining and optimizing the {\it dynamic aperture}, which is defined as the maximum amplitude of a stable particle motion \cite{wiedemann2012nonlinear}.  

Neglecting for now all non-linear effects and considering only the linear dynamics, the beta-function is well defined and satisfies the following equation: 
\begin{equation}
\label{beta}
   2\beta_x \beta_x''-\beta_x'^2+4\beta_x^2K_x = 4 \, .
\end{equation}

In a region free of magnetic fields, such as in the neighborhood of a collider interaction point (IP), usually occupied by particle detectors, a symmetric solution of Eq.~(\ref{beta}) is a parabola: 
\begin{equation}
\label{betastar}
\beta_x(s)=\beta_x^*+\frac{s^2}{\beta_x^*} \, ,
\end{equation}
where, in this case, $s$ denotes the longitudinal distance from the IP. The location of the beam waist usually coincides with the IP and corresponds to the minimum value of the $\beta$-function $\beta_x^*$; 
the asterisk is used to indicate IP parameters. 
Of course, a focusing force $K_x(s)$ is needed to prevent the amplitude from growing. In the case of the widely used alternating gradient periodic focusing lattice, consisting of a sequence of equally-spaced quadrupoles with a magnetic field gradient equal in magnitude, but alternating in sign (``focusing quadrupole - drift space - defocusing quadrupole - drift space'' --- known as a  {\it FODO cell}), Eq.~(\ref{beta}) has stable periodic solutions $\beta_x(s), \beta_y(s)$ in both planes provided that the focal length of the quadrupoles is longer than half the lens spacing $L$, i.e., $f=p/(eB_2l)> L/2$ (where $l$ is the length of the quadrupole magnet, here assumed to be much shorter than the cell length $L$). In that case, the amplitude functions have maxima at the focusing quadrupoles and minima at the defocusing ones, equal to, for example, $(2\pm \sqrt{2})L$ in the case of $f=L/\sqrt{2}$, which corresponds to a betatron phase advance $\Delta \psi_{x,y}=90^\circ$ per FODO cell. 
 
   Expressing $J_x$ in terms of $x,\ x'$ yields  
\begin{equation}
\label{CSparam}
J_x=\frac{1}{2} \left( 
\gamma_x x^2 + 2\alpha_x x x' + \beta_x x'^2 \right) 
= \frac{x^2 + (\alpha_x x + \beta_x x')^2}{2 \beta_x} \,
\end{equation}
with $\gamma_x=\gamma_x(s)\equiv (1+\alpha_x^2(s))/\beta_x(s)$. In a periodic system,  these {\it Courant-Snyder parameters} \cite{courantsnyder1958} $\alpha(s),\,\beta(s),\,\gamma(s)$ are usually defined by the focusing lattice; in a single pass system such as a linac, the parameters may be selected to match the $x$-$x'$ distribution of the input beam.

 For a given position $s$ in the ring, the transverse particle motion in $\{x,x'\equiv dx/ds\}$ phase space describes an ellipse, the area of which is $2 \pi J_x$, where the horizontal action $J_x$ is a constant of motion and independent of $s$.  If the interior of that ellipse is populated by an ensemble of non-interacting particles, that area, given the name $emittance$, is constant over the trajectory as well and would only change with energy. In a typical case of the particle's energy change rate being much slower than betatron motion, and considering a Hamiltonian system (i.e., a hadron collider or a linear collider, either without synchrotron radiation), the adiabatic invariant $\int p_x dx$ is conserved, and given that for small angles $p_x=  x'\cdot \beta \gamma mc^2$, it is common practice to consider an energy-independent {\it normalized emittance} that is equal to the product of the emittance and relativistic factor $\beta \gamma / \pi$ and denoted by $\varepsilon_n$. For a beam with a Gaussian distribution in $\{x,x'\}$, average action 
value $\langle J_{x}
\rangle$ and standard deviations $\sigma_x$, and $\sigma_{x'}$, the definition of the normalized emittance is
\begin{equation}
\label{emitt}
\varepsilon_{nx} \equiv \beta \gamma 
\langle J_x \rangle
= 
\beta \gamma {\sigma_x^2(s) \over \beta_x(s)} 
= 
\beta \gamma {\beta_{x} (s) \sigma_{x'}^2 (s) \over 1+ \alpha_x^2(s)  } 
\, ,
\end{equation}
with a corresponding expression for the other transverse direction, $y$. The angular brackets denote an average over the beam distribution.  
For 1D Gaussian beam, 95\% of the particles are contained within $\{x,x'\}$ phase space area of $6\pi \varepsilon_n /(\beta \gamma)$. Normalized beam emittances are conserved over the acceleration cycle in linear, static focusing lattices $K_{x,y}(s)$, and consequently, one would expect the same $\varepsilon_n$ at the hadron (or linear) collider top energy as the one coming from the very initial low energy particle source, such as the duoplasmotron \cite{brown2004physics, wolf2017handbook} (or photoinjector / damping ring). 
Unfortunately, that is rarely the case as many time-varying or non-linear phenomena come into play; some of the more important ones are briefly discussed in Subsection \ref{imped}. 

In an $e^-/e^+$ storage ring, the normalized emittance is not preserved during acceleration, but at each energy the  
beam's equilibrium emittance is determined by the effect of synchrotron radiation as a balance between radiation damping and quantum excitation \cite{Sands:1970ye}. 
In this case, for a constant accelerator optics, the normalized emittance increases with the third power of the beam energy.

As for the description of a particle's longitudinal motion, one takes the fractional momentum deviation $\delta p/p$ from that of the reference particle as the variable conjugate to $z$. The factors $K_{x, y}$ and $\rho$ in Hill's equations (\ref{betatron}) are dependent on momentum $p$, leading to a number of effects: first, the trajectory of off-momentum particles deviates by $\Delta x(s)=D_x(s) (\delta p/p)$, where the {\it dispersion function} $D_x(s)$ is determined by the magnetic lattice and is usually positive, periodic, and of the order of $\sim \rho / Q_x^2$. Second, the radius of curvature and orbit path-length $C$ vary with the momentum and, to first order, are characterized by the momentum compaction factor $\alpha_c$:
\begin{equation}
\label{alphac}
\alpha_c \equiv \frac{\Delta C/C}{\delta p/p} = \frac{1}{C} \oint
\frac{D_x(s)}{ \rho (s)} ds \, .
\end{equation}
Energy deviations also result in changes of machine focusing lattice properties and variations of the particle tunes, characterized by the {\it chromaticity}  $Q_{x,y}'=\Delta Q_{x,y}/(\delta p/p)$. The natural chromaticity due to energy dependence of the quadrupole focusing is large and negative ($\sim -Q_{x,y}$). Corresponding chromatic tune variations even for relatively small energy deviations $(\delta p/p) \sim 10^{-4}$--$ 10^{-3}$ can become unacceptably large.    
To assure transverse particle stability, usually the chromaticity is partially or fully compensated by additional sextupole magnets placed at locations of non-zero dispersion.   

RF electric fields (Eq.~(\ref{Eacc})) in the $s$ direction provide a longitudinal focusing effect. This is also known as the {\it phase stability principle}, which  
historically was important for the development of  
the synchrotron concept \cite{veksler1944, mcmillan1945synchrotron}.
The frequency $f_s$ of such longitudinal {\it synchrotron oscillations} is (expressed in units of revolution frequency $f_{rev}$, to become the  synchrotron tune $Q_s$)
\begin{equation}
\label{synchrotrontune}
Q_s \equiv \frac{f_s}{f_{rev}} = \sqrt {\frac{(\alpha_c-1 / \gamma^{2}) h eV_{acc} \sin(\phi_s) }{2 \pi \beta c p}}\; ,
\end{equation}
where $h$ again denotes the RF harmonic number.
The synchrotron tune $Q_s$ determines the amplitude of longitudinal oscillations for a particle with an initial momentum offset $\delta p/p$ via 
\begin{equation}
\Delta z = \left( \frac{\delta p}{p} \right)  \frac{E_b Q_s}{eV_{acc} h} C \, .
\label{sigmaz} 
\end{equation}
Similarly to the case of transverse oscillations, 
the area of the longitudinal phase space $\{\Delta E,\Delta t\}$, or
$\{\gamma \beta \delta p/p =(1/\beta) \Delta \gamma, 
z=\beta c \Delta t\}$,
encircled  by a moving particle is 
an adiabatic invariant, and the corresponding normalized {\it longitudinal emittance} proportional to the product of rms bunch length $\sigma_z$ and rms momentum spread $\varepsilon_{n,L}=\beta \gamma mc \sigma_z (\delta p/p)$ is a generally conserved quantity in hadron accelerators and also in linear accelerators.  
In the case of lepton storage rings, synchroton radiation determines the relative momentum spread, which grows with the square of the beam energy \cite{Sands:1970ye}, and the
corresponding bunch length follows from Eq.~(\ref{sigmaz}). 
In hadron synchrotrons, the longitudinal emittance is often intentionally blown up during acceleration, so as to preserve longitudinal beam stability \cite{Baudrenghien:2011zz}.

Longitudinal oscillations are the slowest of all the periodic processes which take place in the accelerators. For example, in the LHC, the frequency of synchrotron oscillations at the top energy of 7 TeV is about $f_s=23$ Hz, the revolution frequency is $f_{rev}=$11.3 kHz, the frequency of betatron oscillations is about $Q_{x,y} f_{rev}$=680 kHz and the RF frequency is $f_{RF}=400.8$ MHz ($h=35640$). 

It should be noted that longitudinal motion is practically absent in linacs. In the absence of bending dipoles, dispersion  $D_x(s)$ is zero and so are the momentum compaction factor $\alpha_c$ and the synchrotron tune $Q_s$. As a result, in a linac ultrarelativistic particles barely change their relative positions during acceleration despite significant energy spread --- see Eq.~(\ref{sigmaz}). 
 
\subsubsection{Beam dynamics impediments to and evolution of luminosity}
\label{imped}

For further discussion, the basic Eq.~(\ref{eq:lumi}) for luminosity is now  re-written in terms of normalized transverse emittances Eq.~(\ref{emitt}) and the amplitude functions $\beta^*$ at the IP as:  
\begin{equation}
\Lumi = f_{0} \gamma n_b { N^2 \over 4\pi\varepsilon_n \beta^*} H\left( {\sigma_z \over \beta^*}, \theta_c \right) \, .
\label{eq:lumi2}
\end{equation}   
Here, $f_0$ signifies either the repetition rate of a linear collider or $f_{rev}$ of a circular one; for simplicity we assume equal bunch populations $N$ in two Gaussian beams with $n_b$ bunches each, 
with equal and round cross sections at the IP $\varepsilon_{nx}=\varepsilon_{ny}=\varepsilon_n$, $\beta_x^*=\beta_y^*=\beta^*$, with
$\sigma_{x}^{\ast}=\sigma_{y}^{\ast}=\sqrt{\beta^* \varepsilon/(\beta \gamma)}$. The numerical factor $H\leq 1$ accounts for geometrical reduction in luminosity \cite{hirata1995prlbeambeam} due to the final bunch length with respect to $\beta^*$ and due to a crossing angle at the IP $\theta_c$.  The former, also referred to as {\it hourglass effect}, is caused by the increase in transverse beam sizes as one proceeds away from the IP, where $\beta(s)$ grows parabolically, as in Eq.~(\ref{betastar}). Thus, for round beams, the hourglass effect lowers the contribution to luminosity from such locations by 
\begin{equation}
H\left( {\sigma_z \over \beta^*}, \theta_c=0 \right) = \sqrt{\pi} \exp(A^2) \textup {erfc} (A)\, ,
\label{hourglass}
\end{equation}    
where $A=\beta^* / \sigma_z$ \cite{sylee2018accelerator}, and also
leads to a harmful modulation of the beam-beam tune shift (see later) at twice the synchrotron frequency. 

In the case of a nonzero crossing angle, assuming small $A$, the factor $H$ is equal to \cite{napoly1993luminosity}:
\begin{equation}
H\left( {\sigma_z \over \beta^*}=0, \theta_c \right) = \frac{1}{\left( 1 + \sigma_z^2 \tan^2(\theta_c/2) / \sigma^{*2} \right)^{1/2}}\, .
\label{crossingangle}
\end{equation} 
The factor $H$ rarely drops below 0.5 for the majority of colliders, unless it is specifically required by physics processes under study, as in \cite{bogomyagkov2018mumutron}. Thanks to the additional focusing effect during the beam-beam collision, the factor $H$ can also be larger than $1$, e.g.~up to about a value of 2 (``dynamic beta'' along with ``dynamic emittance'' effects in circular colliders \cite{Furman:1994sz,PhysRevSTAB.2.104401}, and ``disruption enhancement'' in linear colliders \cite{Yokoya:1991qz}).

Naturally, to achieve high luminosity, one has to maximize the total beam populations $n_b N$ with the lowest possible emittances, and collide the beams at high frequency at locations where the focusing beam optics provides the lowest possible values of the amplitude functions  $\beta^*$, the so-called {\it low-beta insertion} \cite{Robinson:1966gt}. The latter requires sophisticated systems of strong focusing elements, sometimes occupying quite a significant fraction of the collider's total length \cite{levichev2014rast}. The lowest $\beta^*$ is determined by the maximum field gradients and apertures in the interaction region (IR) magnets and the effectiveness of compensation of chromatic and non-linear aberrations. The quest for maximum intensities and lowest emittances is limited by a number of important and often interdependent effects which affect either incoherent (single particle) dynamics or the dynamics of the beam as a whole (coherent effects). 

Examples of incoherent effects are particle losses caused by scattering at a large angle or with a large energy loss, 
so that either the particle's amplitude $\sqrt{2 J_{x,y} \beta_{x,y}(s)}$, or its dispersive position deviation $\Delta x=D_x(s) (\delta p/p)$ physically exceeds the available transverse aperture, usually set by collimators (otherwise, by the vacuum chamber and magnet apertures).  This can be due to residual vacuum molecules near the beam orbit or Compton scattering off thermal photons \cite{telnov1987scattering}, due to Coulomb scattering on other particles within the same bunch ({\it Touschek effect}) \cite{bernardini1963lifetime}, or due to collisions with opposite beam particles and fields, such as inelastic interaction of protons, Bhabha scattering $e^+e^- \rightarrow e^+e^-$, or radiative Bhabha scattering $e^+e^- \rightarrow e^+e^-\gamma$ \cite[Ch.3]{chao2013handbook}. 

Particles can also get lost on the aperture as a result of much slower mechanisms of diffusion caused either by the above processes with smaller scattering amplitudes, but stochastically repeated many times, such as multiple Coulomb {\it intrabeam scattering} \cite{piwinski1988intra, bjorken1983jd, piwinski2018wilson}, by external noises such as ground motion or magnetic field fluctuations \cite{lebedev1994emmgrowth}, or via chaotic mechanisms like Arnold diffusion, modulational diffusion, or resonance streaming in non-linear fields, enhanced by minor tune modulations \cite{zimmermann1994transverse}. Diffusion is characterized by the action dependent coefficient $D(J)=D(J_x, J_y)$ and leads to a slow evolution of the beam distribution function $f(J_{x,y}, t)$ according to the diffusion equation 
\begin{equation}
{\partial f  \over \partial t}= {\partial  \over \partial t} \left( D(J) {\partial f \over \partial J} \right) \, ,
\label{diffusion}
\end{equation} 
and, consequently, to a change (normally an increase) in the average action $<J>$. 
For an ensemble of particles, the corresponding beam emittance growth is given by 
\begin{equation}
{d \varepsilon_n \over dt}  = \beta \gamma {dD(J) \over dJ} - {\varepsilon_{n}  \over \tau_{cool}} \, 
\label{emmcooling}
\end{equation} 
where the additional second term accounts for {\it beam cooling} or damping of particle oscillations. This term appears in the presence of a reaction force opposite to particle momentum if, on average, the corresponding dissipative particle energy loss is compensated for by external power \cite{skrinsky1981cooling, parkhomchuk2008cooling}.   

In the case of electron or positron storage rings, such cooling occurs naturally due to synchrotron radiation (Eq.~(\ref{SR})) and it  fully determines equilibrium emittance according to Eq.~(\ref{emmcooling}) through a balance between radiation damping and excitation of oscillations by random radiation of individual photons \cite{Sands:1970ye,wiedemann2003synchrotron}. Four other methods of beam cooling have been developed and successfully employed to attain low emittances, namely {\it electron cooling} \cite{budker1967,parkhomchuk2000electron, nagaitsev2006experimental} and {\it stochastic cooling} of heavy particles (ions and antiprotons) \cite{vandermeer1985}, \cite[Ch.7]{lebedevshiltsev2014tevatron}, {\it laser cooling} of ion beams \cite{schroder1990first, hangst1991laser,PhysRevLett.81.2052}, and,
in a proof-of-principle experiment, the {\it ionization cooling} of muons \cite{budker1969, neuffer1983principles, mice2019first}. 

\begin{figure}[htbp]
\centering
\includegraphics[width=0.99\linewidth]{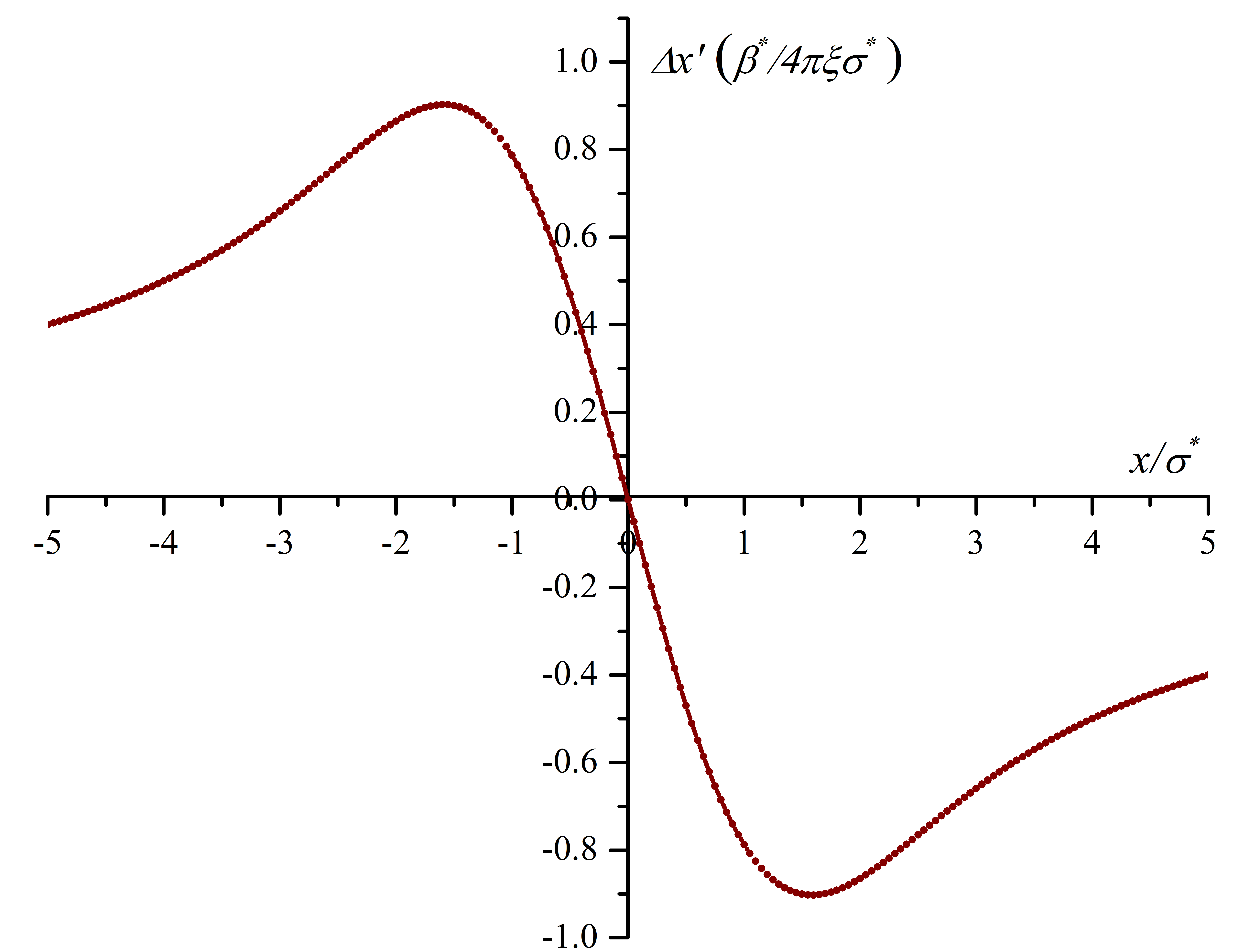}
\caption{Angular kick due to beam-beam force.}
\label{beambeamf}
\end{figure}
The most prominent coherent effects arise from electric and magnetic forces of the opposite bunch at the IPs, characterized by a dimensionless {\it beam-beam parameter} : 
\begin{equation}
  \xi_{x,y}= \frac{r_0 N\beta^*_{x,y}}{2\pi\gamma\sigma_{x,y}^* (\sigma_{x}^*+\sigma_{y}^*)} \, ,
\label{beambeam}
\end{equation} 
where $r_0=e^2/(4\pi\varepsilon_0 mc^2)$ is the classical radius of the  colliding particle (with charge $e$ and mass $m$)  \cite{chao1985beambeam}. 
The beam-beam parameter is roughly equal to the betatron tune shift experienced by small amplitude particles --- positive in the case of opposite charge beams, like $e^+e^-$, and negative for same charge beams as in $pp$ collisions \cite{pieloni2013beambeam}. It also describes the maximum angular beam-beam kick experienced by particles at the IP, which, in the case of round beams, equals $\Delta x'_{max} \approx 0.9 \cdot 4 \pi \xi (\sigma^* / \beta^*)$, with $\xi=r_0 N/(4 \pi \varepsilon_n)$, 
and occurs at $x \approx 1.6 \sigma^*$ \cite{shiltsev1996beambeamkick}. As seen in   Fig.~\ref{beambeamf}, electromagnetic fields of a  Gaussian beam ``1'' present a nonlinear lens to particles of the opposite beam ``2'', 
resulting in changes to the transverse tunes of these  particles in beam 2 by an amount varying between $\xi$ and 0, as particles at large amplitudes experience minimal 
beam-beam force. 

beam-beam forces can lead to coherent effects, such as unstable beam oscillations 
\cite{dikanskij1982collidingcoherentinst,chao1984coherentbeambeam,Yokoya:200504,Alexakhin:1996bc} 
or blow-up of one beam's size while the other beam remains small or even shrinks (beam-beam ``flip-flop'' effect) \cite{krishnagopal1991coherentnlbeambeam,PhysRevSTAB.2.104401}.
In addition, the tune spread arising from $\xi$ and the non-linear nature of beam-beam interactions results in strong diffusion along high-order transverse resonances $k\nu_x+l\nu_y=m$ and, ultimately, in beam size growth and beam losses. Accordingly, it has been concluded operationally that the aforementioned effects are tolerable below certain {\it beam-beam limits} of $\xi_{x,y}\approx 0.003-0.012$ in hadron colliders \cite{shiltsev2005beambeameff}. Thanks to strong synchrotron radiation damping, the beam-beam limit is about an order of magnitude larger in $e^+e^-$ colliders, with maximum $\xi_{x,y} \approx 0.03-0.12$ \cite{seeman1986bbobservations}. 

From Eqs.~(\ref{eq:lumi2}) and (\ref{beambeam}), one can note that  the path to higher luminosity via higher beam intensity and smaller beam sizes almost automatically calls for a higher beam-beam parameter as $\Lumi \propto \xi$. Several methods have been implemented over the decades to get around the beam-beam limit, including: a) carefully choosing working tunes $(Q_x, Q_y)$ away from the most detrimental resonances; b) operation with very flat bunches (wide in the horizontal plane and narrow in the vertical --- see Eq.~(\ref{beambeam})); and more recently, c) compensation of the beam-beam effects using electron lenses \cite{shiltsev2015electronlenses}; d) reduction of the strength of the beam-beam resonance in the {\it round beam} scheme with strongly coupled vertical and horizontal motion \cite{danilov1996roundbeamsconcept,Young:909804,  berkaev2012vepp2000, shatunov2016statusvepp2000}, and e) by using the so-called {\it crab-waist} collision method that beneficially modifies the geometry of colliding bunch profiles only at the IPs so as 
to minimize the excitation of harmful resonances 
\cite{raimondi2006carbwaistworkshop, raimondi2007arxivcrab,zobov2010crabwaisttest}.  

The focusing of the beams during the collision changes the beam optics, especially for low-amplitude particles. 
With a properly chosen working point, e.g., 
just above the half integer resonance in case of $e^+e^-$ collisions with a single IP, this leads to a reduction of the 
effective beta function at the collision point, the {\it dynamic beta} effect \cite{Furman:1994sz}. 
In circular $e^+e^-$ colliders, this optics change in collision, propagating all around the ring, also modifies the equilibrium  horizontal emittance, which is known as {\it dynamic emittance} \cite{Hirata:1989zz,PhysRevSTAB.2.104401}. 
The net IP beam sizes then  
follow from the combined change of $\beta^*$ and $\epsilon_x$. Parameters are normally chosen so that the overall dynamic effect increases the luminosity.
 
In linear colliders, where each bunch collides only once, with typically much smaller
beam size and experiencing much stronger
forces, the strength of
the collision is measured by the ratio 
of the rms bunch length $\sigma_{z}$ to the approximate (linear, thin-lens)
beam-beam focal length.  
This ratio, called {\it disruption parameter} $D_{y}$ \cite{Yokoya:1991qz}, 
is related to $\xi_{y}$ via $D_{y}= 4  \pi \sigma_{z} \xi_{y}/\beta^*_{y}$. 
Significant disruption leads to effectively smaller beam size and a resulting 
luminosity enhancement; it also makes the 
collision more 
sensitive to small offsets, resulting in a ``kink
instability'' \cite{Yokoya:1991qz}. 
Additional beam-beam 
effects arising in the collisions at  
linear colliders are the emission of beamstrahlung (synchrotron radiation in the field of the opposing beam), along with $e^+e^-$ pair creation, and  depolarization by various mechanisms '' \cite{Yokoya:1991qz}. 

Notably, self-fields of an ultrarelativistic beam are such that the electric force $F_E$ and magnetic force $F_M=-\beta^2 F_E$ on its own particles effectively cancel each other out. This is not the case at $\gamma = (1-\beta^2)^{-1/2} \sim 1$, e.g., in the low-energy machines of the injector chain of colliders, where,  similarly to beam-beam phenomena, the beam's own forces set the limit on the {\it space-charge tuneshift parameter}:
\begin{equation}
  \Delta Q_{SC}= \frac{r_0 N} {4\pi\beta \gamma^2\varepsilon_n } \, .
\label{sc}
\end{equation} 
For most rapid cycling proton synchrotrons $\Delta Q_{SC} \leq 0.2-0.3$ \cite{weng1987spacecharge, reiser2008bookSCbeams, hofmann2017spacecharge}.  Space-charge effects at injection usually also determine the ultimate beam phase-space brightness $N / \varepsilon_n$ at top energy. 

With the single bunch brightness set by either space-charge or beam-beam limits, further increases in luminosity require an increase in the number of bunches $\Lumi \propto n_b$.  The beams need to be separated in all but a few head-on IPs, otherwise multiple $2n_b$ collisions points would immediately lead to unacceptable total beam-beam tuneshift parameters $\xi=2n_b \xi_{x,y}$. Such separation can be implemented either by the use of HV electrostatic separators in single-aperture proton-antiproton colliders like in the Tevatron ($n_b=36$), or by having an independent aperture and two magnetic systems for each beam, like in RHIC ($n_b=111$), most modern $e^+e^-$ colliders ($n_b=1584$ in SuperKEKB), or in the LHC ($n_b=2808$). In the latter cases, by necessity, certain regions exist near the main IPs where the colliding beams have to join each other in a common vacuum chamber; here a significant number of parasitic long-range beam-beam interactions between separated bunches can still take place. These parasitic collisions may produce significant, sometimes dominant effects on beam dynamics. The separation of the two beam orbits, by at least $\sim 9 \sigma_{x,y}$ typically, allows avoiding troublesome operational issues. Other complications of beam-beam interactions can result from the fact that bunch dimensions at the IPs are not always the same between the two colliding beams or between vertical and horizontal planes, or that beam intensities are sometimes significantly mismatched. All in all, despite many advances and inventions, beam-beam effects remain one of the most critical challenges, setting a not yet fully resolved limit on the performance of all colliders. 

Higher luminosities within beam-beam limits are possible via increase of the beam current $I_b =e f_0 n_b N$. Three major related difficulties include growing RF power demands in synchrotron radiation dominated $e^+/e^-$ beams, the advent of so-called {\it coherent (or collective) beam instabilities}, and growing demands for minimization of radiation due to inevitable particle losses. Many types of single- and multi-bunch instabilities \cite{chao1993bookcollectiveinst, ng2006bookinstab} are caused by beam interactions with  electromagnetic fields induced by the beam itself due to the {\it impedance} of the vacuum chambers and RF cavities \cite{heifets1991impedancesRMP, zotter1998bookimpedances}, or caused by unstable clouds of secondary particles, like electrons or ions, which are formed around the circulating beams \cite{ohmi1995ecloud, raubenheimer1995fastbeamion, zimmermann2004reviewecloud, flanagan2005kekbecloud}. These instabilities can develop as quickly as within tens to thousands of turns and definitely need to be controlled. Mechanisms that are routinely employed to avoid coherent instabilities include the use of nonlinear magnets to generate sufficient spread of the tunes and therefore, provide ``Landau damping'' \cite{sessler1966instab, metral1999Landau}, fast beam-based transverse and longitudinal feedback systems \cite{karliner2005fb, burov2016feedback}, 
and electron/ion clearing (either by weak magnetic or electric fields or by modulation of the primary beam current profile rendering secondaries unstable, or by reducing the yield of secondary electrons via either special coating or extensive ``beam scrubbing'' of the vacuum chamber walls \cite{kulikov2001electroncloud, fischer2008electroncloudrhic,PhysRevSTAB.14.071001,arduini2013lhcecloud}). To avoid damage or excessive irradiation of accelerator components so that these remain accessible for maintenance in the tunnel, sophisticated collimation systems are utilized.
These systems usually employ a series of targets or ``primary collimators'' (which scatter the halo particles)  and numerous absorbers (sometimes as many as a hundred, which intercept particles in dedicated locations) \cite{mess1994heracollimators, vonHoltey1987LEPcollim, mokhov2011tevatroncollim, schmidt2006lhcprotection}. In the highest energy modern and future colliders, extreme total beam energies $n_b N E_b$ ranging from MJs to GJs and densities reaching many GJs/mm$^2$ pose one of the biggest challenges for high efficiency and robust particle collimation  \cite{assmann2012lhccollim}, \cite[Ch.8.8]{myers2013accelerators}. Novel sophisticated techniques like collimation by bent crystals \cite{mokhov2010crystal, scandale2016lhccrystal} or by hollow electron beams \cite{stancari2011collimation} are therefore being developed. 
	
Finally, operation of the colliders with progressively smaller and smaller beams brings up many issues relevant to alignment of magnets, vibrations, and long-term tunnel stability \cite{fischer1987groundmotion, rossbach1987gmwaves, parkhomchuk1993slowATL, sery1996gm, shiltsev2010ATLreview}. Radiation backgrounds in physics detectors necessitate careful designs of the interaction region and of the accelerator-detector interface in  high-energy high luminosity colliders \cite{mokhov2012mdi, boscolo2017mdi}. HEP demands for polarized beam collisions and very precise c.m.e.~calibration of about $\delta E/E \sim 10^{-5}$ or
even $\delta E/E \sim 10^{-6}$ have been largely satisfied by the development of polarized particle sources  
coupled with sophisticated methods to maintain beam polarization along the acceleration chain,
or, for $e^+$/$e^-$ storage rings, by dedicated spin
matching procedures to enable self polarization, 
combined with 
the well-established method of ``resonant depolarization'' \cite{derbenev1978polarization, shatunov1980depolarization, roser1994agssnakes, barber1995firstspin, bai2006polarized, blinov2009energydepolzarization}. 

\subsection{Past advances of $e^+e^-$ colliders}
\label{e+e-colliders}

In this and following sections, we briefly present key milestones of past colliders and their major breakthroughs and contributions to accelerator science and technology, as well as to particle physics. 
Extended reviews and many additional details can be found in \cite{voss1996history, pellegrini1995developmentofcolliders, sessler2007engines, scandale2014protoncollders}, and \cite[Ch.10]{myers2013accelerators}, \cite[Part 2]{bruning2016challenges}. 

Though the trio of the very first colliders --- AdA at Frascati/Orsay, VEP-I in Novosibirsk and CBX at Stanford --- constituted mostly proof-of-principle machines, they were used for initial studies of quantum electrodynamics (QED) processes (elastic scattering, single and double bremsstrahlung) at their range of center of mass energy $\sqrt{s}$. Technological challenges addressed at these machines included development of ns-fast injector kickers, attainment of ultra-high vacuum in the range of 10s to few nTorr, and reliable luminosity monitoring and other beam diagnostics. Beam physics advances included first observations and studies of the Touschek effect, luminosity degradation due to beam-beam effects at $\xi_{x,y} \sim 0.02-0.04$, complex beam dynamics at non-linear high order resonances, and coherent instabilities due to resistive vacuum pipe walls \cite{bernardini1963lifetime, bernardini1964, bernardini2004, budker1967, levichev2018, barber1966testQED, gittelman1965recentCBX}. 

In the late 1960s to mid-1970s, VEPP-2 in Novosibirsk \cite{skrinsky2002binphistoryHEP}, ACO in Orsay \cite{marin1965ACOstatus}, and ADONE in Frascati \cite{ADONE1971status}, were the first electron-positron colliders with an extended particle physics program, which included studies of $\rho$, $\omega$ and $\phi$ mesons, two-photon pair production, $e^+e^- \rightarrow e^+e^-e^+e^- $, and multi-hadronic events  \cite{balakin1971vepp2phimeson, cosme1972hadronicACO, bacci1972multihadronicADONE, kurdadze1972vepp2multihadron}. With a maximum energy of 2$\times$1.5 GeV, ADONE just missed the discovery of the $J/\psi$ particle (confirming its existence later). Beam instabilities, including bunch lengthening at high intensity, were the most important beam effects studied and a longitudinal phase feedback system was developed and installed in ADONE to control them. Measured luminosity was mostly set by the beam-beam limit together with synchrotron radiation effects, i.e., beam emittances defined by the balance between quantum excitation and radiative damping, and scaled approximately as the fourth power of energy $\Lumi \propto \gamma^4$ \cite{haissinski1969Adone}. VEPP-2 and ACO were also the first storage rings in which the build-up of electron spin polarization through synchrotroton radiation ({\it Sokolov-Ternov effect} \cite{sokolovternov1964polarization}) could be observed and studied \cite{baier1972radiative}. 

At the Cambridge Electron Accelerator (CEA) facility,  electron and positron beams were  collided in a special bypass interaction region  with two quadrupole magnet doublets on both sides of the IP, demonstrating for the first time a {\it low-beta insertion} optics with a small $\beta_y^* \approx  2.5$ cm \cite{Robinson:1966gt}, representing almost two orders of magnitude of reduction compared to more traditional designs. The CEA also measured an unexpectedly large ratio of the hadronic cross section to the muon cross section in electron$–$positron collisions $R=\sigma (e^+e^- \rightarrow {\rm hadrons})/\sigma (e^+e^- \rightarrow \mu^+\mu^-)$ at $\sqrt{s}$ above 3 GeV, hinting at a new decay channel via charm quarks \cite{voss1996history}. 

SPEAR (Stanford Positron-Electron Asymmetric Rings) at SLAC was very productive in particle physics, enabling co-discovery of the $J/\psi$ meson at $\sqrt{s}$=3.1 GeV consisting of a charm quark and a charm antiquark (1976 Nobel Prize in Physics, Burton Richter) and discovery of the tau lepton with mass of $1.7$ GeV/$c^2$ (1995 Nobel Prize in Physics, Martin Perl). Transverse horizontal and vertical {\it head-tail} instabilities were observed at about 0.5 mA of current per bunch and were successfully addressed through a positive chromaticity $Q'_{x,y}>0$ \cite{paterson1975spear, chao1993bookcollectiveinst}. 

Several innovative ideas were tried at the DCI and VEPP-2M. The DCI (Dispositif de Collisions dans l'Igloo) team at Orsay attempted to compensate beam-beam effects by having 4 collinear beams $e^+e^-e^+e^-$ of equal size and current at the IP. However, the machine never fulfilled its expectations and the beam-beam limit was not significantly different than with two beams \cite{DCIgroup1979status, leduff1980dcicompensation}, 
due to higher-order coherent beam-beam instabilities
\cite{Derbenev:1973ws,krishnagopal1991coherentnlbeambeam,Podobedov:1994ef}. VEPP-2M at Novosibirsk reached a luminosity two orders of magnitude above its predecessor VEPP-2, which served for a while as the injector \cite{tumaikin77vepp2m}. The ring operated at the beam-beam limit $\xi_y \approx 0.05$, 
and luminosity was thus proportional to beam current, 
\begin{equation}
\Lumi = f_{0} \gamma \frac{I_b \xi_y} { 2e r_e \beta_y^*} \left(1+{\sigma_y^* \over \sigma_x^*} \right) \, ,
\label{eq:lumiee}
\end{equation} 
as follows from  Eqs.~(\ref{eq:lumi}) and 
(\ref{beambeam}). At the VEPP-2M's low energy and high currents, intrabeam scattering played a major role, leading to emittance growth and momentum-spread increase. As a countermeasure, 
a 7.5 T SC wiggler was used to increase the horizontal emittance and, in parallel, the beam current, and to decrease the damping time, allowing for a higher beam-beam tune shift.
In consequence, a significant gain in luminosity was obtained \cite{levichev2018}. Also, over decades of operation, the VEPP-2M team mastered the control of beam polarization; it used the resonant depolarization method \cite{derbenev1978polarization} to achieve a beam energy calibration at the level of  $\sim$10$^{-5}$--10$^{-6}$ and carried out the most precise measurements of the masses of $\rho, \omega, K^{\pm}$, and $K^0$ mesons \cite{skrinskiishatunov1989precisionmass}. 

A huge boost in colliding beam physics came with the next generation of $e^+e^-$ colliders: DORIS at DESY (Hamburg) \cite{nesemann1983DORISfirst} which had started its operation almost simultaneously with SPEAR, 
the Cornell Electron-Positron Storage Ring (CESR) \cite{mcdaniel1981cesrcommissioning}, and VEPP-4 at Novosibirsk \cite{vepp1980statusVEPP4}. Following the 1977 discovery of $\Upsilon$ at the Fermilab fixed target experiment E288 \cite{herb1977observationUpsilon}, their particle physics programs were aimed at the $b$-quark states and decays, $B-$meson mass and lifetime measurements, $B-\bar{B}$ mixing, and determination of CKM matrix parameters \cite{finocchiaro1980cesrphysics, bohringer1980observation, artamonov1984upsilon, baru1992totalvepp4hadrons, patrignani2013bottommonium}. DORIS initially started as a two ring collider with 480 bunches in each ring, but it was soon realized that in such a regime its total current was very much limited by coherent instabilities due to the  impedance of the RF cavities and beam-beam effects in presence of a vertical crossing angle. DORIS was therefore subsequently converted to a one bunch per beam, single ring collider with head-on collisions.

The history of CESR spans almost three decades \cite{berkelman2004cesrhistory} and witnessed an impressive increase in luminosity by two orders of magnitude \cite{shiltsev2004cpt} thanks to a number of important beam physics and technology advances, including operation with up to 45 bunches per beam in a single ring separated in  accelerator arcs by six 3-m long $\pm$85kV electrostatic separators which generated closed-orbit displacements ({\it pretzels}),  weaving  back  and  forth around  the  ring, and allowing  the  electrons  and positrons  to  simultaneously  be  stored  in  the  same  vacuum chamber without destructive unwanted beam-beam collisions \cite{rubin1989cesr}. Single cell SC RF cavities with damping of detrimental higher order modes (HOMs) excited by the beams \cite{padamsee2008srf, belomestnykh2012srfreview} allowed up to 0.37 A beams each of $e^-$ and $e^+$ to be stored. Tight vertical focusing with $\beta_y^*$=1.8~cm was provided by a pioneering combination of permanent-magnet and SC technologies for quadrupole magnets in the interaction region. 
Over many years, CESR held, and continually improved on, the world record for collider luminosity, from
about $3\times 10^{32}$~cm$^{-2}$s$^{-1}$ with 9 bunches per beam in the early/mid 1990s to
$1.25\times 10^{33}$~cm$^{-2}$s$^{-1}$ with 36  bunches per beam around the year 2000.
CESR also studied the possible implementation of a Moebius ring-collider \cite{moebius}, by colliding round beams with a beam-beam parameter $\xi$ as high as 0.09 \cite{cesryoung}.

The next triplet of high-energy colliders was made of 2$\times$23 GeV c.m.e.~PETRA at DESY \cite{voss1979petra}, 2$\times$15 GeV c.m.e.~PEP at SLAC \cite{helm1983pepcollider}, and 2$\times$32 GeV c.m.e.~TRISTAN at KEK (Japan) \cite{nishikawa1983tristan}. PETRA is known for the discovery of the gluon and for QCD studies. The first measurement of the tau lepton lifetime and accurate measurements of $B$ and $D$ meson lifetimes were carried out at PEP, while the search for high mass resonances (e.g., of the top quark) in TRISTAN was in vain. TRISTAN collided 2$\times$2 bunches in four IPs and was the first large accelerator to extensively use SRF technology with its 104 nine-cell 508 MHz cavities providing a total RF voltage of 0.4 GV RF  \cite{kimura1986tristan}. The {\it transverse  mode coupling instability} 
(TMCI), a sort of 
single-bunch head-tail instability, 
was extensively studied at both PETRA 
\cite{Kohaupt:1980yz} 
and PEP, and effective solutions were found.  

\begin{figure}[htbp]
\centering
\includegraphics[width=0.99\linewidth]{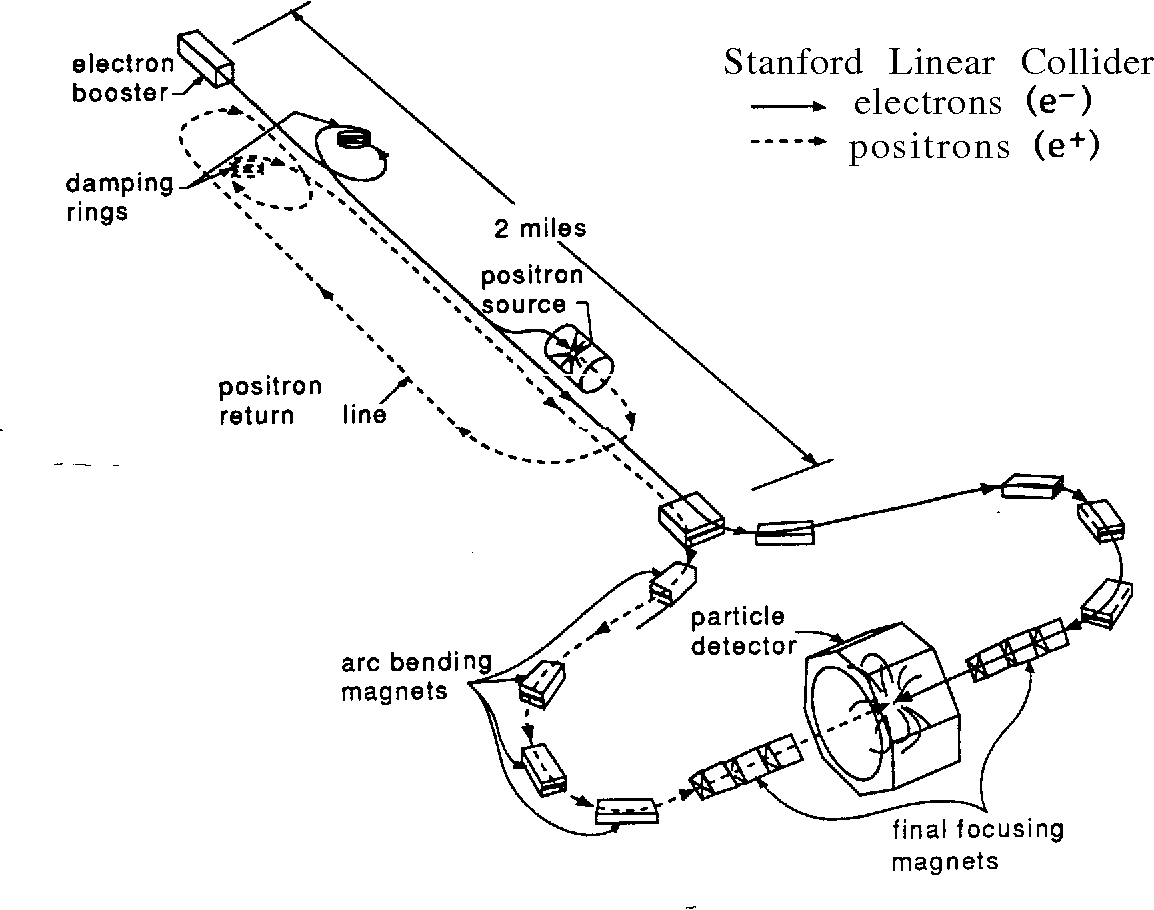}
\caption{The Stanford Linear Collider (SLC). 
Polarized electrons are produced by photoemission from a Ti:sapphire laser and a GaAs photocathode at the electron gun, accelerated to 1.2 GeV, injected into a damping ring (DR) to reduce $e^-$ beam emittance (size), then kicked back into the 3 km long linac to be accelerated together with positrons to 46.6 GeV, then separated magnetically and transported along two arcs and collide head-on at the IP. The positrons are produced by a fraction of 30 GeV $e^-$ beam which is stopped in a target. $e^+$'s are then collected and returned to the upstream end of the linac for many-fold emittance reduction in another DR (from \cite{friedsam1990slcalignment}).}
\label{slc}
\end{figure}

\begin{figure}[htbp]
\centering
\includegraphics[width=0.95\linewidth]{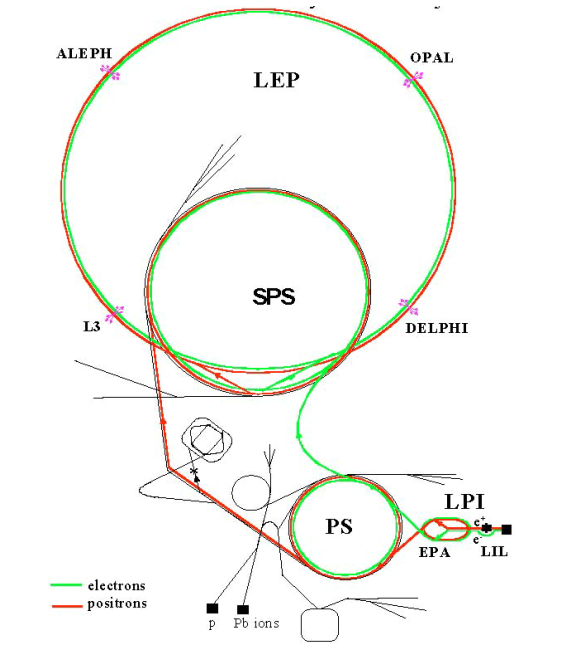}
\caption{
Schematic view of the LEP injector chain of accelerators and the LEP storage ring \cite{pcpb}
with the four experiments ALEPH, DELPHI, L3, and OPAL.
The first part of the chain of injectors, the LEP pre-injector (LPI), 
consisted of two LEP injector linacs (LIL) and an 
electron/positron storage ring (EPA). 
Eight positron bunches, followed by eight electron bunches, were ejected from EPA to the Proton Synchrotron (PS), and then accelerated plus extracted to the Super Proton Synchrotron
(SPS) for further acceleration.  
Positrons and electrons were injected into LEP from the SPS, 
initially at a beam energy of 
20 GeV, and later (since 1995) at 22 GeV, 
to boost the bunch current, which was limited at injection by the TMCI.   
In its last year of operation (2000), the
LEP reached a maximum e$^+$e$^-$ collision energy of 209 GeV. 
} 
\label{lep}
\end{figure}

The highest energy lepton colliders ever built were the Stanford Linear Collider (SLC) \cite{phinney2000slc}, running on the $Z$ pole at a c.m.e.~of 91 GeV, and the Large Electron-Positron (LEP) Collider at CERN \cite{myers2002lep}, the c.m.e.~of which was steadily increased from the $Z$ pole over the $WW$ threshold (160 GeV), to a  highest energy of 209 GeV in a search for the then still elusive Higgs boson. The SLC complex is shown in Fig.~\ref{slc}; the LEP tunnel, including later additions for the LHC, is shown in Fig.~\ref{lep}. 

The LEP and SLC operated simultaneously in the 1990s and were rivals in tests of the Standard Model of electroweak physics. In the seven years that LEP operated below 100 GeV, it produced around 17 million $Z$ particles 
(and later, at 160 GeV, some 40,000 $W^{\pm}$ pairs)  
collected over four experiments. Accurate determination of the parameters of $Z^0$ resonance at $\sqrt{s}=$91 GeV led to a rather precise measurement of the number of light neutrino families $N_{\nu}=2.9840 \pm 0.0082$
\cite{ALEPH:2005ab},
a value that, in 2019, could still
further be improved to 
$N_{\nu}=2.9963 \pm 0.0074$ \cite{Janot_2020}, 
and to an indirect determination of the mass of the top-quark as $M_t=173\pm 23$~GeV. Beam energy calibration with the resonant depolarization method was good to about 0.001\% and the combined error of the resonance scans of 1.9 MeV on
$m_Z$ and of 1.2 MeV on $\Gamma_Z$ were obtained 
after identifying and correcting for 
various small, subtle effects, including 
magnetic field drifts, earth tides, and ground currents induced by a nearby railway
\cite{assmann1999lepcalibration, myers2000lep}.
The LEP magnets contained very little steel so as to provide a relatively small bending field of 1.1 kG needed to circulate 
100 GeV particles in a 27 km ring.
At the highest energy of operation (beam energy of 104.5 GeV), the synchrotron radiation loss per turn was some 3\% of beam energy. That explains the need for LEP's powerful SRF system based on 352 MHz SC niobium-on-copper cavities, which in the last years of operation provided a total RF voltage of about 3.5 GV --- see Section \ref{technology} for further details.  
Without collisions, at top energy the LEP beam lifetime was limited by the  scattering of beam particles 
off thermal photons (black-body radiation inside the beam pipe) \cite{telnov1987scattering}, a new effect observed for the first time \cite{dehning1990lepblackbody}. The TMCI 
 \cite{Kohaupt:1980yz,besnier1984TMCIlep} 
 limited single bunch current at an injection energy of 20 GeV (later 22 GeV) to about 1 mA. A feedback system to address the TMCI was proposed and attempted \cite{danilov1997feedbackLEP}. 
In collision, the luminosity was limited by beam-beam effects at a record high value for the beam-beam tune shift, namely $\xi_y=0.083$ per collision point \cite{Assmann:2000xr}, or $n_{IP}\xi_{y}=0.33$ for the total tune spread.  

The SLC --- see Fig.~\ref{slc} --- was the world's first linear collider of single electron and positron bunches. It operated at 120 Hz rate and provided 80\% longitudinal $e^-$ polarization at the IP coming from a strained GaAs photo-gun \cite{kulikov1995slsGaAsgun}. Other accelerator advances at the SLC included the application of 
{\it BNS damping} \cite{balakin1983bns} to suppress
the single-bunch beam break up \cite{Chao:1980gr} 
(a kind of head-tail instability occurring in 
linear accelerators) 
and the corresponding emittance growth \cite{seeman1992bnsslc}, a pulse-by-pulse IP position feedback system, implementation of 
sophisticated nonlinear optics knobs,  
procedures for the frequent tuning of various  
IP optics aberrations 
\cite{Emma:1997cp,Hendrickson:1999rv},  
and a high-efficiency positron source 
\cite{Clendenin:1988np}, providing more than
$5\times 10^{12}$ $e^+$ per s for injection into the 
SLC linac \cite{Krejcik:1992aa}. 
The SLC also pioneered the beam-beam deflection 
scans for IP beam-size diagnostics \cite{Bambade:1989pb}
and, for the first time, observed 
beamstrahlung \cite{Bonvicini:1989ci}, 
i.e., the synchrotron radiation emitted
during the collision in the electromagnetic 
field of the opposing bunch, and exploited
it for diagnostics purposes. 
The SLC also demonstrated a significant 
increase of 
luminosity, by more than a factor of two, 
due to ``disruption enhancement,'' i.e.,
the mutual focusing of the colliding electron
and positron bunches at the interaction point \cite{Barklow:1999rs}.
During the decade of its operation, from 1989 to 1998, the SLC produced close to 600,000 $Z$ bosons -- about 3\% of LEP production -- but with a longitudinally polarized electron beam,
allowing the SLC's experiment SLD to perform the
world's single most precise measurement of the 
weak mixing angle $\sin^{2} \theta_{W}^{\rm eff}$ \cite{PhysRevLett.84.5945}. 

\begin{figure}[htbp]
\centering
\includegraphics[width=0.99\linewidth]{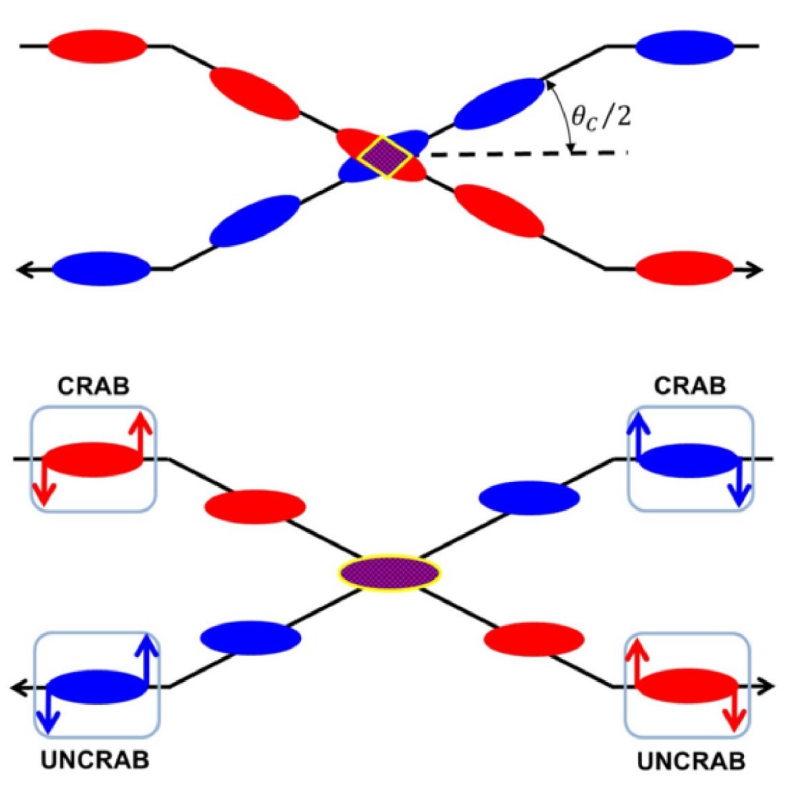}
\caption{Top: beam collision scheme with crossing angle suffers from geometric luminosity reduction. Bottom:  crab-crossing scheme that results in full bunch overlapping and thus maximum luminosity. Deflecting RF cavities generate a null kick 
to the center of the bunch while
its head and tail receive opposite transverse kicks (from \cite{verdu2016crabcavities}).}
\label{crabcavities}
\end{figure}

Two particle factories that aimed for precision measurements with luminosities far exceeding those of its predecessors (in particular CESR) operated 
during the first decade of the 21st century.  
These were the two $B$-factories, PEP-II at SLAC \cite{zisman1993pepcdr} and KEKB at KEK \cite{toge1995kek}. They were conceived as asymmetric (unequal energy) two-ring electron-positron colliders, constructed to measure the properties of  the $b$ quark sector, the $CP$ violation and to confirm the CKM matrix \cite{bevan2014bfactoryphysics}. The energy of positrons was much lower than that of electrons, so that the created $B$ and $\bar {B}$ mesons had significant forward momentum away from the collision point, making it easier for detectors to pinpoint the origin of the $B$ particles' decay products. Table \ref{bfact} presents 
the beam parameters achieved at these colliders.  
After a few years of operation, 
both PEP-II and KEKB introduced a so-called {\it top-up injection} mode of operation \cite{Seeman:2015gpa,Satoh:2010zz}, where 
small amounts of beam are injected quasi-continually, so as to keep the beam current and luminosity constant over long periods of time, e.g., a day, until the occurrence of a technical failure. The particle detectors remained active and continued data collection during, or shortly after, each beam injection. PEP-II and KEKB had sophisticated transverse and longitudinal bunch-by-bunch feedback systems to suppress coherent instabilities and other measures to allow storage of very high currents \cite{seeman2008lastpepii, oide2009kekb}. PEP-II holds the world record of stored positrons (at 3.2 A) and electrons (at 2.1 A). KEKB set the world record for highest luminosity at 2.1$\times$10$^{34}$ cm$^{-2}$s$^{-1}$. KEKB was also the first collider to use SRF {\it crab cavities} \cite{oide1989crabs} to tilt the bunches at the IP and avoid the geometric luminosity reduction due to the crossing angle $\theta_c$,  Eq.~(\ref{crossingangle}) --- see Fig.~\ref{crabcavities}. Luminosity improved by a modest 10--20\%; the vertical beam-beam parameter $\xi_y$ increased from 0.06 to 0.09, less than what had been expected from simulations (0.15) \cite{oide2014rastreview,funakoshi2014}. 
One possible explanation for the discrepancy is residual nonlinear optics aberrations at the collision point \cite{funakoshi2014}.

\subsection{Past advances of hadron colliders}
\label{ppcolliders}

The Intersecting Storage Rings (ISR) at CERN \cite{johnsen1973isr} was the world's first $pp$ collider. It was made up of two independent, interleaved normal$-$conducting synchrotron rings intersecting  at  eight  points,  five  of  which  were  used  for experiments. The ISR physics program aimed at achieving an understanding of proton structure at the c.m.e.~levels, exceeding the most powerful fixed target machines of the SPS at CERN and the Main Ring at Fermilab \cite{giacomelli1979ISRphysics}, both of which were constructed after the start of ISR operation. 
The machine relied on a process called  {\it momentum  stacking} to accumulate record high  currents (up to 60 A) and achieved luminosities in hadron collisions surpassed only two decades later \cite{myers2010isrsps}. The discovery of {\it Schottky noise} resulting  from  the  discrete  nature  of   particles   in   the   beam led to its extensive use for diagnostics of unbunched (coasting or DC) beams and allowed the first successful demonstration of stochastic cooling and reduction of beam emittances \cite{vandermeer1972stochasticISR, bramham1975stochastic}. 

S$p \bar{p}$S, the next collider at CERN, was built as a modification of the Super Proton Synchrotron (SPS), with the goal of discovery of the massive neutral intermediate vector bosons \cite{rubbia1977producing}, successfully achieved in 1983 (1984 Nobel Prize in Physics, Carlo Rubbia) \cite{rubbia1985nobellecture}. Most critical for the success of the S$p \bar{p}$S was the stochastic cooling of antiprotons (1984 Nobel Prize in Physics, Simon van der Meer), which took place in a specially constructed 3.5 GeV Antiproton Accumulator ring and allowed accumulation of up to 6$\times$10$^{10}$ $\bar{p}$ per hour \cite{vandermeer1985}. 

\begin{figure}[htbp]
\centering
\includegraphics[width=0.99\linewidth]{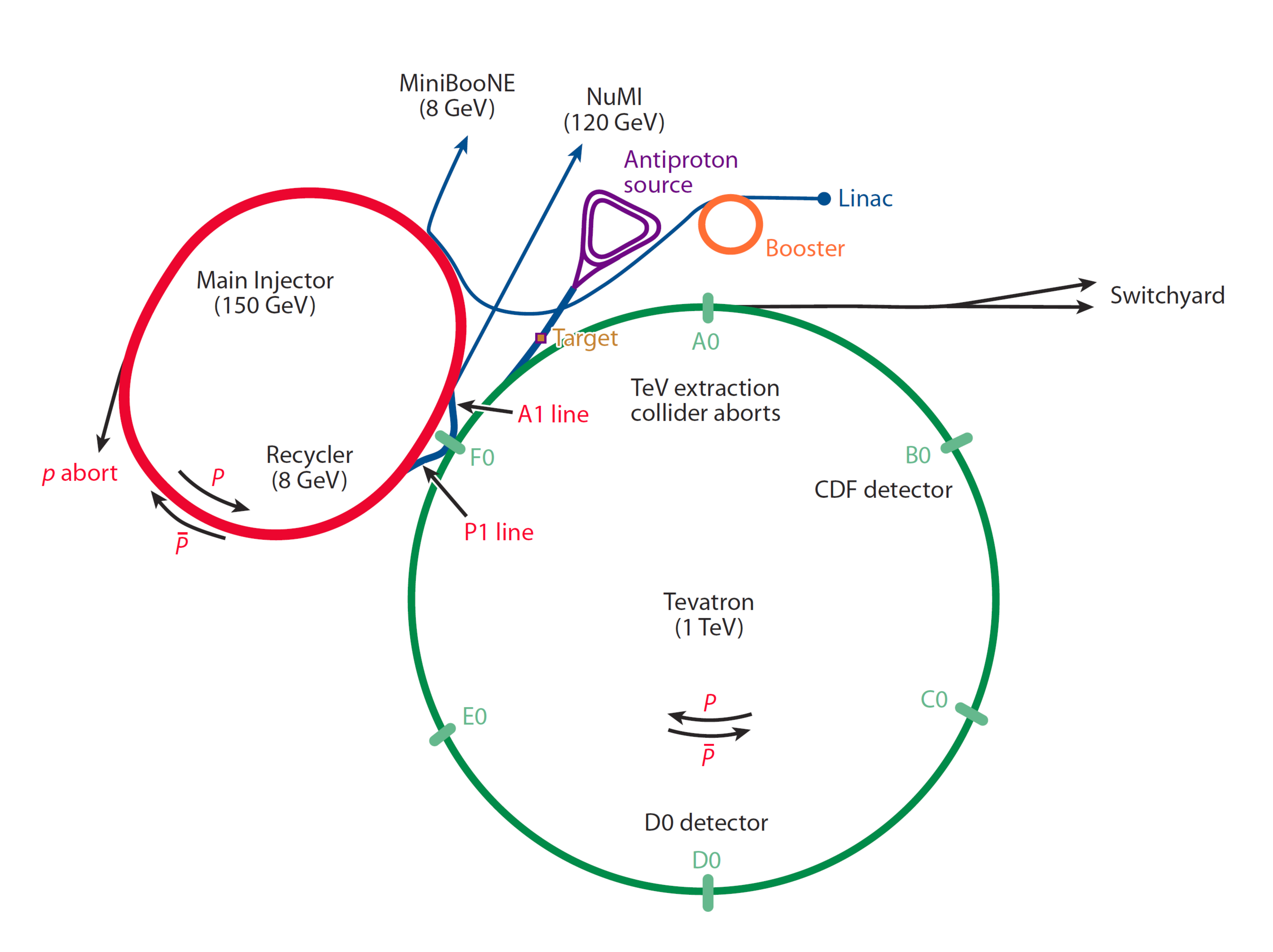}
\caption{Layout of the Fermilab accelerator complex. The accelerators are shown to scale; the radius of the
Tevatron is 1.0 km. Proton beam energy out of the linac is 400 MeV and 8 GeV out of the Booster synchrotron; the energy of antiprotons in the antiproton source (triangular shape Debuncher and Accumulator) is also 8 GeV (from \cite{holmes2013legacy}).}
\label{Tevatron}
\end{figure}

The first superconducting synchrotron in history, the Tevatron \cite{edwards1985tevatron} was also converted into a $p \bar{p}$ collider in 1985 \cite{dugan1989tevatron}. It was the highest energy collider for 25 years and its legacy includes many results for which the high energy of $\sqrt{s}$=1.96 TeV was decisive, such as the discovery of the top quark in 1995, and precise measurements of the masses of the top quark and W boson \cite{quigg2011Tevatron}.  It was also a pioneering instrument that advanced the frontiers of accelerator science and technology \cite{holmes2013legacy, lebedevshiltsev2014tevatron}. Its 4.5 T dipole magnets employed Nb-Ti superconducting cable operating at 4.5 K \cite{tollestrup2008}, requiring what was then the world's largest cryogenic system \cite{fowler1990tevatroncryo, norris1989tevatroncryo}. The antiproton production complex  \cite{church1993antiprotonsource} consisted of three 8 GeV $\bar{p}$ accelerators (the Accumulator, Debuncher, and Recycler --- see Fig.~\ref{Tevatron}), and employed 25 independent stochastic cooling systems and one high-energy electron cooling set-up \cite{nagaitsev2006experimental} to  accumulate up to a record high 
value of 25$\times$10$^{10}$ $\bar{p}$ per hour. Over the years, some 10$^{16}$  antiprotons were
produced and accumulated at Fermilab (about 17 nanograms), more than 90\% of the world’s total man-made production of nuclear antimatter \cite{shiltsev2012Tevatronstory}. Despite severe parasitic long-range interactions of the two beams, each consisting of 36 bunches placed on helical orbits by two dozen $\pm$150 kV high-voltage (HV) separators, a total beam-beam tuneshift parameter of $n_{IP}\xi \approx$0.025-0.03 was achieved, a record for hadron beams \cite{shiltsev2005beambeameff}. Other notable advances included the first high energy accelerator built with permanent magnets  --- the 3.3 km 8 GeV Recycler \cite{jackson1996recycler},  
advanced longitudinal beam manipulation techniques of {\it slip-stacking} and {\it momentum mining} \cite{kiyomi2002slipstacking, bhat2004momentummining}, and the first operational use of electron lenses \cite{shiltsev2008teldesign,  shiltsev2015electronlenses} for beam collimation \cite{zhang2008DCbeam, stancari2011collimation} and for  compensation of long-range beam-beam effects \cite{shiltsev1999tels, shiltsev2007bbcdemowithtels}.  
The Tevatron ultimately achieved luminosities a factor of 430 higher than the original design specification. 

\subsection{Past advances of lepton-hadron colliders}
\label{epcolliderspast}

The first lepton-proton collider, the 6.4 km long Hadron-Elektron-Ring-Anlage (HERA) at DESY in Germany \cite{voss1994hera}, was the first facility to employ both applications of superconductivity: 5 T magnets in the 920 GeV proton ring and SRF accelerating structures to provide about 12 MW of RF power to compensate for synchrotron radiation losses of 30 GeV lepton beams (positrons or electrons). 
With proper orbit and optics control, 
the HERA lepton beam would naturally become transversely polarized to about 60\% 
(within about 40 minutes) thanks to the Sokolov-Ternov effect \cite{barber1994herapolarization}. Special magnets called {\it spin rotators} were implemented on either side of the collider IPs to produce 30--45\% longitudinal polarization at the experiments \cite{buon1986heraspinrotators, barber1995firstspin}. HERA operated from 1992 to 2005 at $\sqrt{s}$ of about 320 GeV and luminosities of up to 3--5$\times 10^{31}$~cm$^{-2}$s$^{-1}$ \cite{hera}, and allowed the investigation of deep-inelastic and photoproduction processes at then highest energy scales \cite{klein2008HERAphyscs}. 

%
%

\section{Modern colliders}
\label{moderncoll}

Naturally, the colliding beam facilities of the present utilize many of the advances of past machines in order to operate at the energy or luminosity frontier, or at both. The challenges they face are quite unique and formidable.   

\subsection{Modern $e^+e^-$ colliders}
\label{moderne+e-}

\subsubsection{VEPP-4M and BEPC-II}
\label{vepp4mbepcii}

Two of the currently operational lepton colliders --- VEPP-4M in Novosibirsk and BEPC-II at IHEP (Beijing, China) --- were originally constructed in the 1980s \cite{Xu1983bepc, blinov1983vepp4}, but went through a long series of optimizations and upgrades, continually  contributing important research in modern particle physics. 
The BEPC (Beijing Electron Positron Collider) was originally built as a single ring collider to produce tau and charm particle physics, but it was later upgraded to a double-ring 
high-luminosity factory, with up to 2.1 GeV per beam and some 90 bunches. The RF system comprises two SC single cavities at 500 MHz. Longitudinal instability in $\sim$1A beams originating from high order modes in the cavities initially limited the luminosity, though this problem was recently resolved  through a bunch-by-bunch longitudinal feedback system. The machine reached a record luminosity of 10$^{33}$cm$^{-2}$s$^{-1}$ \cite{qin2012bepcii} at the $\psi$ resonance with $\sqrt{s}=3.77$GeV \cite{yifang2008besIII}. As of early 2019, the BESIII experiment at BEPC-II finished accumulating a sample of 10 billion $J/\psi$ events - the world's largest dataset produced directly from $e^+e^-$ annihilations \cite{besIII2019record}.

The latest of several upgrades of the single ring VEPP-4M collider operating in a wide beam energy range of 
0.92--5.2 GeV is a new injection  complex \cite{emanov2018injectionvepps} that now comprises a  270  MeV  $e^-$  linac,  a  510  MeV
$e^+$  linac,  injection  channels,  and a  damping
ring; this is followed by the 350 MeV to 2 GeV  booster-accumulator VEPP-3 ring (which by itself was a $e^+e^-$ collider for a short time in the mid 1970s \cite{pestov1976vepp3}). Single bunch currents were originally limited at about 7 mA by beam-induced wakefields in the vacuum chambers, but commissioning of the transverse feedback system allowed a many-fold increase to about 25 mA \cite{blinov2014statusvepp4}. Eight pairs of electrostatic separation plates allow operation of $2\time2$ bunches in the pretzel orbit scheme. Unique to VEPP-4M is its ability to operate over a wide range of energies and precise determination  of beam  energy using the resonant depolarization method, with a record high absolute accuracy of 10$^{-6}$. The former is used in studies of two-photon processes such as $\gamma \gamma \rightarrow {\rm hadrons}$, while the latter allows measurements of the masses of the $J/\psi$, $\psi(2s)$, and $\psi(3770)$ mesons, and the tau lepton with record 
accuracy \cite{aulchenko2003Jpsimass, shamov2009taumass, anashin2010D0mass}.

\subsubsection{VEPP-2000}
\label{vepp2000}

Significant innovations in beam physics led to the latest of the Novosibirsk colliders, VEPP-2000 which consists of a single ring, 
with two detectors and two-fold symmetry \cite{shatunov2000vepp2000project}. 
The VEPP-2000 physics program in the 
range $\sqrt{s}$=0.3 – 2 GeV includes precise measurements of the total hadronic cross section, exclusive hadronic channels, two photon physics, tests of higher order quantum electrodynamics processes, and studies of nucleon form factors at the threshold of the reaction   $e^+e^- \rightarrow nn, pp$  \cite{shemyakin2016vepp2000, achasov2017vepp2000}. As in other beam-beam  limited machines, the VEPP2000 luminosity for a fixed machine lattice scales as $\Lumi \propto \gamma^4$. The collider  design exploits  the  {\it round beam  concept} \cite{danilov1996roundbeamsconcept} that provides additional stability to particle dynamics, even in the presence of non-linear beam-beam forces via conservation of angular momentum $M=xy'-yx'$. This scheme requires equal emittances $\varepsilon_x=\varepsilon_y$, equal fractional tunes $Q_x=Q_y$, equal amplitude functions at the IPs $\beta_x=\beta_y$, and no betatron coupling in the collider  arcs. This scheme was implemented in VEPP2000 by placing two pairs of 13 T superconducting final focusing solenoids into two interaction regions symmetrically with respect to the IPs \cite{shatunov2016statusvepp2000}. Observations showed that an extremely high beam-beam parameter $n_{IP}\xi_y$=0.25 (similar to LEP in the presence of strong radiation damping) was achieved and that round beams led to significant luminosity enhancement \cite{berkaev2012vepp2000, shatunov2018vepp2000}. 

\subsubsection{DA$\Phi$NE}
\label{dafne}

\begin{figure}[ht]
\begin{center}
\includegraphics[width=0.9\columnwidth]{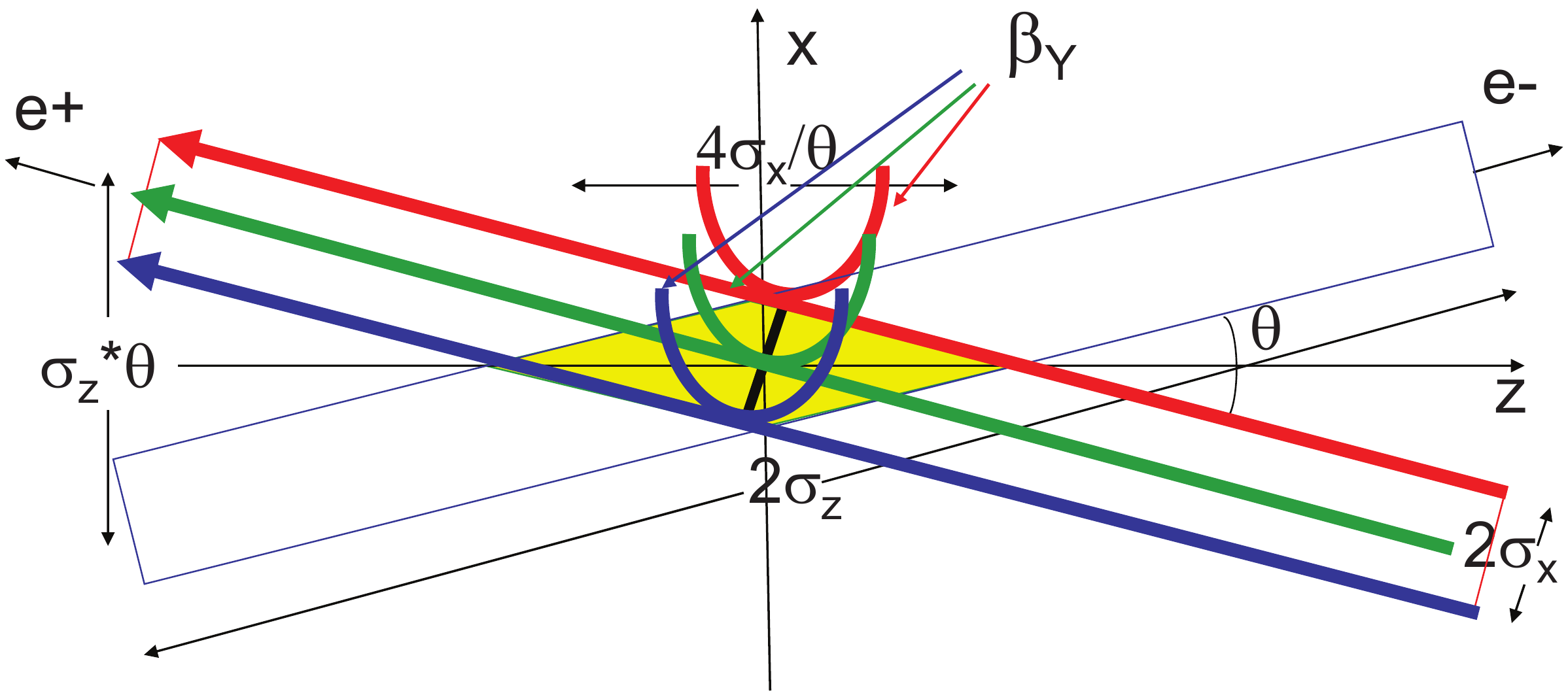}
\end{center}
\centerline{a) Crab sextupoles OFF}
\begin{center}
\includegraphics[width=0.9\columnwidth]{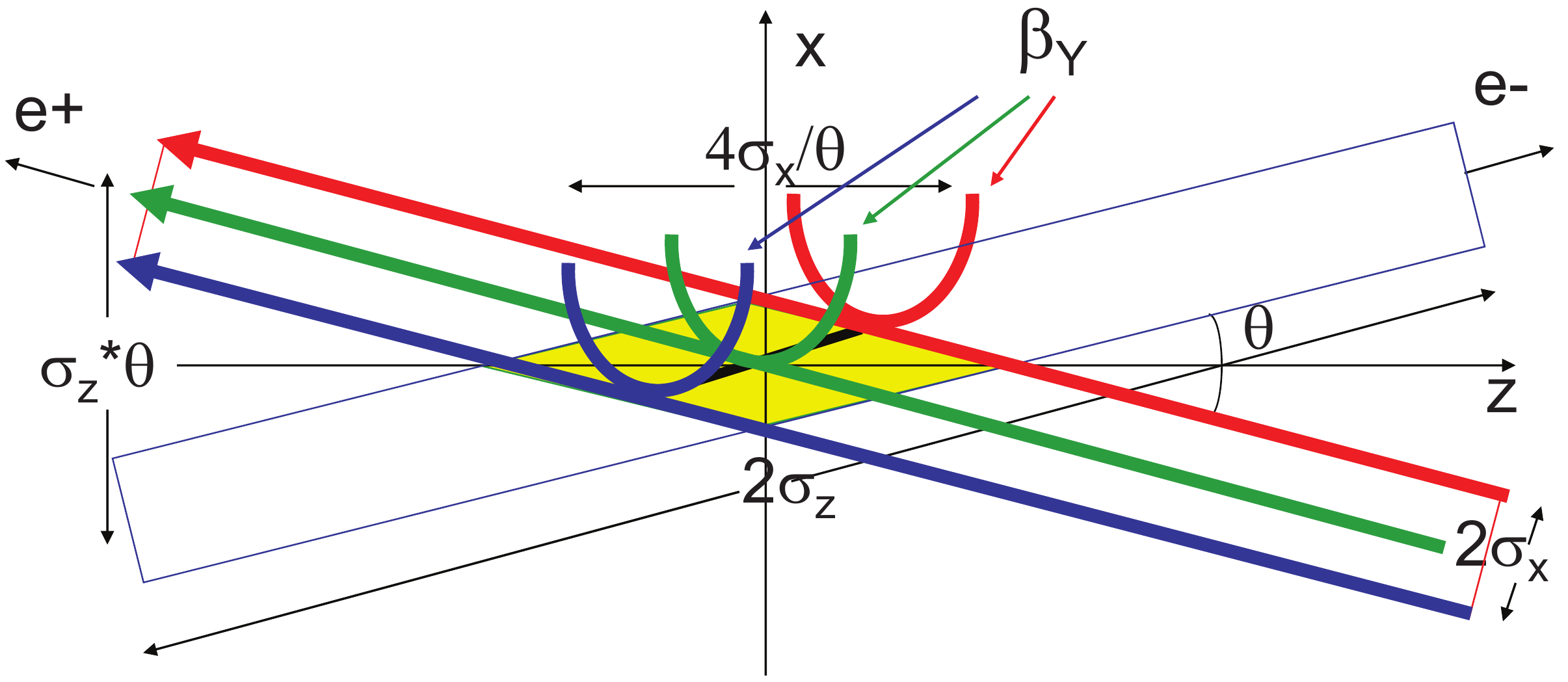}
\end{center}
\centerline{b) Crab sextupoles ON}
\caption{Crab waist collision scheme 
\protect\cite{zobovhandbook}.}
\label{fg:Zobov1_252}
\end{figure}

 DA$\Phi$NE at Frascati \cite{vignola1996dafne} was built in the late 1990s to operate at the  energy  of  the $\phi$-resonance (1.02 GeV c.m.e.), which with high probability decays to $K^+K^-$, enabling precision measurements of $K$-meson physics \cite{ambrosino2006kloe, bazzi2011siddharta}.  
In 2008, a new collision  {\it crab-waist} scheme proved to be effective for decreasing $\beta_y^{\ast}$ without shortening the bunch length, while also 
reducing the strength of beam-beam resonances at DA$\Phi$NE, tripling collider luminosity \cite{zobov2010crabwaisttest,zobov2016crabwaist}. 
The crab-waist collision combines a large {\it Piwinski angle} $\Phi=\sigma_z \tan(\theta_c/2)/\sigma_x^*$ --- see Eq.~(\ref{crossingangle}) --- with the cancellation of the resulting 
synchro-betatron resonances which occur under conditions of $kQ_x+lQ_y+mQ_s=n$, where $k, l, m$ and $n$ are integers \cite{piwinski1986synchro} by means of electromagnetic sextupoles with a special betatron phase advances to the collision point 
\cite{raimondi2006carbwaistworkshop, raimondi2007arxivcrab}. 
The crab waist collision scheme had first been proposed in 2006 for the former   
Italian SuperB project \cite{raimondi2006carbwaistworkshop}. 
Its key concepts and resulting merits  \cite{zobovhandbook},
can be understood from 
Fig.~\ref{fg:Zobov1_252}, which presents two bunches colliding under a horizontal crossing angle $\theta$. 
The {\it first ingredient} is a large Piwinski angle $\Phi=(\sigma_{z}/\sigma_{x})\tan(\theta_c/2)\gg 1$, as it had earlier been proposed for hadron colliders \cite{frfz}. 
In the crab-waist scheme, $\Phi$  is
increased by decreasing $\sigma_{x}^{\ast}$ and increasing $\theta_c$. In this way, the luminosity increases and the horizontal tune shift decreases; the effect of any parasitic collisions around the primary collision point becomes negligible.
However, the most important effect is that the overlap area of the colliding bunches is reduced, since it is proportional to $\sigma_{x}/\theta_c$ (see Fig.~\ref{fg:Zobov1_252}).
As a {\it second ingredient}, the vertical $\beta$-function $\beta_{y}$ is made comparable to the overlap area size (i.e.~$\ll\sigma_{z}$):
\begin{equation}
\beta_y^\ast\approx\frac{2\sigma_x}{\theta_c}\cong\frac{\sigma_z}{\Phi}\ll\sigma_z \, .
\label{eq:cw1}
\end{equation}   
Reducing $\beta_{y}^{\ast}$ at the IP yields a luminosity increase at the same bunch current. 
In addition, if the bunch current is 
limited by $\xi_{y}$ --- which decreases when  $\beta_{y}^{\ast}$ is lowered --- the bunch current can be raised to further push up the luminosity.
The vertical synchrobetatron resonances are also suppressed \cite{Zobov4_252}.  
With a finite overlap region, decreasing  $\beta_{y}^{\ast}$ does not require 
an associated decrease in the bunch length, as would be required in the standard collision scheme because of the hourglass effect. The possibility of a longer bunch length also improves local higher-order-mode heating and any effects of  coherent synchrotron radiation. 
However, implementation of the above two ingredients excites  new beam-beam resonances, which may strongly limit maximum achievable tune shifts. For this reason, 
the crab-waist transformation was introduced \cite{zobov2010crabwaisttest, zobov2016crabwaist}, as a final and {\it third ingredient}. 
As seen in Fig.~\ref{fg:Zobov1_252} (bottom), the $\beta$-function waist of one beam is now  oriented along the central trajectory of the other beam. In practice, the rotation of the vertical $\beta$-function is accomplished by sextupoles placed on both
sides of the IP in phase with the IP (modulo $\pi$) horizontally and at $\pi/2$ betatron phase difference (modulo $\pi$)  
vertically. 
The integrated strength $B_3l$ of these sextupoles should satisfy the following condition, which depends on the crossing angle 
$\theta$ and on the $\beta$-functions at the IP  (superindex $^{\ast}$) and sextupole
locations (subindex ``sx''), 
\begin{equation}
B_3l=\frac{p}{e}\frac{1}{2\theta_c}
\frac{1}{\beta_y^\ast \beta_{y,{\rm sx}}} \sqrt{\frac{\beta_x^\ast}{\beta_{x,{\rm sx}}}} \, ,
\label{eq:cw2}
\end{equation} 
where $e$ is the particle (electron) charge and $p$ the design momentum.
The main effect of the crab waist transformation is the suppression of betatron and synchrobetatron 
resonances arising (in collisions without crab waist) due to vertical tune modulation by horizontal betatron oscillations. The collision of flat beams with $\sigma_y^* \ll \sigma_x^*$ is
an essential condition for resonance suppression with the crab sextupoles \cite{Zobov6_252}.
The implementation of crab-waist collisions at DA$\Phi$NE  provided an increase in luminosity by a factor of 3, in good agreement with numerical simulations \cite{Zobov7_252,Zobov6_252}. 
All ongoing or proposed projects for next-generation circular 
lepton factories are based on the crab-waist scheme: SuperKEKB,   
the Super $\tau$-charm factories SCT \cite{sct} 
and HIEPA \cite{hiepa}, finally yet importantly 
the Higgs and
electroweak factories FCC-ee \cite{fccee,cwcoll,oideoptics}, and CEPC \cite{cepc}. 


\subsubsection{SuperKEKB}
\label{superkekb}
\begin{figure}[htbp]
\centering
\includegraphics[width=0.99\linewidth]{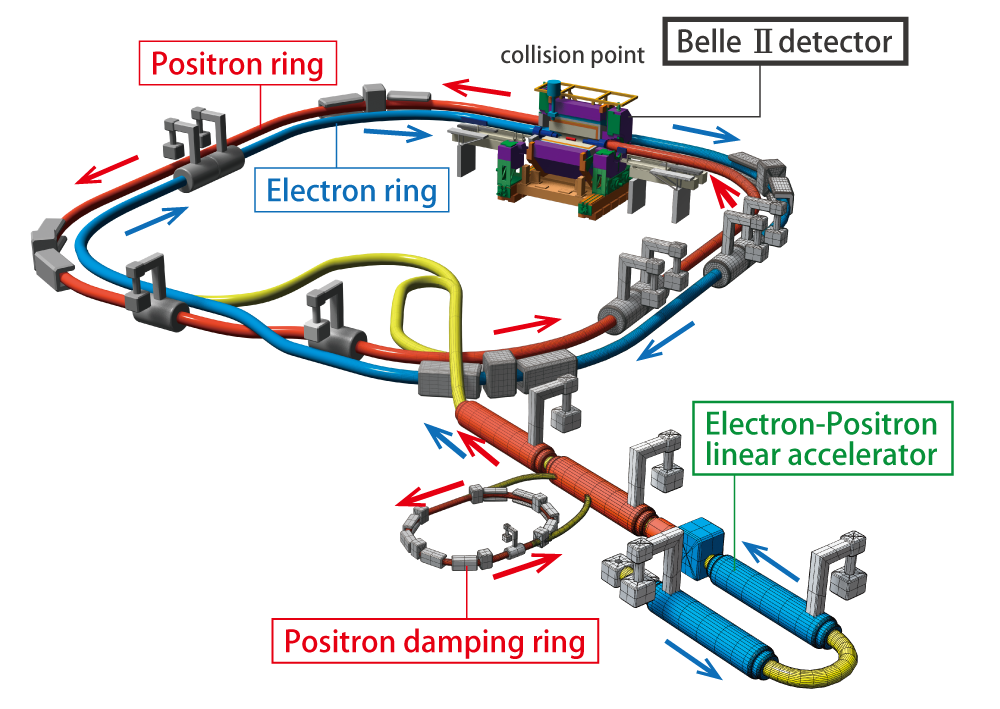}
\caption{Schematic of SuperKEKB (Courtesy: KEK).}
\label{SKEKBschematic}
\end{figure}

The SuperKEKB at KEK is an $e^+e^-$ collider with a design peak luminosity of 8$\times$10$^{35}$cm$^{-2}$s$^{-1}$ \cite{ohnishi2013superKEKB} (40 times that of the KEKB B-factory, see above) at  $\sqrt{s}$ close to the mass of the $\Upsilon(4S)$ resonance making it a second-generation B-factory for the Belle II experiment \cite{abe2010belledesign}. SuperKEKB is an asymmetric-energy and double-ring collider with a 7 GeV electron ring and a 4-GeV positron ring; see Fig.~\ref{SKEKBschematic}.  
Its mission is to seek new physics beyond the Standard Model with a target integrated luminosity of 50 ab$^{-1}$. Referring to  Eq.~(\ref{eq:lumiee}), the luminosity gain of 40 with respect to its predecessor KEKB is achieved with the same beam-beam parameter $\xi_y \approx 0.09$, twice as high a beam current $I_e$, an order of magnitude smaller transverse emittances, and a vertical IP beta function 20 times smaller than that of KEKB, namely $\beta_y^*$=0.3~mm. The latter realizes the {\it nanobeam scheme}, named such because the vertical beam sizes at the IP can be squeezed to $\sim$50 nm.
The SuperKEKB nanobeam scheme is an adaptation of the original 2006/2007 crab-waist proposal for the Italian SuperB project \cite{raimondi2007arxivcrab} to SuperKEKB. The scheme involves a large horizontal crossing angle between two colliding beams $\theta_x \approx 83$mrad, such that the Piwinski angle is very large $\Phi=\sigma_z \tan(\theta_x/2)/\sigma_x^* \approx 20$, with a bunch length much longer than the beta function at the IP  $\sigma_z=6$~mm$\gg \beta_y^*$ and small horizontal and vertical emittances. Unlike a head-on collision scheme, here the  bunches intersect one another only at the short and narrow central parts close to the IP.
 The first $e^+e^-$ collisions in SuperKEKB occurred in April 2018, eight years after the end of the KEKB operation, and successful collider commissioning is underway \cite{ohnishi2018superKEKB,superkekbakai}. 
For the year 2020, SuperKEKB operation is planned with crab-waist collisions (see Ch.~\ref{dafne}) 
at $ \beta_y^*$ values in the mm range.

Table \ref{bfact} compares the achieved
parameters of  PEP-II and KEKB, with the design values of SuperKEKB \cite{superkekbakai}.

\begin{center}
\begin{table*}[ht]
{\small
\hfill{}
\begin{tabular}{|l|c|c|c|c|c|c|c|}
\hline
\hline
Parameter &  Unit & 
\multicolumn{2}{c|}{PEP-II} & 
\multicolumn{2}{c|}{KEKB} & 
\multicolumn{2}{c|}{SuperKEKB}   \\
& & \multicolumn{2}{c|}{(achieved)} & 
\multicolumn{2}{c|}{(achieved)} & 
\multicolumn{2}{c|}{(design)}   \\
\hline
ring &  & LER & HER & LER & HER & LER & HER \\ 
\hline  
Species & 
& $e^{+}$ & $e^{-}$ 
& $e^{+}$ & $e^{-}$ 
& $e^{+}$ & $e^{-}$ \\
Beam Energy & GeV &
 3.1 & 9.0 &  
 3.5 & 8 & 4 & 7 \\
 Circumference & m & 
 \multicolumn{2}{c|}{2199}
 &
  \multicolumn{2}{c|}{3016} &
   \multicolumn{2}{c|}{3016} 
 \\ \hline
Horizontal IP beta function $\beta_{x}^{\ast}$ & mm &
1050 & 400 & 
1200 & 1200 &  32 &  25 
\\
Vertical IP beta function $\beta_{y}^{\ast}$ & mm &
9--10 & 9--10 & 
5.9 & 5.9 & 0.27 & 0.30
\\
Hor. rms norm. emittance $\varepsilon_{nx}$ & $\mu$m & 182
& 880 & 
123 & 376 &  25 &  63 
\\
Vert. rms norm. emittance $\varepsilon_{ny}$ & $\mu$m &
4.8 & 14  & 
1 & 3.4 & 0.07 & 0.3
\\
\hline
Beam current & mA &
3213  & 2069 & 
 1640  & 1190 & 3600 &  2600
 \\ 
Bunches / beam &  & 
 \multicolumn{2}{c|}{1658} &
 \multicolumn{2}{c|}{1584} &
  \multicolumn{2}{c|}{2500} \\
Bunch current & mA & 
1.94 & 1.25  & 
1.04 & 0.75 & 1.44 & 1.04 \\
Rms bunch length & mm & 
10--12 & 10--12 & 
7 & 7 & 6.0 & 5.0
\\ 
Full Crossing Angle & mrad & 
 \multicolumn{2}{c|}{$<0.05$} &
  \multicolumn{2}{c|}{0 (crab crossing)} &
   \multicolumn{2}{c|}{83}
\\ 
Vert.~beam-beam parameter $\xi_{y}$ & 
& 0.047 & 
0.062 & 
0.098 & 0.059 & 0.069 & 0.060 
\\ \hline
Luminosity & $10^{34}$~cm$^{-2}$s$^{-1}$ &
\multicolumn{2}{c|}{1.2} &
 \multicolumn{2}{c|}{2.11} &
 \multicolumn{2}{c|}{80} 
\\
\hline\hline
\end{tabular}}
\hfill{}
\caption{Design parameters of SuperKEKB \cite{superkekbakai} compared with past achieved parameters
in PEP-II \cite{seeman2006, pepiiseeman} 
and KEKB, distinguishing the respective Low and High Energy Rings (LER and HER). The beam-beam parameter in this table is computed without the hourglass factor or any geometric factors.}
\label{bfact}
\end{table*}
\end{center}

\subsection{Modern hadron colliders}
\label{modernhadron}


Two hadron colliders are presently in operation ---  
the Relativistic Heavy Ion Collider (RHIC) at 
Brookhaven National Laboratory (BNL)
and the Large Hadron Collider (LHC) at the
European Organization for Nuclear Research (CERN).
Both collide either protons (polarized protons in the case of the RHIC) or other, heavier ions, or protons with ions.
For the LHC, a high-luminosity upgrade (HL-LHC) has been approved and will come into operation around 2026.
Typical parameters of the RHIC and LHC, and design parameters for the HL-LHC are compiled in Table \ref{lhcrhic}.

\begin{center}
\begin{table*}[ht]
{\small
\hfill{}
\begin{tabular}{|l|cc|cc|cc|}
\hline
\hline
Parameter &
\multicolumn{2}{c|}{RHIC} & 
\multicolumn{2}{c|}{LHC} & 
\multicolumn{2}{c|}{HL-LHC}   \\
& \multicolumn{2}{c|}{ } & 
\multicolumn{2}{c|}{(2018)} & 
\multicolumn{2}{c|}{(design)}   \\
\hline
Species & $pp$ & $Au-Au$ & $pp$ & $Pb-Pb$ & $pp$ & $Pb-Pb$ \\ 
Max. beam energy (TeV) &
 0.255 & 0.1/n &  
 6.5 & 2.72/n & 7 & 2.76/n \\
 Circumference (km) & 
 \multicolumn{2}{c|}{3.834}
 &
  \multicolumn{2}{c|}{26.659} &
   \multicolumn{2}{c|}{26.659} 
 \\ 
Polarization  &
 55\% &  n/a & n/a 
  & n/a & n/a & n/a \\ 
\hline
Beta function at IP $\beta_{x,y}^{\ast}$ (m) &
0.65 & 0.7 & 
0.30$-$0.25 & 0.5 &  0.15 &  0.5 
\\
Transverse emittance $\varepsilon_{n}$ ($\mu$m, rms, norm.) & 3
& 2.2 & 
1.9 & 2.3  &  2.5 & 1.7    
\\
IP beam size ($\mu$m) & 85 
& 115 & 8 & 19 
& 7 &  17
\\
Beam current (mA) &
257  & 220  & 
 550 & 24  & 1100 & 33
 \\ 
Bunches per beam & 111 & 111 & 2556 & 733 & 2760  & 1232 \\
Bunch population ($10^{10}$) & 
18.5 &  0.2  & 
10$-$12.5 & 0.02  & 22 &  0.02 \\
Bunch length (rms, cm) & 
60 & 30 & 
8 & 7$-$10 & 9 & 8
\\ 
Full crossing angle ($\mu$rad) & 
 0 & 0 &
320--260 & 300 &
500 & $>200$ 
\\ 
Beam-beam parameter / IP, $\xi$ ($10^{-3}$) 
& 7.3 & 4.1 & 
4.5 & 1.1 & 8.6 & 1.1
\\ \hline
Luminosity ($10^{30}$~cm$^{-2}$s$^{-1}$) &
245 (peak) & 0.016 (peak) & $2.1\times 10^{4}$ & 0.007
& $5\times 10^{4}$ &
0.006
\\
& 
 150 (avg.) & 0.009 (avg.)  &  &  & (leveled) & (leveled) 
 \\ 
Max. int'd NN lum./experiment (fb$^{-1}$) & 
 1.3 at 250/255 & 0.03  &  169 & 0.04 & 250 / y &  0.43 \\ 
\hline\hline
\end{tabular}}
\hfill{}
\caption{Typical proton-proton and heavy-ion parameters of
the RHIC and LHC, and design parameters for the HL-LHC
upgrade.}
\label{lhcrhic}
\end{table*}
\end{center}

\subsubsection{RHIC}
\label{rhic}

\begin{figure}[htbp]
\centering
\includegraphics[width=0.95\linewidth]{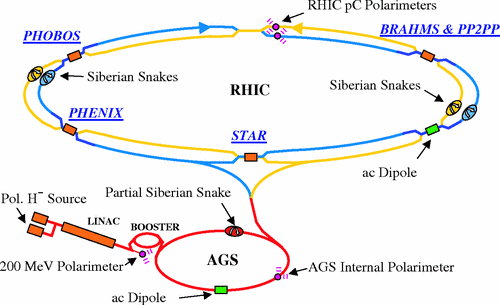}
\caption{Layout of the RHIC collider with its injector complex \cite{ranjbar}.
The two RHIC rings cross in six points. The two principal experiments still running are PHENIX and STAR. The smaller experiments PHOBOS, BRAHMS and PP2PP have been completed. The LINAC is the injector for polarized protons into the Booster/AGS/RHIC chain. A jet target used for precision beam polarization measurements. A tandem injector for ions has been replaced by an Electron Beam Ion Source (EBIS) starting with the 2012 run.
} 
\label{rhicphoto}
\end{figure}

The RHIC is a double ring which collides heavy ions and/or polarized protons circulating in opposite directions. It is based on SC Nb-Ti dipole magnets with a field of 3.45 T and housed in the 3.84 km tunnel previously built for the abandoned ISABELLE project. The two RHIC rings cross at six IPs. Two large experiments, STAR and PHENIX, are located at the interaction points referred to as six and eigth o'clock, respectively --- see Fig.~\ref{rhicphoto}. The PHENIX experiment is presently undergoing a major upgrade to become sPHENIX.

The RHIC brings into collision  combinations of fully-stripped ions such as {\it H-H (p-p), p-Al, p-Au, d-Au, h-Au, Cu-Cu, Cu-Au, Au-Au}, and {\it U-U} over a wide energy range. The high charge per particle (+79 for gold, for instance) makes IBS of particles within the bunch a special concern, even for moderate bunch intensities. 

3D stochastic cooling of bunched ion beams was successfully implemented in the RHIC in 2012 \cite{accel:Brennan} and is now routinely used. With stochastic cooling, steady increases in bunch intensity, and numerous other upgrades, the RHIC now operates with average luminosity in Au-Au collisions of $90 \times 10^{26}$~cm$^{-2}$s$^{-1}$ which is 44 times the design value. Another special feature of accelerating heavy ions in the RHIC is that the beams cross the ``transition energy'' during acceleration --- a point at which  $\gamma = \gamma_t \equiv 1/\sqrt{\alpha_c}$ and the derivative of the revolution period 
with respect to the momentum 
is zero, leading to zero synchrotron tune and temporary formation of very short and potentially unstable bunches according to Eqs. (\ref{alphac}, \ref{synchrotrontune}, \ref{sigmaz}). This is quite typical for  low-energy accelerators, where the necessary phase jump required of the RF system is implemented rapidly and little time is spent near this condition. In the case of RHIC accelerating heavy ions, SC magnets cannot ramp very quickly and the period of time spent crossing transition is relatively long and must be dealt with carefully. For $p-p$ operation the beams are always above their transition energy and so this condition is completely avoided. 

The RHIC physics program greatly relies on the machine's ability to accelerate and collide polarized proton beams \cite{bunce2000prospects}. Proton beam polarization is produced in a low-energy source \cite{zelenski2010review} and must be  maintained through numerous depolarizing resonances during the acceleration cycle \cite{bai2006polarized}. A proton beam energy of 255 GeV with 55\% final polarization per beam has been realized \cite{roser2008polarized, rhic2017status}. As part of a scheme to compensate for the head-on beam-beam effect, two electron lenses were installed; in 2015, these operated routinely during polarized proton operation at 100 GeV beam energy and doubled both peak and average collider luminosity \cite{accel:Fischer,rhic-bbc}.

RHIC physics searches for 
a critical point in the nuclear matter phase diagram \cite{stephans2006critrhic} required operation below the nominal injection energy of 10 GeV/u. In order to reach the integrated luminosity goals, the first bunched beam electron cooler, with electrons from a high-current high-brightness RF accelerator \cite{PhysRevAccelBeams.23.021003}, 
was successfully commissioned for the lowest RHIC energies \cite{fedotov2019bunchedbeamecool}.

\subsubsection{LHC}
\label{LHC}
The superconducting Large Hadron Collider is the world's highest energy collider \cite{evanslhc, bruning2012lhc}. 
It supports a broad particle-physics program at the energy frontier \cite{gianotti2005lhcphysics}.  

\begin{figure}[htbp]
\centering
\includegraphics[width=0.95\linewidth]{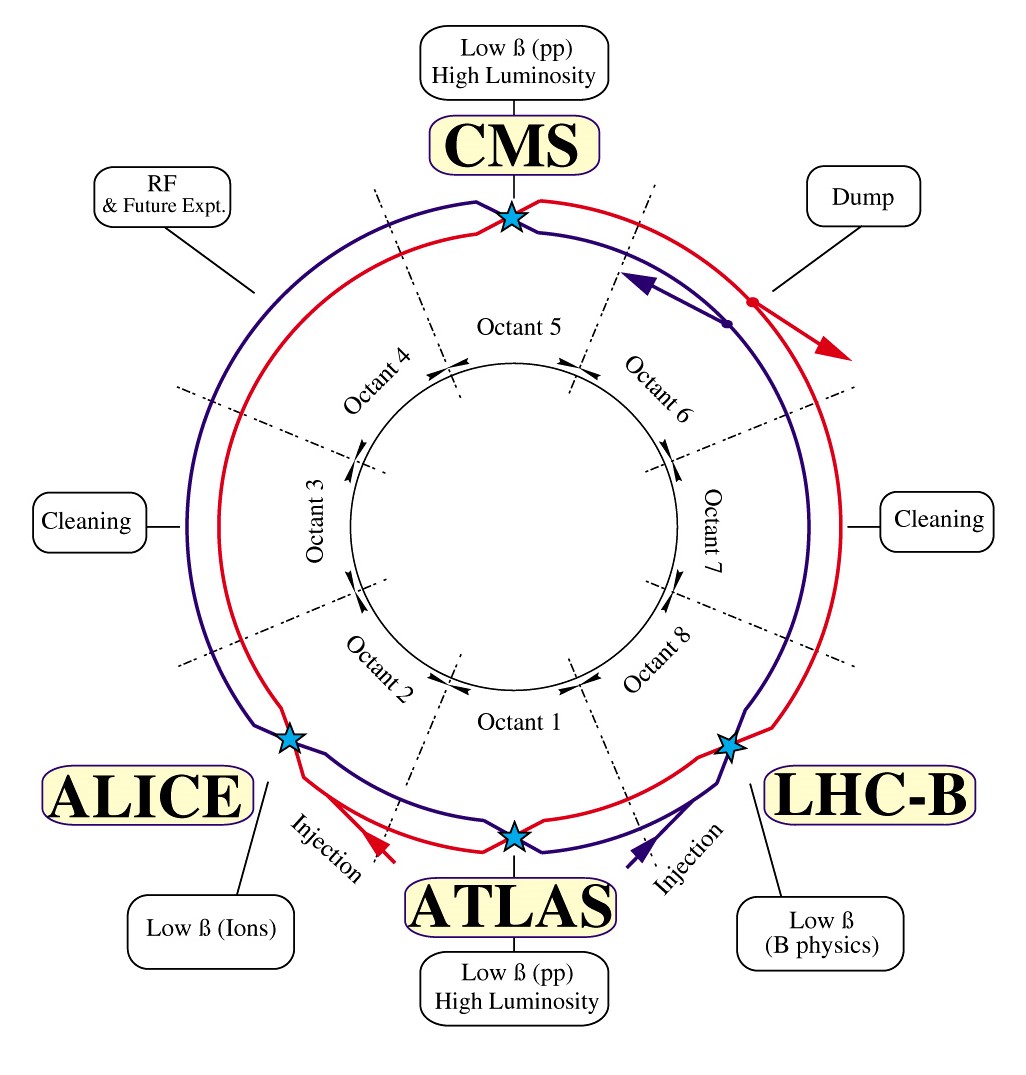}
\caption{Layout of the LHC double ring, with its eight long straight sections hosting two general and two special-purpose experimental detectors and/or devoted to specific accelerator functions, such as betatron collimation (cleaning), momentum collimation, beam extraction, RF systems and diagnostics, and injection.       
(Image credit: CERN). 
} 
\label{lhclayout}
\end{figure}

Over most of the LHC's 26.7 km circumference, the two counter-rotating hadron beams are contained in two separate vacuum pipes passing through the same 
superconducting  
twin-aperture Nb-Ti accelerator magnets.    
The LHC beams cross at four IPs, which host  two multipurpose high-luminosity experiments, ATLAS and CMS, and two special purpose experiments, ALICE (mainly 
devoted to heavy-ion physics) and 
LHCb (B meson physics). With four crossings, as shown in Fig.~\ref{lhclayout}, each beam passes half of a revolution on the outer side, the other half on the inner, so that the circumferences of the two beams are identical. Construction of the LHC technical components and their subsequent installation took more than a decade (1995--2007), and the machine's cost to CERN's budget was 3756 MCHF plus 1224 MCHF of labor cost \cite{evans2009large}; colliding beam operation started in 2010. Operation of the LHC machine requires some 120 MW of AC wall plug power that is about half of 230 MW for the entire CERN, whose annual electric energy consumption is about 1.3 TWh (2015) \cite{lhc2017faq}.

Table \ref{lhcrhic} shows the LHC luminosity performance in $pp$ and $Pb-Pb$ collisions. 
In $pp$ collisions the LHC has so far reached 
a world record luminosity of $2.1\times 10^{34}$~cm$^{-2}$s$^{-1}$, which, within the  measurement accuracy, equals the record for $e^+e^-$ colliders still held by KEKB.   
For the LHC's ATLAS and CMS experiments, 
in the first ten years of the LHC operation, 
the $Pb-Pb$ luminosity well exceeded the design value of $10^{27}$~cm$^{-2}$s$^{-1}$, 
 while for the ALICE experiment the luminosity needed to be ``levelled'' around this 
 value  \cite{Jowett:2648704}.  
The LHC can also provide $Pb-p$ collisions as it did in 2013 and 2016, and other ion-ion or  
ion-proton collisions, at different energies.

\begin{figure}[htbp]
\centering
\includegraphics[width=0.95\linewidth]{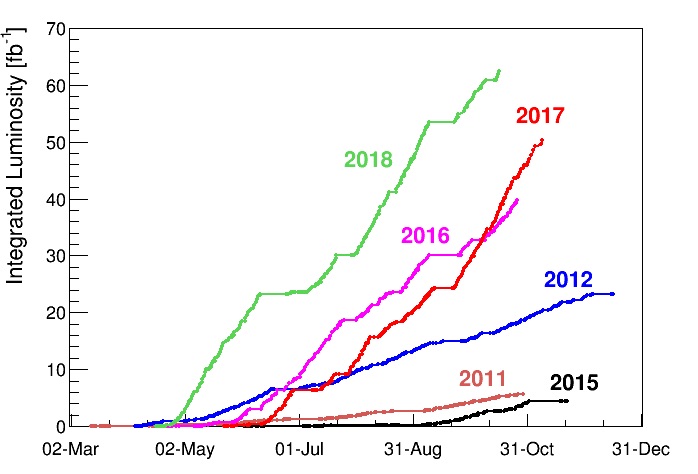}
\caption{Integrated yearly luminosity between 2011 and 2018 for proton operation (from \cite{lhcipac2019operation}). 
} 
\label{lhclumi}
\end{figure}

In the LHC Run 2 (from 2015 through 2018), operation for HEP was conducted with 6.5 TeV protons in each beam. The LHC has set many records for both peak and annual integrated luminosities of hadron colliders (see Fig.~\ref{lhclumi}), largely surpassing the total integrated luminosity of all previous hadron colliders combined. 
It is predicted that the final-focusing quadrupoles
around the ATLAS and CMS experiments 
will be destroyed by radiation from collision debris after a total integrated luminosity of around 300 fb$^{-1}$.
More than half of this value has already been delivered.
This provides motivation and guides timing for the 
High Luminosity LHC (HL-LHC) upgrade \cite{hllhc}, 
scheduled for around 2025, when the final quadrupole
triplets will be exchanged with new ones of larger aperture.

Initial luminosity measurements were conducted by sweeping beams transversely across each other (so-called {\it van der Meer scan} \cite{balagura2011vdm}) as was done long ago on the ISR \cite{vdm1968scan}. Both total and inelastic cross sections for $pp$ collisions were measured with high precision in the first years of LHC operation, e.g.~\cite{lhccross}. These are important for beam lifetime and for {\it event pile-up} (the number of interactions per bunch crossing) in the detectors. Notably, the LHC beam energy is known to 0.1\% and the orbit circumference slowly varies due to Earth tides by some 1.1 mm \cite{todesco2017lhcenergy}.

%
%

The extremely high $pp$ luminosities at the LHC, of up to $2.1 \times 10^{34}$~cm$^{-2}$s$^{-1}$,  \cite{lhcipac2019operation} are achieved by 
(1)  operating high-quality beams from the injector complex, presently comprising a 50 MeV proton linac (to be replaced, in 2020, 
by a 160 MeV H$^-$ linac), 1.4 GeV PS Booster (being upgraded to 2 GeV), the 
26 GeV Proton Synchrotron (PS), and the 450 GeV Super Proton Synchrotron (SPS), with transverse emittances 
that are more than 
$40\%$ lower than the design;  (2) smaller $\beta^{\ast}_{x,y}$, which got decreased from the design value of 55 cm down to 25 cm in 2018, also  thanks to the lower emittance allowing to avoid the aperture limitation in the final focus quadrupoles where the maximum beta-function grows according to Eq.(\ref{betastar}); and (3) by
a large number of bunches and a high beam current $I_b \ge$0.5 A.
An outstanding optics control and 
excellent optics reproducibility at the LHC 
\cite{tomas2012,persson2017,maclean2019} 
enabled achieving the aforementioned low value of 
$\beta^{\ast}$ with excellent beam-beam performance, 
and while guaranteeing the safety of the machine.  

In the LHC at energies of 5--7 TeV per proton, 
for the first time synchrotron radiation transforms from a curiosity 
to a challenge in a hadron accelerator.  
At design beam current, the system must remove roughly 7 kW 
due to synchrotron radiation. 
As photons are emitted, their interactions with the vacuum chamber wall can generate free electrons, with consequent {\it electron cloud} development \cite{arduini2013lhcecloud}. 
The heat load due to synchrotron radiation, electron cloud, and 
also beam image currents,  is intercepted by a special 
``beamscreen'' installed inside the magnets. 
The beamscreen temperature of $\sim$5--20 K 
is higher than the 1.9 K of the cold bore, which allows for efficient heat removal and for
cryopumping through numerous slots in the screen's top and bottom.  
 Overall, the LHC vacuum system comprises 
 150 cubic meters of beam vacuum and 9000 cubic meters of cryogenic vacuum; the LHC beam lifetime due to interaction with residual vacuum molecules is larger than 100 hrs \cite{vacuum2011lhc}. 

The LHC beam currents translate into stored energies of several hundreds of MJ per beam.  Component protection, beam collimation, and controlled energy deposition are consequently of high priority \cite{assmann2012lhccollim}. Of particular concern is the possible failure mode of an asynchronous beam dump, where a single extraction kicker module accidentally fires. This would trigger the firing of all other kickers (with some delay), but some bunches will be swept across the aperture.
These errant bunches would be intercepted on primary collimators, made of robust carbon-fibre-reinforced carbon to withstand such a catastrophic scenario. The LHC collimation system consists of more than 100 collimators, organized hierarchically. The measured cleaning performance, beam loss rates, and loss distributions have been consistent with expectations,  even during the delicate phase of the  $\beta^{\ast}$ squeeze \cite{lhccoll2019}. 
Beam particles scattered off the  short primary collimators (TCPs) are caught by longer secondary collimators (TCLs) placed at slightly larger apertures, with appropriate phase advances behind the primaries.
Tertiary collimators (TCTs) are placed in front of the final quadrupole triplets around the collision points. 
Special collimators protect against errors at injection sites and, especially, at the entrance of the beam extraction channel.
Still other collimators catch large amplitude debris particles coming from the collision point.
The system is designed such as to keep any beam loss in cold magnets to a minimum, and to ensure no magnet quenches even for a proton beam lifetime as low as 12 minutes.

At the LHC, where the two beams are brought together into a single common beam pipe at each of the four IPs, the large number of bunches, and subsequent short bunch spacing (25 ns), would lead to approximately 30 head-on collisions through 120 m of common beam pipe at each IP.  A small crossing angle is thus employed, which reduces the luminosity by about 15\%, with a similar reduction in the total beam-beam tune shift \cite{frfz}.  
Still, in the common beam-pipe section around each 
IP, the bunches moving in one direction will experience about 30 parasitic long-range encounters with counter-rotating bunches. At the LHC, the beams are  crossed alternatingly in the horizontal and vertical planes, so that to  first order the tune shift induced by the long-range collision cancels between the IPs \cite{peggsneuffer}.
For the HL-LHC, it is considered to compensate for residual perturbations of particle motion due to long-range parasitic collisions, e.g., with the help of current-fed wire compensators ~\cite{ypfz}. In that scheme, 1 m long thin current-carrying wires will be placed parallel to both beams at a normalized distance not much larger than the beam-beam separation at long-range collision points and, therefore, will provide similar but opposite action \cite{jpkwire,dorda}. Prototype wire compensation tests at the LHC have been successful \cite{ipac19sterbini}. 

As had been predicted \cite{fz97,ob98,mf98,fzfrgr}, 
an intensity limit has indeed  arisen at the LHC from the build up of an electron cloud inside the vacuum chamber \cite{arduini2013lhcecloud}. This electron cloud may drive different types of beam instabilities and creates additional 
significant heat loads on the beam screen inside the cold magnets. Indeed, the electron cloud is a primary source of beam instability in the LHC, especially with a proton bunch spacing of 25 ns. Beam performance tends to improve in 
time thanks to beam-induced surface conditioning (``scrubbing''). In addition, occasional losses of transverse or longitudinal Landau damping arise due to classical machine impedance with contributions from the resistive vacuum chamber, RF cavities, and chamber transitions. With regard to instability mitigation, the following lessons have been learned 
in operating the LHC 
\cite{emetral1,emetral2,emetral3,xbeam17}: 
(i) there exists a narrow range of machine settings for which the beam remains stable all along the cycle; (ii) instabilities occur if transverse betatron coupling exceeds a certain threshold value (different at different stages of operation); (iii) chromaticity settings are crucial along the cycle and cannot be relaxed; 
(iv) second-order chromaticity can contribute to beam stabilization \cite{schenk}; (v) octupole-magnet settings have to be adapted according to beam emittance; and (vi) the transverse damper is indispensable to preserve beam stability all along the acceleration cycle.


Interestingly, the electron cloud can drive coherent instabilities even when beams are in collision, with associated strong Landau damping. Simulations and earlier measurements at the SPS show that, for lower bunch intensities, the electron cloud in the dipoles tends to form a central stripe.  At the LHC, the central density threshold of the electron-cloud driven single-bunch head-tail instability
 ($\sim 5\times 10^{11}$~m$^{-3}$ at a chromaticity of  
$Q'\approx 15$) is crossed when the bunch intensity decreases; 
for $Q'>20$, the threshold becomes much higher. This explanation of beam instabilities observed towards the end of LHC physics fills is also consistent with the disappearance of the phenomenon after scrubbing.

Heavy-ion luminosity at the LHC can be limited 
by the so-called bound-free pair production (BFPP) during the collision 
of $Pb$ nuclei:
\begin{equation}
^{208}Pb^{82+} + ^{208}Pb^{82+}\rightarrow 
^{208}Pb^{82+} + ^{208}Pb^{81+} + e^+ \; .
\end{equation}
This process, with a large cross section of about $\sigma\sim 280$~barn, generates a secondary beam of $^{208}Pb^{81+}$ ions, 
with a fractional rigidity change equivalent 
to a relative momentum deviation of $\delta = 0.0124$, that can potentially quench superconducting magnets downstream of the IP
(see Ref.~\cite{Jowett:2648704} and 
references therein).  
In 2015, orbit bumps were introduced to displace the BFPP losses safely into a connection cryostat, thereby avoiding magnet quenches \cite{Jowett:2648704}.

In the coming years, the ambitious HL-LHC upgrade program \cite{hllhc} aims at an order-of-magnitude increase in integrated proton-proton luminosity. 
The heavy-ion physics program of the LHC will also continue during the HL-LHC period,
with approximately ten times higher peak 
luminosities in $Pb-Pb$ and $Pb-p$ collisions
than available at the present LHC. 

In the HL-LHC, the $\beta^{\ast}$ will be squeezed even further, to as low as 10 cm, with the help of a novel Achromatic Telescopic Squeeze (ATS) optics \cite{fartoukhats} (which is presently being tested and commissioned in the LHC), along with new larger aperture Nb$_3$Sn final quadrupoles \cite{rossi2019rast} and crab cavities.
Additional collimators  will be installed inside the dispersion suppressors around the main collimation (cleaning) insertion, and around some of the experiments.
The purpose of adding these collimators is to absorb off-energy particles generated during collisions (especially heavy-ion collisions) or by scattering off one of the existing primary or secondary collimators. 
The new collimator installation 
requires the replacement of several 8.3 T Nb-Ti dipoles by stronger and shorter 11 T dipoles made from Nb$_{3}$Sn superconducting cable to provide 
the necessary space without altering the overall geometry.

According to a recent proposal \cite{krasny2020}, the collision of low-emittance calcium ion beams in the HL-LHC promise partonic luminosities similar to, or higher than, the HL-LHC $pp$ operation, at a lower event pile up \cite{krasny2020}. 
The low-emittance beam would be produced by fast transverse laser cooling ($\sim 10$ s) of partially stripped calcium ions in the SPS, based on the {\it Gamma Factory} 
concept \cite{Krasny:2015ffb,Krasny:2018alc} (also see Ch.~\ref{muoncolliders}).

\section{Future Colliders}
\label{futurecolliders} 

Both nuclear physics and particle physics face critical questions which require next-generation colliding-beam facilities. Understanding of protons and neutrons, or nucleons --- the building blocks of
atomic nuclei --- has advanced dramatically, both theoretically and experimentally, over the past half a century. It is known that nucleons are made of fractionally charged valence quarks, as well as dynamically produced quark-antiquark pairs, all bound together by gluons, the carrier of the strong force. A central goal of modern nuclear physics is to understand the structure of the proton and neutron directly from the dynamics of their quarks and gluons governed by the quantum chromodynamics --- see more details in, e.g., Ref.~\cite{eic2018assessment} and references therein. 

For HEP to make significant advances, major new machines of two types will be required.
The first type is Higgs factories with a c.m.e.~of 240--250 GeV for precision studies of the Higgs boson ($m_H=125$ GeV) and exploration of the Higgs sector in greater detail, including measurements of Higgs couplings to fermions and vector bosons, self coupling, rare decays, mass and width. These Higgs factories could also furnish important complementary electroweak precision measurements at other $e^+e^-$ collision energies, e.g., on the $Z$ pole, above
the $WW$ boson threshold, and at energies sufficient for $t\bar{t}$  production.
The second type is colliders exploring the energy frontier for potential discoveries through direct searches with c.m.e.~levels significantly beyond those of the LHC.
The next energy-frontier colliders would aim at 
producing and discovering new particles/phenomena beyond the Standard Model, reaching mass-scales in the range of tens of TeV and offering a widely extended discovery reach for new gauge bosons $Z’$ and $W’$, colorons, diquark scalars, SUSY, heavy Higgs, test for compositeness of the Standard Model particles, etc.  

In addition, precision physics at future high-luminosity factories operating at the $\tau$-charm energy also provides sensitivity to new physics at multi-TeV energies and beyond. Reference~\cite{eppsu2020granada} presents a comprehensive review of the  emerging  particle physics landscape  and  its  potential  future.

Below we comprehensively detail colliders which are believed to be feasible (both technically and cost-wise) for construction over the next several decades. All of these rely mostly on currently available technologies, such as NC or SC RF and/or NC or SC magnets, and generally require either no or limited R\&D to assure energy reach and performance.  

\subsection{Ion, $e$-$A$ and $e$-$p$ colliders 
}
\label{ioneA}

\subsubsection{NICA}
\label{NICA}
NICA (Nuclotron-based Ion Collider fAсility) is a new accelerator complex under construction at the Joint Institute for Nuclear Research (JINR, Dubna, Russia) \cite{kekelidze2012nica}. Its purpose is to study properties of hot and dense baryonic matter,  spin physics, properties of the strong interaction vacuum and QCD symmetries, to explore the nature and properties of strong interactions between quarks and gluons, and to search --- for signs of the phase transition between hadronic matter and quark-gluon plasma plus new phases of baryonic matter \cite{sissakian2009nuclotron, senger2016nica, brodsky2016nica}. 

\begin{figure}[htbp]
\centering
\includegraphics[width=0.99\linewidth]{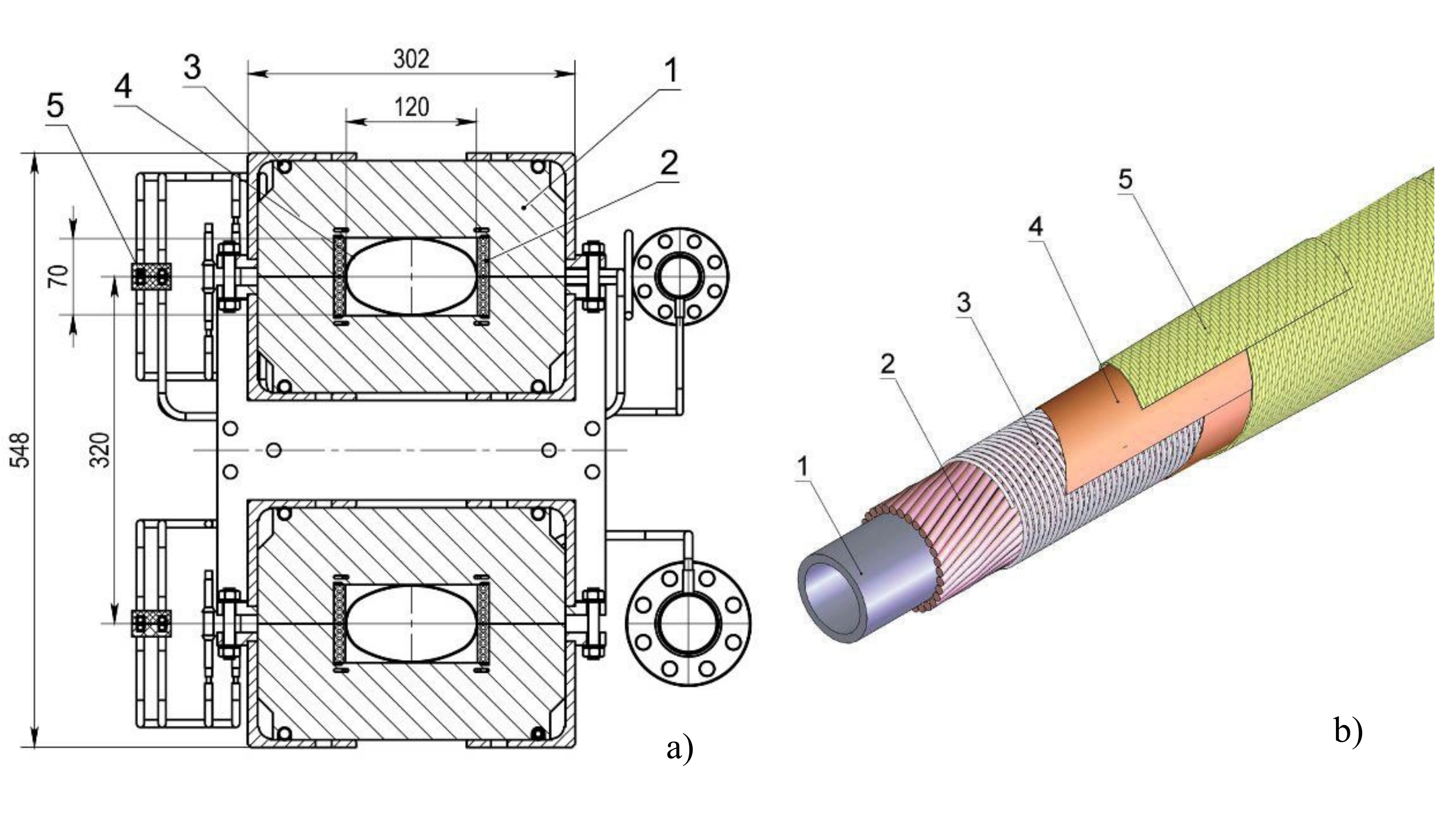}
\caption{Superferric 1.8 T magnets of the NICA collider: a) (left) cross-section of the magnet, based on a cold, window-frame iron yoke and a hollow superconductor winding: 1 --- lamination, 2 --- SC cable, 3 --- yoke cooling tube, 4 --- beam pipe, 5 --- current carrying bus bars. The magnets are placed in a 4.5 K cryostat (not shown); b) (right) 10.4 kA SC cable: 1 --- 3 mm diameter cooling tube, 2 --- Nb-Ti SC wire, 3 --- Ni-Cr wire, 4 and 5 --- insulation tapes (adapted from \cite{khodzhibagiyan2019nicamagnets}).}
\label{nicadipole}
\end{figure}

NICA will provide a variety of beam species, ranging from protons and polarized deuterons to massive gold ions. The collider average design luminosity in heavy ion and light ion collisions at $\sqrt{s_{NN}}$=4--11 GeV is $\Lumi$=1$\times$10$^{27}$cm$^{-2}$s$^{-1}$ for a variety of nuclei up to  $^{197}Au^{79+}$ and should be in the range $\Lumi$=(1--10)$\times$10$^{31}$cm$^{-2}$s$^{-1}$  for polarized proton and deuteron collisions in an energy range of  $\sqrt{s}$=12--27 GeV. The  facility employs some existing injectors such as light  ion  sources, an Alvarez type linac LU-20 based source  of  polarized  protons  and  deuterons, a new electron string ion source (ESIS) that will provide up to 2$\times$10$^9$ gold ions per 7 $\mu$s pulse at 50 Hz rate, and a linear accelerator  consisting  of  RFQ  and  RFQ  Drift Tube  Linac (RFQ DTL) sections. The linac accelerates ions with mass-to-charge ratio $A/Z \leq $8 up to an energy of 6 MeV/u, with efficiencies higher than 80\%.
A new 600 MeV/nucleon Booster synchrotron ring with a circumference  of  211 m will be housed inside the historical JINR Synchrophasotron yoke. Its  maximum magnetic rigidity of $B\rho=$25 Tm is provided by 40 1.8 T SC dipole magnets operating at 4.5 K, which can be ramped at 1.2 T/s \cite{kostromin2016boostermag}. The 60 keV electron cooling  system of the Booster, needed for ion accumulation and rapid (3--4 sec) reduction of the beam emittance up to 100 MeV/u energies, has been built and commissioned \cite{bubley2017commissioning, zinovyev2018start}. Ions, protons and deuterons  are then further accelerated, up to the  energy of the collider experiments,  using the upgraded Nuclotron synchrotron --- a 251.52 m circumference superconducting magnet ring, operational since 1993 \cite{issinsky1994nuclotron}. This ring has a maximum field of 2 T, ramping rate of 1 T/s for a 4 s cycle, and a maximum magnetic rigidity of 45 Tm. The collider itself will consist of two SC rings of racetrack shape, with maximum  magnetic  rigidity  of  45 Tm  and a circumference of 503.04 m. Two IPs are foreseen in opposite  straight  sections  of  the NICA collider --- one for heavy-ion studies with the multi-purpose detector (MPD) \cite{golovatyuk2016mpd} and another for polarized beams, housing the spin physics detector (SPD) experiment \cite{savin2015spd}. The maximum field of the collider dipole {\it superferric magnets}, which use iron to shape the field and superconductors to excite it, is 1.8 T \cite{khodzhibagiyan2019nicamagnets} --- see Fig.~\ref{nicadipole}. Intrabeam scattering is predicted to result in short emittance growth times of about 3 min at 1 GeV/u and about 40 min at 4.5 GeV/u --- see Table \ref{iieic}. Effective stochastic and electron cooling systems are required to counteract the emittance growth and to assure operation with a high average luminosity. 
In the energy range of 1 to 3 GeV/u, a 2.5 MeV, 0.5 A electron cooling system should provide a comparatively short 10 s cooling time and will allow operation of the collider at the space-charge limit of about $\Delta Q_{SC} \sim -0.05$. In the ion beam energy range of 3 GeV/u to 4.5 GeV/u, a stochastic cooling system will ensure characteristic cooling times of about 500 s \cite{kostromin2012beam}.

The NICA project cost is about \$500M. NICA   construction started in 2013. The first beam run of its injectors is scheduled for 2020 and the first colliding beams are expected in 2022 \cite{kekelidze2016nicanew, syresin2019nica}. 

\subsubsection{Low energy electron-ion collider proposals: ELISe at FAIR, EicC at HIAF}
\label{eicmedium}

\begin{figure}[htbp]
\centering
\includegraphics[width=0.99\linewidth]{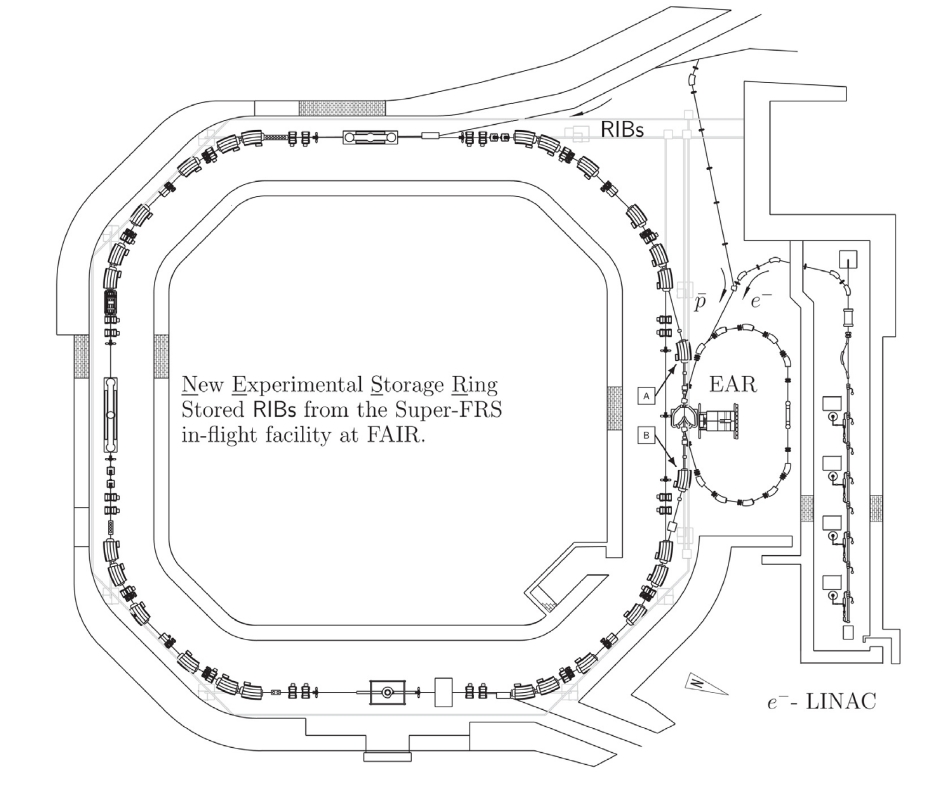}
\caption{Conceptual layout of the planned electron(antiproton)-ion collider hosting the ELISe experiment. The intersection region $A-B$ is situated in a bypass section to the New Experimental Storage Ring (NESR) and hosts a dedicated spectrometer  (from  \cite{antonov2011elise}).}
\label{elise}
\end{figure}

The electron-ion collider (EIC) experiment ELISe \cite{antonov2011elise} is part of the experimental program envisaged at the international Facility for Antiproton and Ion Research (FAIR \cite{gutbrod2006fairbtr}) in Darmstadt, Germany. 
It will offer the unique opportunity to scatter electrons with an energy of up to 0.5 GeV off short-lived exotic nuclei with energies up to 0.74 GeV/nucleon \cite{simon2007elise, suda2017elise}, in order to investigate the structure of radioactive isotopes. Figure~\ref{elise} presents the schematic layout of the New Experimental Storage Ring (NESR, circumference 222.9 m) \cite{dimopoulou2007nesrdesign} for Rare Isotope Beams (RIB) and the Electron Antiproton Ring (EAR, circumference 53.7 m, selected such that the revolution frequency of the EAR is 5 times that of one of the ions). Electrons with energies ranging from 125 to 500 MeV will be provided by an electron linac and stored in the EAR. Antiprotons of similar momentum can be directed from a dedicated collector ring (not shown in Fig.~\ref{elise}) into the EAR via a separate beam line. The electron ring is placed outside the NESR, so that a bypass beam line connects them and provides sufficient space for the electron spectrometer and a recoil detector system. The ion and electron (or antiproton) beam trajectories intersect at an IP surrounded by an electron spectrometer; auxiliary detectors for measuring reaction products are also included in the ELISe plan. 

Experiments require high resolution of transferred energy and momentum in electron-ion scattering.  A momentum spread of the electron beam (8 bunches, 5$\times$10$^{10}e^-$ each) of about 0.036\% can be achieved; its value mainly depends on intra-beam scattering (IBS) and statistical fluctuations due to synchrotron radiation. IBS also causes the beam size to grow and limits both luminosity and lifetime. Collision focusing optics with $\beta_{y,x}=$15/100 cm allows for
luminosity values ranging from $\Lumi$=10$^{28}$~cm$^{-2}$s$^{-1}$ to 10$^{30}$~cm$^{-2}$s$^{-1}$ for a wide variety of isotopes from He to U. The number of ions in each of 4 NESR bunches varies between 7$\times$10$^7$ and 8$\times$10$^9$, depending on the optimization of production and preparation of secondary beams, maximum yield, and the acceptance of the Super FRagment Separator (Super-FRS). At high intensities, the ion population is expected to be limited by space-charge effects at a tune shift parameter of $\Delta Q_{SC} \sim -0.08$ --- see Table \ref{iieic}. 

Construction work on the FAIR project began in the summer of 2017. The final scope of the project, consisting of many rings, and the construction schedule will depend on cost. In 2005, this was estimated to be 1,262 M Euros but was recently reconsidered; additional funding needed amounts to 850 M Euros, not including contingencies \cite{FAIRcost}.

Conceptually similar is a proposal \cite{chen2018eicc2} from the Institute of Modern Physics (IMP CAS, Lanzhou, China) to build a high luminosity polarized electron-ion collider in China (EicC) with $\Lumi$=4$\times$10$^{33}$cm$^{-2}$s$^{-1}$ at $\sqrt{s}=$12-24 GeV, based on the capabilities offered by the Heavy Ion High Intensity Accelerator Facility (HIAF \cite{yang2013hiaf}). The Huizhou HIAF project was approved in 2015, with construction commencing in 2018; facility commissioning is expected in 2025. The 2.5 billion Chinese yuan complex will operate a 180 m long superconducting 17 MeV/nucleon linac and a 569 m 34 Tm Booster Ring capable of accumulating, cooling and accelerating ions to 4.25 GeV/nucleon or protons to 9.3 GeV. The first stage of the complex extension to the electron-ion collider calls for an additional high current (3--4 A) 3.5-5 GeV electron ring to collide with up to 20 GeV HIAF protons and ions, an SRF 4--5 pass linac-ring electron injector, a polarized ion source, and Siberian snakes for the existing HIAF accelerators --- see Table \ref{iieic}. The final stage of the EicC,
called EicC-II, assumes the new construction of 1.5--2 km long, figure-8 shaped 60--100 GeV proton and 5--10 GeV electron rings in the same tunnel \cite{chen2018eicc}. Construction cost, as well as details of the design and schedule of the Chinese electron-ion collider, require further study.  

\subsubsection{High-energy electron-ion collider (EIC)  proposals: JLEIC at TJNAF and eRHIC at BNL}
\label{eichighenergy}

Higher energy electron-ion colliders can answer scientific questions which are central to completing our understanding of nuclear matter as well as being integral to the agenda of nuclear physics today. For example, the 2018 National Academies of Science assessment of US-based EIC science \cite{eic2018assessment} emphasized the priority of constructing a new facility that will be flexible over a multi-decade operating lifetime, and which can support exploration of nuclear physics over a wide range of center-of-mass energies and ion species with highly polarized electrons and light ions. According to the White Paper \cite{accardi2016eic}, the requirements of an EIC include highly polarized ($P_{e,n}\sim$70\%) electron and nucleon beams (as the precision of measurements of interest scales as $\Lumi P_e^2 P_n^2$), a spectrum of ion beams from deuterons to the heaviest nuclei (U or Pb), variable c.m.e.~values from $\sqrt{s}=$20 GeV to 100 GeV, upgradable to $\sim$140 GeV, high luminosities of 10$^{33-34}$ cm$^{-2}$s$^{-1}$, as well as possibilities of having more than one interaction region. 
Significant accelerator R\&D is needed to attain the required energy, luminosity, and polarization, including development of SRF crab-cavities and advanced SC magnets for collider rings and interaction region focusing, ERL based electron cooling of hadron beams, essential to attain luminosities two orders of magnitude beyond the predecessor HERA $ep$ collider, and polarized particle sources beyond the state-of-the-art, augmented by the development of special magnets and operational techniques to preserve the polarization through the acceleration process to the collisions.

Two multi-laboratory collaborations have evolved in the United States,
each of which proposes site-specific conceptual EIC designs based on infrastructure already available: the Jefferson Laboratory Electron Ion Collider (JLEIC) led by the Thomas Jefferson National Accelerator Facility (JLab), and eRHIC led by Brookhaven National Laboratory.  

\begin{figure}[htbp]
\centering
\includegraphics[width=80mm]{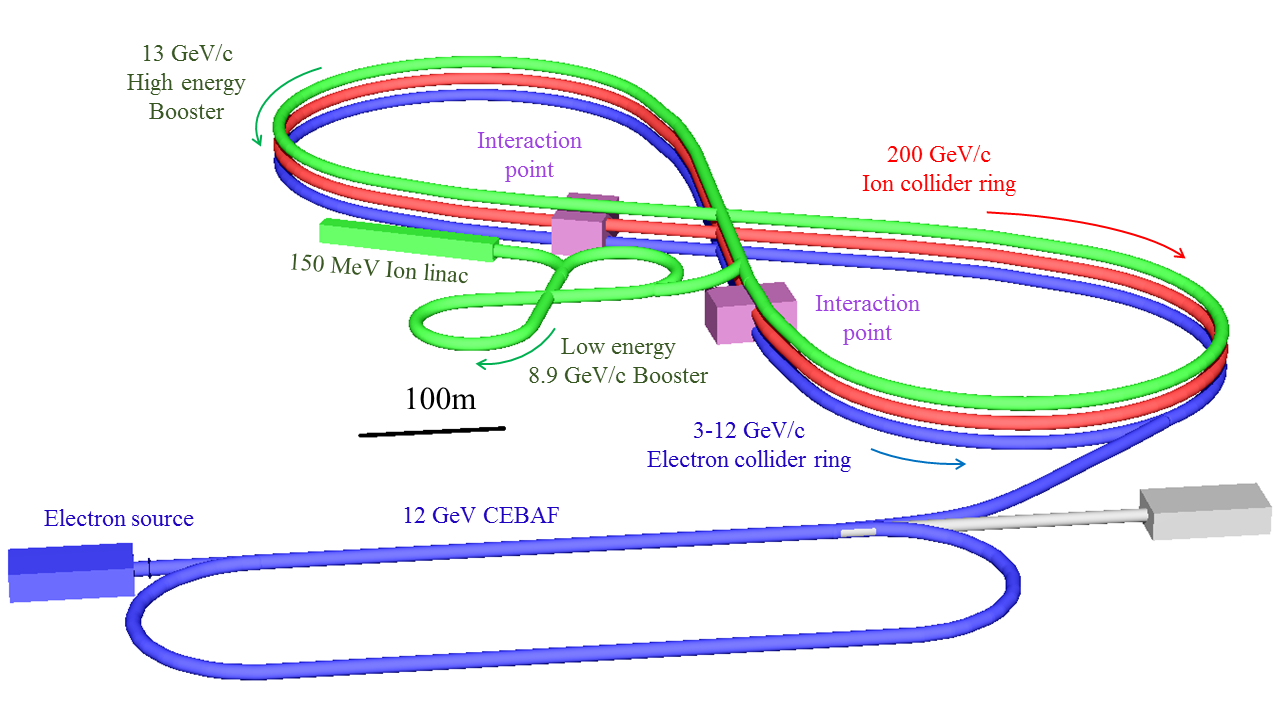}
\caption{Layout of the Jefferson Lab Electron-Ion Collider (JLEIC) (from \cite{zhang2019jleic}).} 
\label{JLEIC}
\end{figure}

The JLEIC is designed to take advantage of the existing 12 GeV electron SRF recirculating linac CEBAF at Jefferson Lab, which will be used to provide an electron beam for the collider.
Both colliding beams will be stored in two figure-8 shaped collider rings. 
One ring, made of NC magnets, stores electrons at 3 GeV to 12 GeV, with an average beam current of up to 3 A (below 7 GeV). 
The second ring, consisting of 6T SC magnets, 
either stores fully-stripped ions, with up to 80 GeV per nucleon, or protons with energies
between ranging from 30 to 200 GeV. 
The two collider rings and the additional 13 GeV/c high energy ion/proton booster ring are stacked vertically, have nearly identical circumferences of 2.3 km, and are housed in the same underground tunnel next to the CEBAF facility, as illustrated in Fig.~\ref{JLEIC}. The unique figure-8 shape of the collider allows complete cancellation of spin precession between the left and right arcs, in which guiding vertical magnetic fields are of opposite directions, thus resulting in zero net spin tune independent of energy. This shape is chosen for optimization and preservation of ion spin polarization during acceleration in the booster and collider rings, as well as during beam storage. The crossing angle of the tunnels is 77.4$^\circ$ and electron and ion beamlines
intersect at an angle of $\theta_c=$50 mrad in two long straights next to the crossing point, allowing accommodation of two detectors. The CEBAF 1.5 GHz linac will serve as a full-energy injector into the 3--12 GeV electron ring, requiring no upgrade for energy, beam current, or polarization. An entirely new hadron-beam complex is required for proton and ion beam generation and acceleration. This complex includes sources for polarized light ions and for non-polarized light to heavy ions; a 150 MeV SRF linac for protons, a compact figure-8 8.9 GeV/c low energy booster ring, and the 13 GeV/c high energy booster ring, which injects into to the main 200 GeV/c proton (or 80 GeV/$n$ ion) ring. Key design parameters of the JLEIC are presented in Table \ref{iieic}. The JLEIC  upgrade to 140 GeV c.m.e.~will require an  increase of the proton energy to 400 GeV 
through the installation of new 12 T SC magnets.

\begin{figure}[htbp]
\centering
\includegraphics[width=80mm]{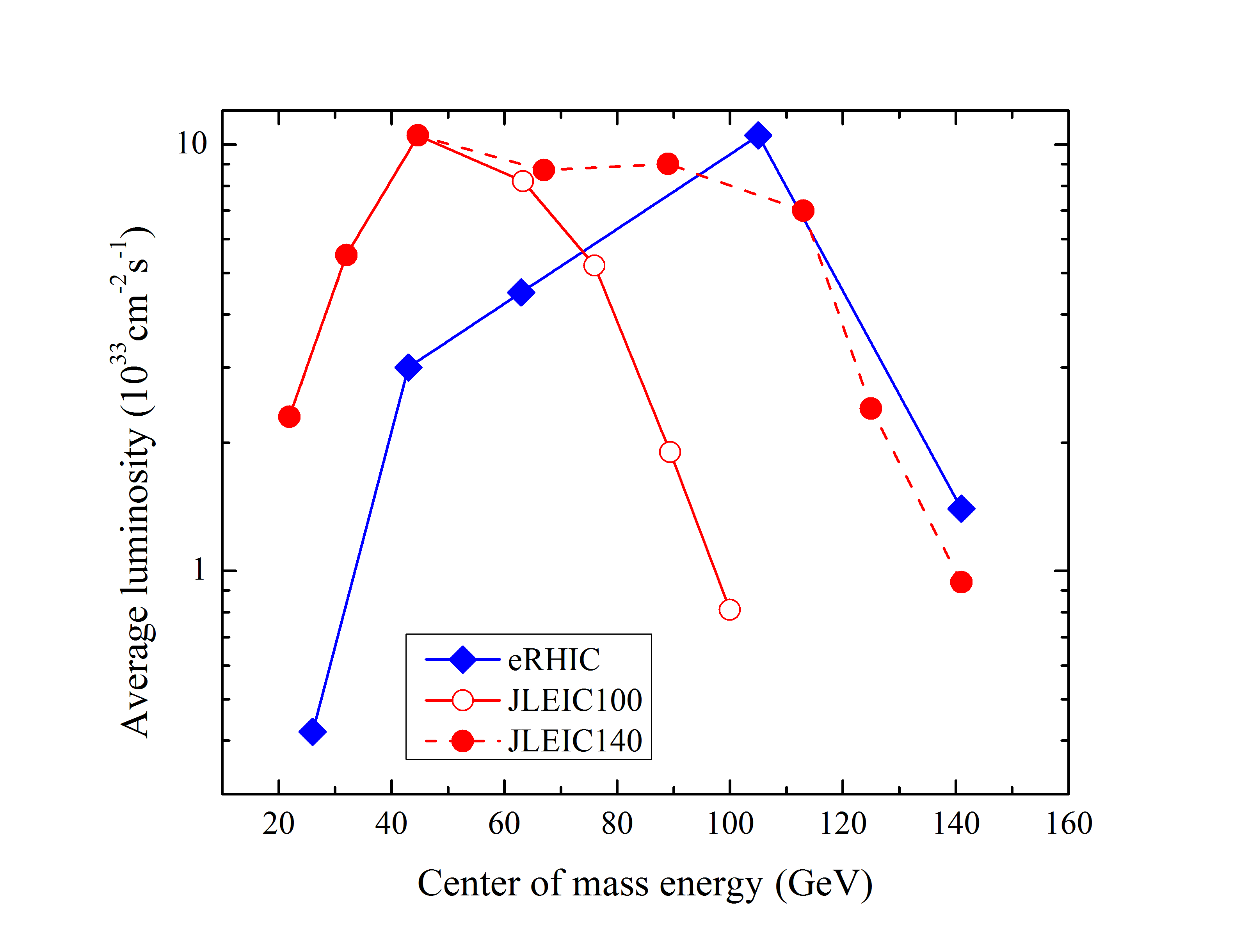}
\caption{Average $e-p$ luminosity of JLEIC and eRHIC as a function of c.m.e. The JLEIC average luminosity takes into account a 75\% operational duty factor and is given for both baseline design (open circles) and for a potential future upgrade with a 400 GeV proton ring (solid circles). The eRHIC luminosity is averaged over the data-taking cycle and equals 95\% of the peak luminosity.} 
\label{eiclumis}
\end{figure}

The JLEIC luminosity performance is determined by different limits, depending on c.m.e.  (Fig.~\ref{eiclumis}). At 20--35 GeV energies it is limited by space-charge effects for the hadron beams, in the range of 35--60 GeV, by beam-beam effects on both beams,  at higher energies by synchrotron radiation of the high energy electron beam \cite{zhang2019jleic}. Synchrotron radiation results in emittance growth and also limits the maximum electron current. The electron current, which is 3 A at energies below 7 GeV, decreases to less than 0.5 A at 12 GeV, if the total radiation power is limited to 10 MW. Hadron beam cooling is needed to combat IBS and to reduce or preserve beam emittances. It, therefore, is critical for the JLEIC $e-p$ and $e-i$ luminosities. A three stage electron cooling is proposed, that includes a conventional 50 kV DC cooler in the low-energy booster ring (for ions), a state-of-the-art 1.1--4.3 MeV DC cooler in the high energy booster ring, and a 
43--109 MeV ERL-based cooler in the collider ring (the electron energy range indicates values for lead ions and protons). The required cooling rates call for electron bunches with 3.2 nC charge supplied at the 476.3 MHz repetition rate of the ion bunches, resulting in a 1.5 A beam current, far higher than what has ever been demonstrated in an ERL. To reduce the average electron cooling current, a Circulating Cooler Ring (CCR) concept is proposed, that circulates the high charge bunches 11 times through  the  cooler  before  returning  them  to  the ERL \cite{benson2018jleicerl}. This novel concept needs further development and testing of its key parts such as the fast harmonic transverse kicker needed to kick electrons in and out of a 60 m circulator ring, and the magnetized electron beam generation and transport to assure a low temperature of the electrons and, thereby, a high cooling efficiency. 

A beam crossing angle of 50 mrad is necessary to avoid parasitic collisions due to short bunch spacing, make space for machine elements, improve detection and reduce detector background. To  prevent  a factor of $\sim$12 luminosity  loss  caused by  the crossing angle, SRF crab cavities will need to be installed on both sides  of each  IP, and for  both beams,   to restore head-on  collisions in the center-of-mass frame.

\begin{center}
\begin{table*}[ht]
{\small
\hfill{}
\begin{tabular}{|l|c|cc|cc|cc|cc|cc|}
\hline
\hline
Parameter &  NICA &  
\multicolumn{2}{c|}{ELISe} & 
\multicolumn{2}{c|}{EicC-I} & 
\multicolumn{2}{c|}{JLEIC} & 
\multicolumn{2}{c|}{eRHIC} & 
\multicolumn{2}{c|}{LHeC}\\
Species & $ii (pp)$ & 
$i$ & $e$ & 
$p$ & $e$ & 
$p$ & $e$ & 
$p$ & $e$ & 
$p$ & $e$ \\
\hline
C.m. energy $\sqrt{s}$ (GeV) & 9 &
\multicolumn{2}{c|}{1.8} & 
\multicolumn{2}{c|}{16.7} & 
\multicolumn{2}{c|}{44.7} & 
\multicolumn{2}{c|}{105} &
\multicolumn{2}{c|}{1300} \\
Beam energy (GeV) & 4.5/u &
0.74/u & 0.5 &
20 & 3.5 & 
100 & 5 &
275 & 10 & 
7000 & 60 \\
Circumference (m) & 503 & 
224 & 53.7 &
600 & 800 & 
\multicolumn{2}{c|}{2336} & 
\multicolumn{2}{c|}{3834} & 
26700 & 9000 \\
Number of bunches & 24 & 
40 & 8 &
\multicolumn{2}{c|}{2000} & 
\multicolumn{2}{c|}{3228} & 
\multicolumn{2}{c|}{1320} & 
2808 & n/a \\
Particles per bunch (10$^{10}$) & 0.22 & 
8.6$\times$10$^{-4}$ & 5 &
0.5 & 3.2 & 
1 & 4.7 & 
6 & 15.1 & 
22 & 0.23 \\
Emittance (H/V,rms norm., $\mu$m) & 1.1/0.8 & 
0.07 & 45 &
1 & 68 & 
0.7/0.13 & 83/17 & 
9.2/1.6 & 20/1.3 & 
1.9 & 53 \\
Beta functions at IP $\beta_{x,y}^{\ast}$ (cm) & 35 &
100/15 & 100/15 &
2/1 & 20/10 & 
8/1.3 & 5.7/1 &
91/4 & 41/5 & 
7 & 5 \\
Bunch length (rms, cm) & 60 & 
15 & 4 & 
3 & 10 & 
2.5 & 1 &
6 & 1.9 & 
7.6 & 0.006 \\
Beam-beam parameter $\xi_{x,y}$ ($10^{-3}$) & 50 &
n/a & n/a & 
3  & 10 & 
15 & 40 & 
14/7 & 70/100 & 
0.4 & n/a
\\
Space-charge param. $|\Delta Q_{SC}|$ & 0.05 &
0.08 & n/a & 
0.01 & n/a & 
0.018 & n/a & 
n/a & n/a & 
n/a & n/a
\\
IBS time (hor./long., min) & 42 & 
\multicolumn{2}{c|}{n/a} & 
8.3/3.3 & n/a &
0.7/2.3 & n/a & 
126/120 & n/a & 
$\sim$4000 & n/a
\\ 
Polarization & 80\% & 
\multicolumn{2}{c|}{n/a} & 
\multicolumn{2}{c|}{n/a} & 
85\% & 80\% & 
80\% & 80\% & 
0 \% & 90\%
\\
\hline 
Luminosity ($10^{30}$~cm$^{-2}$s$^{-1}$) & 0.001 & 
\multicolumn{2}{c|}{0.01} &
 \multicolumn{2}{c|}{$10^{3}$} &
 \multicolumn{2}{c|}{1.5$\times 10^{4}$} & 
  \multicolumn{2}{c|}{1.05$\times 10^{4}$}  & 
   \multicolumn{2}{c|}{0.8$\times 10^{4}$} 
\\
\hline\hline
\end{tabular}}
\hfill{}
\caption{Key design parameters of future ion-ion and electron-ion/electron-proton colliders.}
\label{iieic}
\end{table*}
\end{center}

The eRHIC design \cite{willeke2019erhic} aims at polarized electron-proton collisions in the c.m.e.~range from 29 to 141 GeV, which are  accomplished by colliding 41 to 275 GeV protons delivered by the existing ``Yellow Ring'' of the RHIC heavy ion collider and the entire existing hadron beam injector chain, with 5 to 18 GeV electrons from a new electron storage ring installed in the RHIC tunnel. The eRHIC peak luminosity reaches about $10^{34}$~cm$^{-2}$s$^{-1}$ at $\sqrt{s}=$100 GeV --- see Fig.~(\ref{eiclumis}). Key parameters of the eRHIC are given in Table~\ref{iieic}. 
Polarized electrons are provided by a full-energy spin-transparent rapid-cycling synchrotron (RCS) using normal-conducting RF cavities, located in the 3.8 km RHIC tunnel.
The RCS is specifically designed to be free of intrinsic resonances over the entire acceleration range from 400 MeV at injection to 18 GeV. Spin transparency is achieved by a high super-periodicity of the RCS focusing lattice of $P$=96, and an integer tune of $[Q_y]$=50. With such parameters, intrinsic spin resonances, which occur under condition \cite{sylee1997spinsnakes}: 
\begin{equation}
G\gamma = nP \pm [Q_y],
\label{electronspin}
\end{equation}
(here, $n$ is an integer, $G$= 0.00115965 is the anomalous
gyromagnetic ratio of the electron, $\gamma$ is the relativistic
Lorentz factor), are avoided over the entire energy range of the RCS and the simulated polarization transmission efficiency is about 97\%, even in the presence of magnet misalignments as large as 0.5 mm rms.  

Focusing for the electron storage ring is achieved through 16 FODO cells in each of the six arcs. To obtain the required design emittance over the entire energy range from 5 to 18 GeV, the ring operates with different betatron phase advances per FODO cell --- 90 degrees at 18 GeV, and 60 degrees at 10 GeV and below. The bending sections in these cells are realized as super-bends, with each section consisting of three individual dipoles, namely two 2.66 m long dipoles with a short (0.44 m long) dipole in-between. At beam energies of 10 GeV and above, all three segments are powered uniformly for a smooth, uniform bend, while at 5 GeV, the polarity of the short center dipole is reversed, resulting in additional synchrotron radiation in this configuration to provide the required fast radiation damping enabling the high electron beam-beam parameter $\xi_y$ of 0.1. The total electron beam intensity is set by a 10 MW power limit on the ring SRF system, which must restore the synchrotron radiation losses. Arbitrary spin patterns in the electron storage ring --- with simultaneous storage of bunches with spin ``up'' and bunches with spin ``down'' in the arcs --- are achieved by injecting polarized electron bunches with the desired spin orientation at full storage energy. Since the Sokolov-Ternov effect \cite{sokolovternov1964polarization} will lead to depolarization of bunches with spins parallel to the main dipole field, a frequent replacement of individual bunches is required, at a rate of about one bunch per second, to keep the time-averaged polarization sufficiently high.

\begin{figure}[htbp]
\centering
\includegraphics[width=80mm]{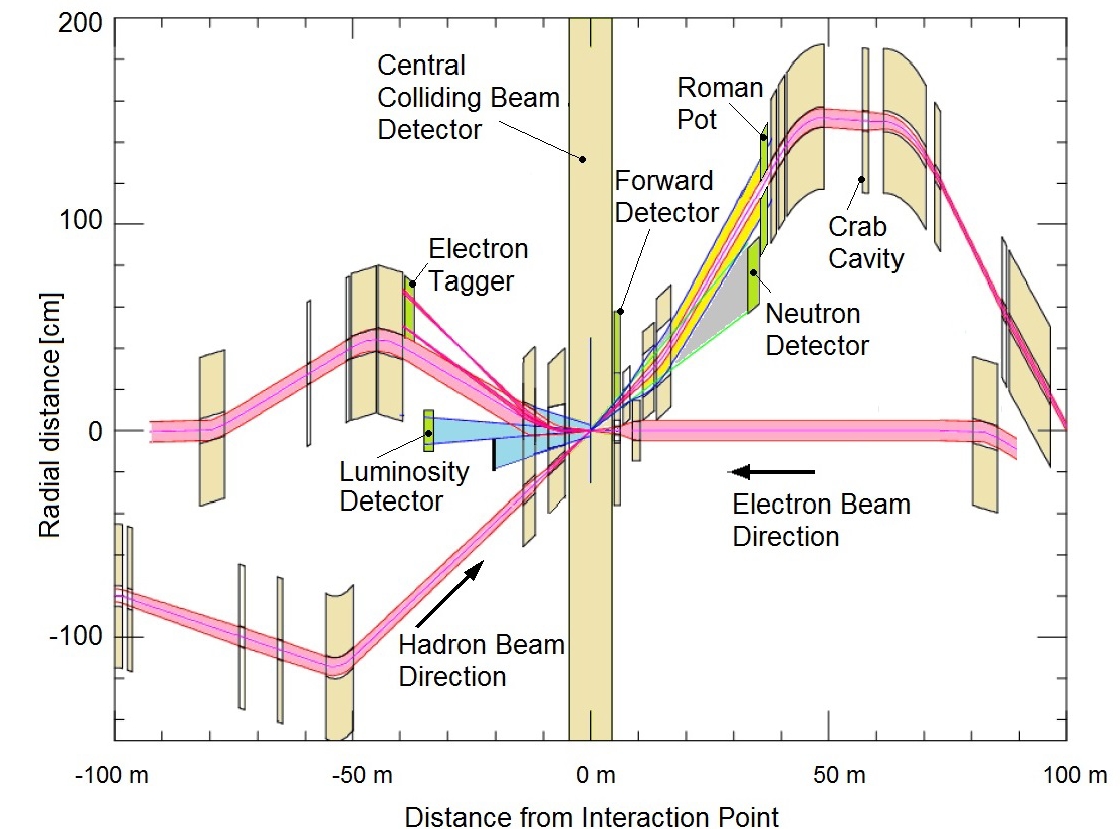}
\caption{Layout of the eRHIC interaction region. The length scales for the horizontal and vertical axis are very different. Beams cross with a crossing angle of 25 mrad.  The IR design integrates focusing magnets for both beams, luminosity and neutron detectors, electron taggers, spectrometer magnets, near-beam detectors (Roman pots for hadrons), crab cavities, and spin rotators for both beams (from \cite{willeke2019erhic}).} 
\label{eRHICIR}
\end{figure}

The beams of electrons and ions (or protons) collide in one or two interaction regions at a 25 mrad crossing angle. Different frequency choice for the eRHIC crab cavity systems have been considered. For example,  
 a combination of 200 and 400 MHz for the long proton bunches, and of 400, 800, 1200, and 1600 MHz for the shorter electron bunches would avoid the luminosity reduction due to the crossing and minimize hadron beam emittance growth to less than 5\%/h. 
 An alternative, technically simpler option would use 200 MHz crab RF systems for both protons and electrons \cite{Verdu-Andres:2019zkz}. 
 A dedicated fill pattern ensures that each bunch collides only once per turn. This way, luminosity is shared equally between the two detectors without exceeding the beam-beam limit. To maximize luminosity, the beams are focused to flat cross sections $\sigma_x^* \gg \sigma_y^*$ at the IP. Sophisticated interacting region optics (see Fig.~\ref{eRHICIR}) provides sufficient separation of the hadron beam from the 5 mrad forward neutron cone, separates the electron beam from the Bethe-Heitler photons used for luminosity measurements, and allows for a safe passage of the synchrotron radiation fan generated upstream of the IP through the detector. 

The hadron beam parameters are similar to what has been achieved in the RHIC, with the exception that the number of bunches will be increased from 110 in the present RHIC up to 1320 (or 1160) in the eRHIC, while increasing the total hadron beam current by a factor of three. The higher current could cause unacceptable heating of the cryogenic stainless-steel beam pipes. A thin layer of copper will, therefore, be applied in-situ, to improve the beam-pipe surface conductivity. A much thinner layer of amorphous carbon will next be deposited on top of the copper coating to reduce secondary electron yield and, thereby, suppress the formation of electron clouds. A broad spectrum of desired center-of-mass collision energies implies operation over a wide range of ion energies and, therefore, substantially different ion velocities. In order to maintain the synchronization between  electron and hadron beams, the circumference of one of the two rings has to be adjusted. This is accomplished by two methods: between 100 and 275 GeV proton energy, a $\pm$ 14 mm radial orbit shift is sufficient to account for variations in velocity of the hadron beam. For proton beam operation at 41 GeV, the beam will travel through the (inner) ``Blue'' arc of the RHIC between interaction regions IR12 and IR2 instead of the (outer) ``Yellow'' arc, thus reducing the circumference by 93 cm.

Usable store length in the eRHIC collider is limited by IBS growth time of approximately 2 hours. Since the turnaround time between stores is of the order of 30 min, average luminosity would only be about half the peak value. To counteract the fast emittance growth due to IBS and, hence, to increase usable store length, strong hadron cooling with some 1 hour cooling time is required. Two cooling schemes are currently under consideration: a somewhat conventional cooling with a bunched electron beam, or variations of {\it coherent electron cooling}, where an electron beam is used as a pickup and kicker in a very high bandwidth stochastic cooling scheme \cite{litvinenko2009coherentecool}. The feasibility of the former, based on the electron RF photoinjector system has recently been  demonstrated \cite{fedotov2019bunchedbeamecool,PhysRevAccelBeams.23.021003}, but the required high beam intensities for a bunched beam electron cooler for eRHIC by far exceed the capabilities of present-day electron guns. An alternative scheme, in which the electron beam is stored in a small storage ring equipped with strong damping wigglers, is being evaluated. Coherent electron cooling has not yet been demonstrated experimentally.
An alternative could, therefore, be to use the existing ``Blue'' ring as a full-energy injector
to cool proton/ion bunches at (or slightly above) 25 GeV injection energy, which is much easier due to strong energy dependence of cooling forces, and then to ramp the ``Blue'' ring and replace the entire fill in the “Yellow” storage ring every 15 min. Yet another possible design scenario, without any hadron cooling, results in optimized machine parameters yielding a (peak) luminosity of 0.44$\times 10^{34}$~cm$^{-2}$s$^{-1}$. In general, hadron cooling is one of the most important topics of the ongoing eRHIC R\&D program, together with the  development of the required crab cavities, efficient SRF for the  electron storage ring, and fast kickers to manipulate the  significantly increased number of bunches. 

Early in 2020, the U.S. Department of Energy (DOE) announced that the Electron-Ion Collider will be built at Brookhaven National Laboratory in Upton, New York \cite{cho2020electronioncollider}.

\subsubsection{LHeC, HE-LHeC and FCC-eh}
\label{LHeC}
Deep inelastic scattering of electrons on protons or nuclei has traditionally proven the best means to probe the inner structure of nucleons and nuclei. A unique opportunity for this can be offered by colliding 7 TeV protons circulating in the LHC 
with 60 GeV electrons from an energy recovery linac (ERL) \cite{kuze2018lhec}. Since such a Large Hadron electron Collider (LHeC) \cite{lhec} uses a beam of the already built hadron collider, 
it can be realized at an affordable cost and can run concurrently with hadron-hadron collision experiments. The LHeC can provide much cleaner collision environment at $\sqrt{s}$=1.3 TeV and would be  another powerful 
tool to study Higgs boson properties \cite{lhechf,GARCIA:2013rwa}.

The electron ERL is of racetrack shape, accommodating two 10 GeV SC linacs  in the straights, connected by arcs with a bending radius of about 1 km. The LHeC linacs employ SC bulk Nb cavities with a frequency of 800 or 400 MHz.  Three acceleration passages are required to  attain the design electron energy of 60 GeV at the IP, followed by three turns of deceleration for energy recovery (the basic 
ERL principle is sketched in Fig.~\ref{fig:erl}). The total circumference of the baseline LHeC is exactly one third of the LHC, easing the introduction of electron bunch gaps for ion clearing without perturbing the proton beam. Smaller circumferences (one fourth, one fifth) are also possible at the price of lower electron beam energy.

The IP beam size of the LHeC is determined by the emittance and minimum $\beta^{\ast}$ of the proton beam --- see Table \ref{iieic}. Luminosity is then set by the electron beam current. A luminosity of order 10$^{34}$~cm$^{-2}$s$^{-1}$  is required for Higgs boson physics studies \cite{lhechf}, demanding a high average electron beam current of 20--30 mA at the collision point, that in turn can only be achieved with energy recovery. The 3-pass beam recirculation including energy recovery implies a six times higher current in the SC linacs, which simultaneously accelerate three beams of different energies and decelerate three other beams. Construction of a high-current multi-turn 500 MeV ERL test facility for the LHeC, called PERLE, is planned at LAL in Orsay \cite{angal2018}. PERLE will demonstrate the technical feasibility of the LHeC concept. At the end of 2019, an already constructed multi-pass ERL test facility of a different type, 
CBETA at Cornell, has achieved 4 turns of acceleration followed by 4 turns of deceleration \cite{CBETA2020},
albeit at a much lower beam current than required for the LHeC
(also see Ch.~\ref{other advanced}). 
The LHeC cost and AC wall plug power estimates are 1.8 BCHF and 100 MW, respectively \cite{bruning2018lheccost}. A 30 GeV electron ERL version of the LHeC will cost below 1 BCHF, at approximately the same luminosity and operation cost. Such a version is studied as a possible first stage assuming that the c.m.e.~reach can later be increased by accepting a slightly higher power consumption and better SRF performance. 

The same or a similar ERL could be used to realize electron-hadron collisions also at the High-Energy LHC (HE-LHeC, with twice the LHC's proton energy) or at the FCC-hh (FCC-eh, with 7--8 times more energetic protons) --- see Chapter \ref{frontiercolliders}. Luminosities for HE-LHeC and FCC-eh are above $10^{34}$~cm$^{-2}$s$^{-1}$ at $\sqrt{s}$=1.8 TeV and $\sqrt{s}$=3.5 TeV, respectively \cite{mkobdsfz}.

\subsection{Lepton colliders studying Higgs boson and  electroweak sector}
\label{leptoncolliders}

Higgs factory proposals generally aim at improving the precision of coupling measurements of Higgs boson, top quark, $W$ and $Z$ by an order of magnitude or more compared with previous studies. 

The International Linear Collider (ILC), with a center of mass energy of 250 GeV in $e^+e^-$ collisions, has been under consideration for more than two decades and could   potentially be upgraded to $\sqrt{s}$=500 GeV 
and further to 1 TeV. CERN's Compact Linear Collider (CLIC) design, developed since the mid-1980s,  
also includes possible upgrades, from an initial 380 GeV c.m.e.~to ultimately 3 TeV, which would enable searches for new particles of significantly higher masses. 

Two proposals for circular $e^+e^-$ colliders have appeared more recently: the ``Future Circular Collider'' (FCC-ee) at CERN \cite{fccee} and the ``Circular Electron-Positron Collider'' (CEPC) in China \cite{cepc}.
These ambitious, large-scale projects based on well-established technologies are not extendable to TeV or multi-TeV energies, but offer several important advantages that include the potential for much higher luminosities and, thus, higher precision, the ability to operate multiple experiments simultaneously, and their 100 km circular tunnels that could later house $O$(100 TeV) hadron colliders.

At lower energies, the main aim of the proposed Super Tau-Charm Factories is the production and precise study of charmonium states and of the tau lepton. 

\begin{figure*}[htbp]
\centering
\includegraphics[width=0.8\textwidth]{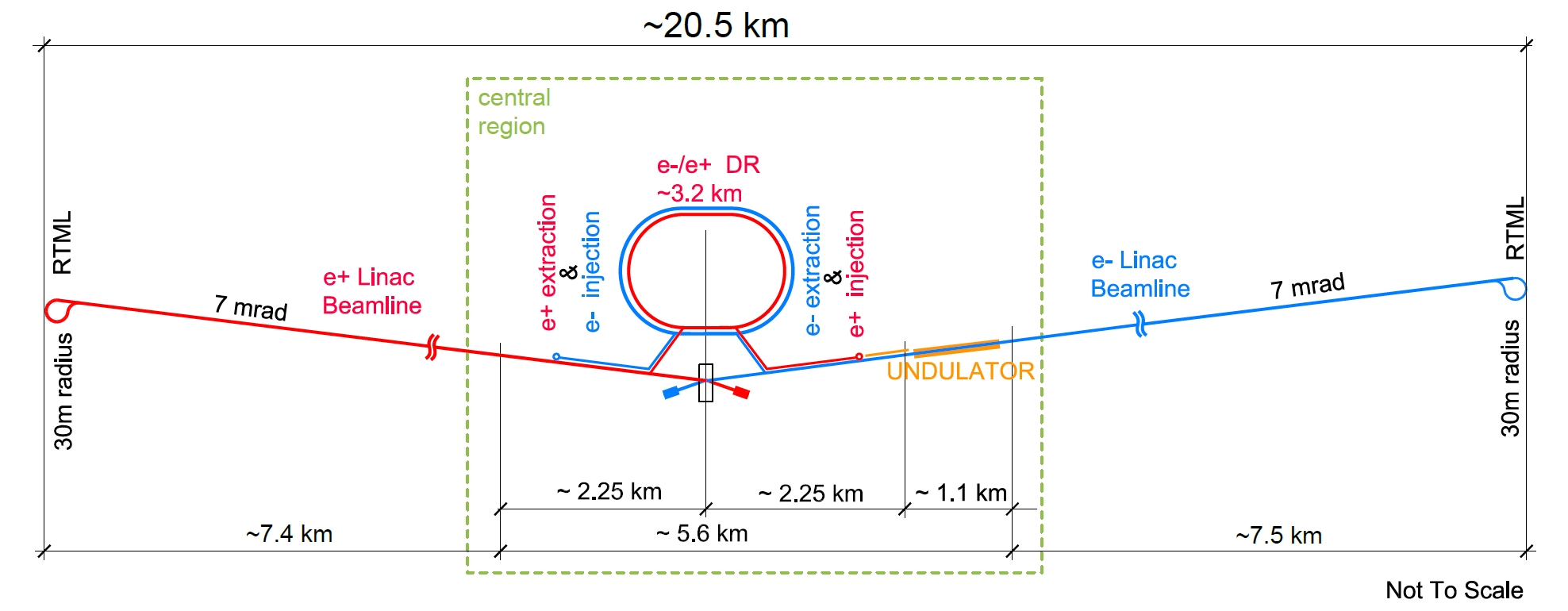}
\caption{Schematic layout of the ILC in the 250 GeV staged configuration (from \cite{ilc2019global}). } 
\label{ILCscheme}
\end{figure*}

\subsubsection{Super $\tau$-Charm Factories}
\label{tcfactories}
Two Super Tau-Charm Factories are being proposed, in Russia (SCT)  \cite{sct} and in China (HIEPA) \cite{hiepa}, respectively. 
They will provide excellent possibilities in search for new physics and for detailed studies of known phenomena. Both proposals consider double ring e$^+$e$^-$ colliders operating at c.m.e.~between about 2 and 7 GeV.
Their design luminosity is 2 or 1 times $10^{35}$~cm$^{-2}$s$^{-1}$, achieved with $\beta_{y}^{\ast}= 0.5$~(0.6)~mm and a crab-waist collision scheme. The expected beam lifetime is about 10 minutes, supported by top-up injection, requiring a positron production rate of up to 10$^{11}$/s. The electron beam can be longitudinally polarized at the collision point. Preliminary conceptual designs are available for both projects.  
Numerous synergies and complementarity exist between the two Super $\tau$-charm factory projects, the already constructed SuperKEKB, and the proposed future higher-energy circular $e^+e^-$ Higgs factories FCC-ee and CEPC --- see Section~\ref{circfutureee}.

\subsubsection{Linear $e^+e^-$ colliders for Higgs sector: ILC and CLIC}

As noted above, linear colliders are (almost) free of  synchrotron radiation losses and their energy scales with the gradient and length of their accelerating sections --- see  Eq.~(\ref{eq:energy_l}). Beam acceleration by a sequential array of RF cavities is by itself a straightforward technique to reach the c.m.e.~required for Higgs boson studies in $e^+e^-$ collisions.  The main challenge in designing a high energy, high luminosity single pass collider is the power requirement. Indeed, luminosity of a linear collider is constrained by three limiting factors (in parentheses):
\begin{equation}
\Lumi = \Big( N_e n_b f_{\rm r} \Big) \left( \frac{1}{\sigma_y^*} \right) \left( \frac{N_e}{\sigma_x^*} \right) \frac{H_D}{4 \pi},
\label{LumiHiggs}
\end{equation}
where the first factor is proportional to the total beam current, set by  particle sources (especially challenging is the positron production; see below), by coherent beam instability concerns, and most importantly, by the available RF power.  
If the total AC wall-plug power of the collider is $P_{\rm wall}$ and efficiency of converting it into beam power is $\eta\equiv P_{b}/P_{\rm wall}$, then $N_e n_b f_{\rm r} = P_{\rm wall}/(\eta eE_{cme})$. The efficiency of the RF system, the largest and most critical part of total efficiency $\eta$, is usually less than 10\% \cite{delahaye2016linearcoll}, 
and it constitutes the biggest technological challenge for linear colliders. 
The second factor in the luminosity equation calls for ultra-small vertical beam size at the IP, $\sigma_y^*$, that in turn requires record small beam emittances formed in dedicated damping rings \cite{emma2001systematic}, precise $O$(10$\mu$m) mechanical and beam-based alignment, stabilization of focusing magnets and accelerating cavities at the nm level  \cite{baklakov1994study, sery1996gm, baklakov1998ground,shiltsev2010ATLreview}, and beam position monitors (BPMs) with 0.1 $\mu$m resolution, 
in order to obtain the rms beam sizes of 8 nm (vertical) and 520 nm (horizontal) at the ILC IP and of 3 nm/150 nm at the CLIC IP \cite{raubenheimer2000estimatesstability, kubo2011emmgrowthlinacs, pfingstner2017cliccorrection}. 

The third factor in Eq.~(\ref{LumiHiggs}) $(N_e / \sigma_x)$  defines the beam energy spread and the degradation of the luminosity spectrum arising from the so-called {\it beamstrahlung} radiation of photons and $e^+/e^-$ pairs in the strong electromagnetic (EM) fields of the tightly compressed opposite bunch \cite{bell1995bs, chen1992beamstrahlung}. 
This effect grows with collision energy. For example, it amounts to $\delta E/E$ $\sim$1.5\% in the 250 GeV c.m.e.~ILC, while in the 380 GeV CLIC it already results in some 40\% of the collider luminosity being more than
1\% away from the maximum c.m.e. 
The management of $P_{\rm wall}$ leads to an upward push on the bunch population $N_e$ and, therefore, on the number of beamstrahlung photons emitted per $e^{\pm}$ which is approximately equal to $N_{\gamma}\approx 2 \alpha r_{e} N_e/\sigma_{x}^{\ast}$, where $\alpha$ denotes the fine-structure constant \cite{zimmermann2001tutoriallc}. Typically, one aims for $N_{\gamma} \stackrel{<}{\sim} 1$, to retain a significant luminosity fraction close to the nominal energy.     
A consequence is the use of flat beams, where $N_\gamma$ 
is controlled by the beam width $\sigma_x^*$, and the luminosity is adjusted through the beam height $\sigma_y^{\ast}$, motivating the extremely small vertical small sizes at both 
ILC and CLIC.  
The final factor in Eq.~(\ref{LumiHiggs}), $H_D$, represents the enhancement of luminosity due to the {\it pinch effect}, i.e., the additional focusing occurring during the collision of oppositely charged bunches;
$H_{D}$ typically assumes values between 1 and 2. 


\begin{figure}[htbp]
\centering
\includegraphics[width=80mm]{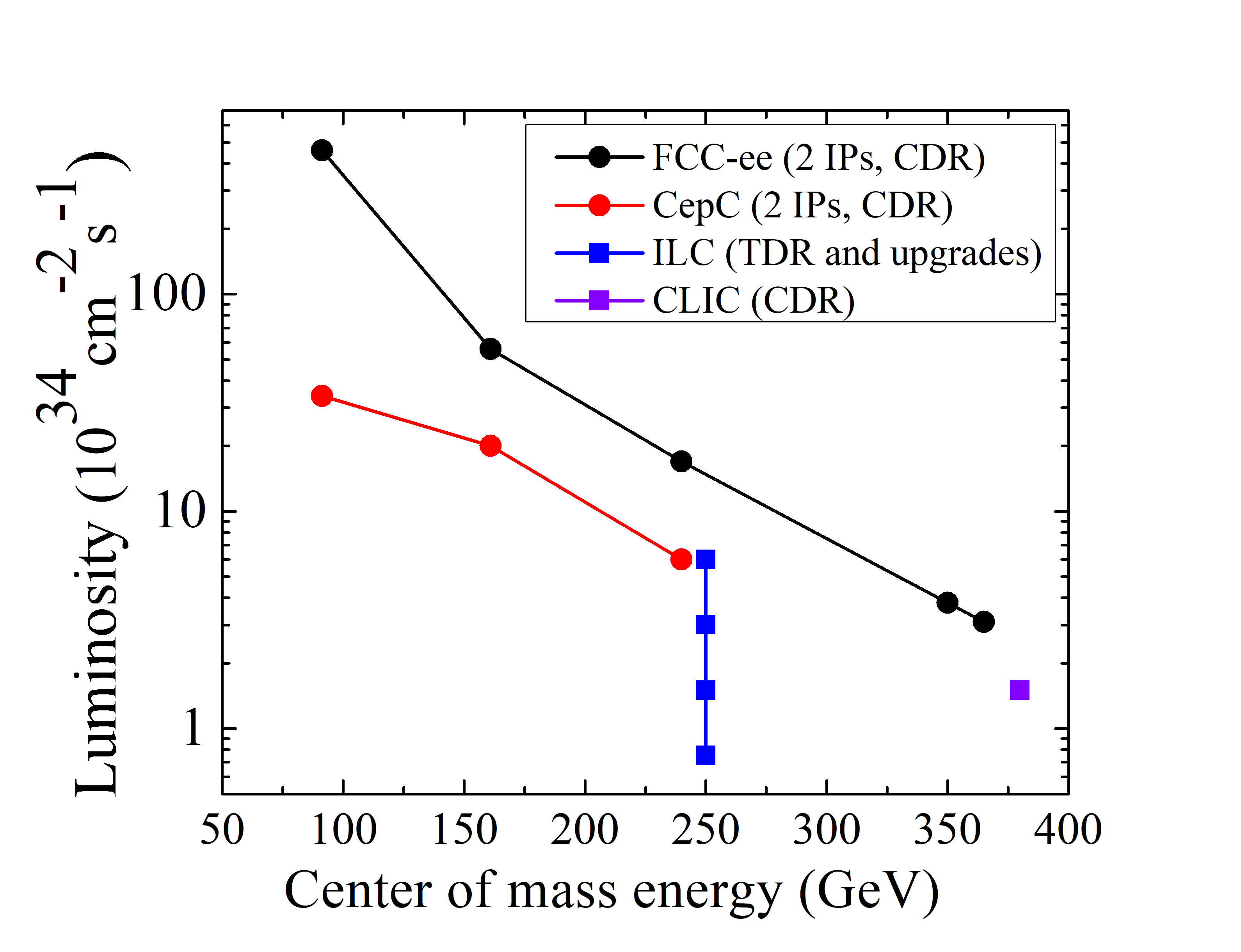}
\caption{Luminosity of the proposed Higgs and electroweak factories vs center of mass energy $\sqrt{s}=2E_b$.} 
\label{HFlumis}
\end{figure}

The {\it ILC Technical Design Report} \cite{accel:ILC} (TDR) foresaw a baseline c.m.e.~of 500 GeV, with a first stage at 250 GeV, and upgrade provision for 1 TeV, at luminosity values comparable to the LHC's.  
Recent revisions, motivated by the low mass of the Higgs boson, have established new optimized configurations for collisions at 250 GeV \cite{accel:ILC2,accel:ILC3}.  The ILC employs 1.3 GHz SRF cavities made of bulk Nb, operating at an accelerating gradient of 31.5 MV/m at 2 K, and it requires some 130 MW of site power --- see the key machine parameters in Table \ref{eefuturetable}. 
The 1.3 GHz pulsed SRF technology of the ILC 
was developed for the TESLA project  \cite{aune2000srf} and successfully applied for the European X-Ray Free Electron Laser \cite{altarelli2006europeanxfel}. Progress toward higher field gradients and $Q$ values of SC cavities continues to be made, with nitrogen doping, nitrogen-diffusion techniques, and Nb$_3$Sn cavities being recent examples \cite{accel:n2dope,accel:n2diff, padamsee2019rastsrf}. 

Figure~\ref{ILCscheme} presents a schematic overview of the ILC with its main subsystems. The accelerator extends over 20.5 km, dominated by the main electron and positron linacs and another $\sim$5 km of beam delivery and final focus system. It consists of two main arms intersecting at a 14 mrad crossing angle. Electrons with 90\% polarization are produced by an electron gun, where a Ti:sapphire laser pulse hits a photocathode with a strained GaAs/GaAsP superlattice structure. The baseline solution for ILC positron production employs a 
320 m long SC helical undulator with 5.85 mm diameter beam aperture, located at the end of the 125 GeV electron main linac \cite{alharbi2019energy}. When the main electron beam passes through this undulator it produces polarized photons that are converted to positrons, in a rapidly rotating target (2000 rounds per minute), resulting in 30\% longitudinal positron polarization. 
An alternative design, which does not require a fully operational main linac, instead utilizes a separate, dedicated 3 GeV electron accelerator to produce positrons via conventional $e^+e^-$ pair production, 
when the electron beam hits a target (no positron polarization provided in that case). After acceleration to 5 GeV, electrons and positrons are injected into the centrally placed 3.2 km-long damping rings, each equipped with 54 SC wigglers, needed to assist fast damping of the beam's initially large normalized emittances to 20 nm (4 $\mu$m) in the vertical (horizontal) plane within the 200 (100) msec time between collider shots with repetition frequency $f_{\rm r}=5$ (10) Hz. Next is the Ring-to-Main-Linac (RTML) system that includes beamlines to transport the  low-emittance beams to the beginning of the main accelerators where two-stage bunch compressors squeeze the longitudinal bunch length to 0.3 mm, and the beam energy increases from 5 GeV to 15 GeV, before the beams are sent into the main linacs to be accelerated to 125 GeV each.

The main linacs accelerate the beams in SC Nb cavities, each 1.04 m long and having 9 cells, with a mean accelerating gradient of 31.5 to 35 MV/m --- see Fig.~\ref{fig:ILC_cavity}. Cryomodules of 12 m length  provide cooling and thermal shielding of the cavities; these contain all necessary pipes for liquid and gaseous helium at various temperatures and house 9 or 8 such cavities plus a quadrupole unit for beam focusing. The RF power for the cavities is generated by commercially available 10 MW klystrons, with a peak efficiency of 65\%. Finally, the beam-delivery system focuses the beams to the required size of 516 nm $\times$ 7.7 nm at 250 GeV c.m.e. It is expected that the electron beam polarization at the IP will be 80\% (i.e., only 10\% off its original level) and that the vertical beam emittance will not be more than $\sim$75\% above its original damping ring value.  A feedback system, which profits from the relatively long train of 1312 bunches with inter-bunch separation of 554 ns, shall ensure the necessary beam-beam pointing stability at the IPs. The ILC is designed to allow for two detectors mounted on movable platforms and operated in a so-called {\it push-pull mode}; these detectors can be moved into and out of the beam within a day or two.

Besides energy upgrades to $\sqrt{s}$=0.5 Tev and 1.0 TeV,  luminosity upgrades are also possible by doubling the number of bunches per pulse to 2625 at a reduced bunch separation of 366 ns (which would require 50\% more klystrons and modulators and an increased cryogenic capacity), and by an increase in the pulse repetition rate $f_{\rm r}$ from 5 to 10 Hz (which would require a significant increase in cryogenic capacity, or running at a reduced accelerating gradient after an energy upgrade).
The corresponding points are indicated in Fig.~\ref{HFlumis}. 

After the discovery of the Higgs boson in 2012, the Japan Association of High Energy Physicists (JAHEP) made a proposal to host the ILC in Japan and the Japanese ILC Strategy Council conducted a survey of possible sites for the ILC in Japan, eventually selecting a suitable one in the Kitakami region of Northern Japan \cite{ilc2019global}. The cost of the 250 GeV ILC project in Japan is estimated at 700 BJPY (with $\pm$25\% uncertainty, including cost of labor). 
%
%

The {\it Compact Linear Collider (CLIC)} is a TeV-scale high-luminosity linear $e^+e^−$ collider proposal, that envisions three stages of construction and operation at c.m.e.~of 380 GeV, 1.5 TeV, and 3 TeV, and a site length ranging between 11 km and 50 km. What makes it distinct from the ILC is its novel two-beam acceleration scheme, in which NC copper high-gradient 12 GHz accelerating structures are powered by a high-current 1.9 GeV drive beam to efficiently enable an accelerating gradient of 100 MV/m, about three times the gradient of the ILC. 
For the first CLIC stage at $\sqrt{s}$=380 GeV, suitable for Higgs boson studies, the optimum gradient turns out to be a little lower, $G=$72 MV/m \cite{robson2018compact}, and for this stage an alternative RF power drive option with 12 GHz klystron powering is also being considered \cite{accel:CLICupdate}. The main parameters of CLIC are given in Table \ref{eefuturetable}.

\begin{figure}[htbp]
\centering
\includegraphics[width=0.99\linewidth]{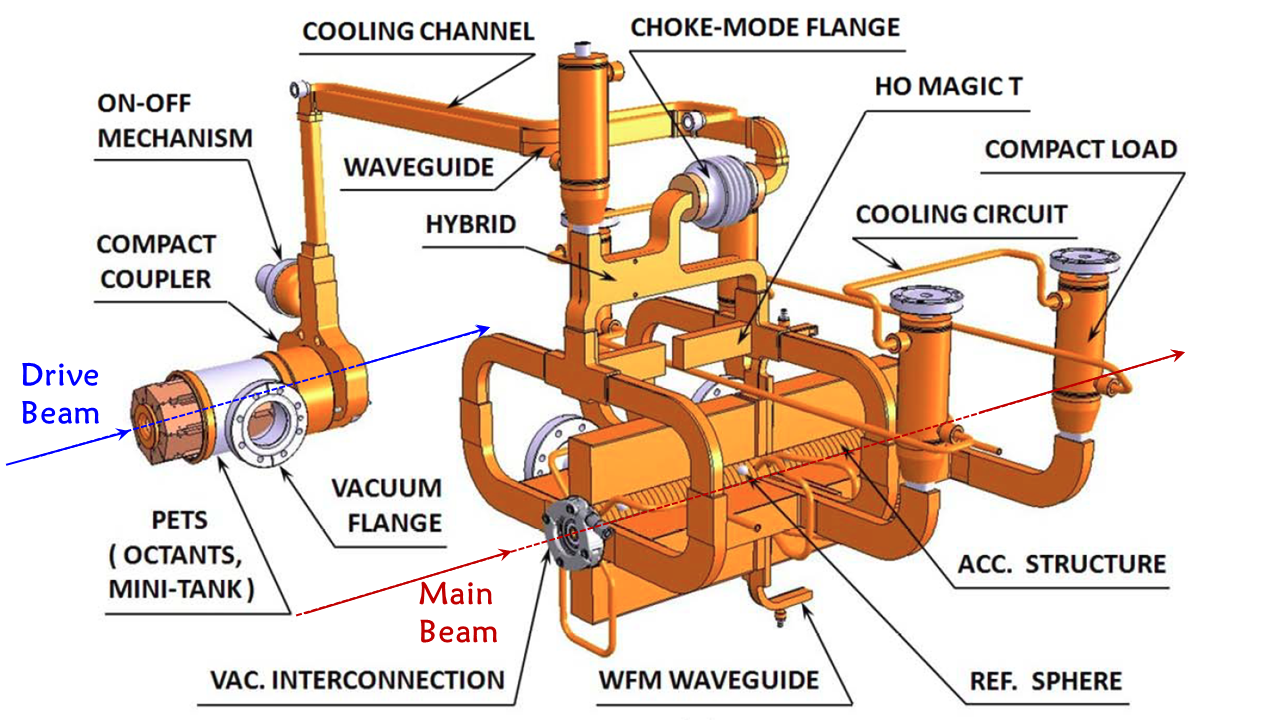}
\caption{3D model of the CLIC two-beam RF module (adapted from \cite{accel:CLIC}).}
\label{fig:CLICtwobeams}
\end{figure}

CLIC two-beam configuration is implemented by using two distinct parallel linear RF accelerating structures some 0.6 m apart,  connected by a waveguide network; see Fig.~\ref{fig:CLICtwobeams}. A low-impedance power extraction and transfer structure (PETS, of about 0.3 m length with 23 mm aperture) resonantly decelerates the drive beam consisting of 12 GHz bunches with an average gradient of about $-6.5$~MV/m. The kinetic energy of the drive beam is efficiently converted into the energy of 12 GHz EM waves which are extracted and sent to power two high-impedance accelerating structures (each 23 cm long, with 5 mm aperture) to accelerate the main beam with gradient up to $G=$100 MV/m. The maximum gradient must be achieved at nominal pulse length and shape (156 ns flat-top, 240 ns full length) and with a breakdown rate (BDR) of  less than $\sim10^{-6}$ --- low enough for the  reliable operation of some 20,000 structures in two linacs. This is one of the CLIC challenges, as an empirical scaling law \cite{grudiev2009scaling} relates the breakdown rate, the gradient $G$, and the RF pulse length $\tau_{RF}$ approximately via 
\begin{equation}
BDR\propto G^{30} \tau_{RF}^5 \; .     
\label{eq:bdr}
\end{equation}

\begin{figure*}[htbp]
\centering
\includegraphics[width=0.8\textwidth]{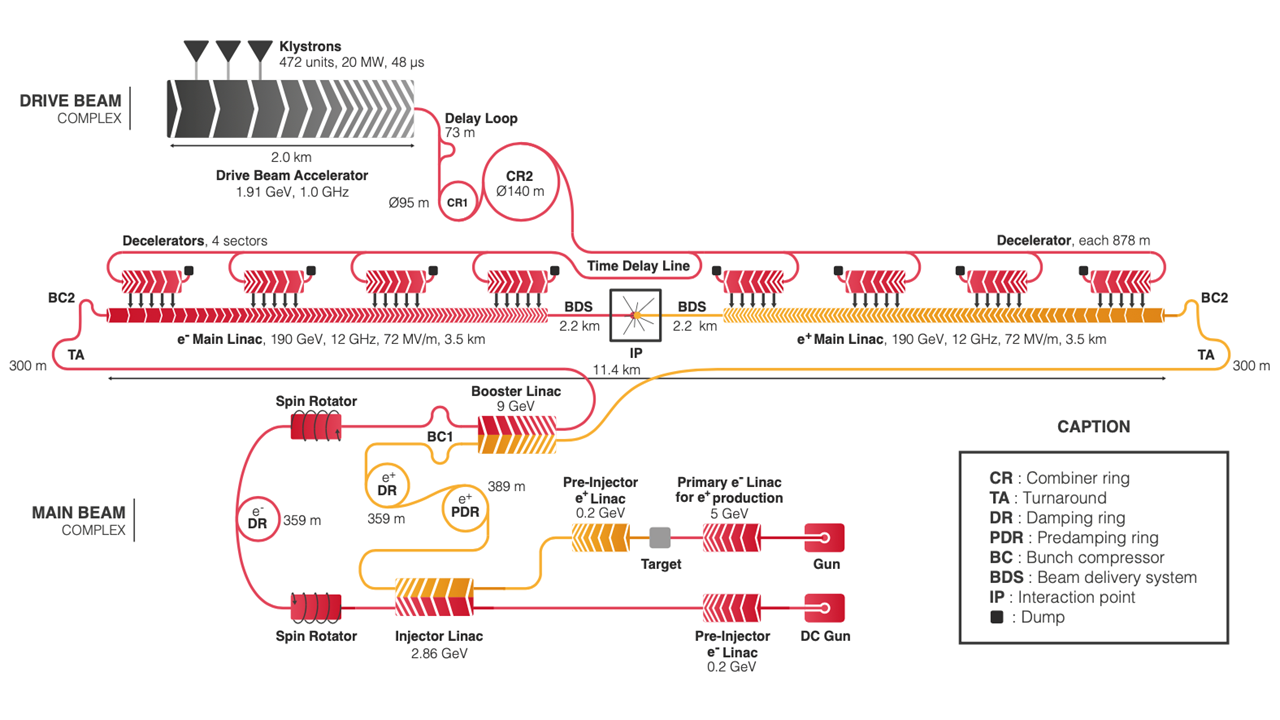}
\caption{CLIC accelerator complex layout at $\sqrt{s}$=380 GeV \cite{stapnes2019layout}.}
\label{fig:CLIC_layout380}
\end{figure*}

Figure~\ref{fig:CLIC_layout380} presents a schematic layout of the CLIC complex. The main spin-polarized $e^-$ beam is produced on a strained GaAs cathode in a conventional RF source and accelerated to 2.86 GeV. The beam emittance is then reduced in a damping ring. For positron beam production, a dedicated 5 GeV linac sends electrons onto a crystal to produce energetic photons, which in turn hit a second target to produce $e^+$. These positrons are captured, accelerated to 2.86 GeV and sent through a series of two emittance damping rings. The CLIC RTML system accelerates 352 bunches, with 0.5 ns bunch spacing, in each electron and positron beam to 9 GeV, and compresses their bunch 
lengths to 70 $\mu$m rms 
(or 44 $\mu$m for higher c.m.e.). 

After the main linacs have accelerated the beams to 190 GeV, collimators in the beam delivery system remove any transverse tails and off-energy particles, and finally the final focus magnets compress the beams to the required small transverse sizes at the collision point. After the collision, the spent beams are transported to a beam dump.

Every $1/f_{\rm r}=$1/(50~Hz)=20 ms, the 20 MW  drive beam (1.91 GeV, 101 A) is generated in a 48 $\mu$s long pulse of bunches spaced by 0.6 m in a central complex with a 1 GHz fundamental frequency of its 472 klystrons. After a sequence of longitudinal manipulations,  involving a delay line and two combiner rings, the initial beam is transformed into two series of four 244 ns long final sub-pulses with a 12 GHz bunch spacing of 2.5 cm (i.e., 24 times the initial beam current) which are sent in opposite directions to power the two linacs. The first sub-pulse in each linac powers the first drive-beam decelerator, running in parallel to the colliding beam.
When this sub-pulse reaches the decelerator end, the second sub-pulse has reached the beginning of the second drive-beam decelerator and will power it, running in parallel to the colliding beam, and so on. 

The CLIC luminosity critically depends on beam emittances (particularly vertical ones) at collision at the IP, requiring generation of $e^+$ and $e^-$ beams with a minimum emittance and their tight preservation during acceleration and focusing. The latter calls for control of all relevant imperfections, such as pre-alignment of all the main linac and beam delivery system components at the 10~$\mu$m level, suppression of vibrations of the quadrupoles due to ground motion to the level of 1.5 nm at frequencies above 1 Hz for the main linac (and to 0.2 nm above 4 Hz in the final focus system) \cite{collette2010activestabilization}, novel beam-based trajectory tuning methods to minimize the effect of dynamic and static imperfections using submicron resolution BPMs \cite{eliasson2008clicatl, balik2013mitigategmfeedback}, and mitigation of the effect of wakefields caused by high current beams passing through misaligned accelerating structures. 
As a net result, between the damping ring and the IP,  
the CLIC vertical normalized emittance increases by less than a factor of 4. 

CLIC accelerator design, technical developments and system tests have resulted in a high energy efficiency, and a correspondingly low power consumption of around 170 MW for the 380 GeV stage, and a machine total cost estimate of approximately 6 BCHF \cite{accel:CLICupdate}.

There has been remarkable progress in linear collider R\&D in recent years. Beam accelerating gradients met the ILC goal of 31.5 MV/m at the Fermilab FAST facility in 2017 \cite{broemmelsiek2018} and at KEK in 2019, and exceeded  CLIC specifications at the CLEX facility at CERN, where the drive beam was used to accelerate the main beam with a maximum gradient of 145 MV/m \cite{robson2018compact}. The Accelerator Test Facility  in KEK has also demonstrated attainment of the required vertical beam emittance in the damping ring and focusing of that beam into 40 nm vertical rms beam size \cite{bambade2010atf}. 

Higgs Factory proposals based on linear $e^+e^-$ colliders offer several advantages: they are based on mature technologies of NC RF and SRF that have been well explored at several beam test facilities. Their designs have been developed to a sufficient level of detail. At present the ILC design is described in a TDR and the design of CLIC in a comprehensive conceptual design report (CDR). Advantageous for HEP research is also beam polarization (80\% for $e^-$ and 30\% for $e^+$ at the ILC;  80\% for $e^-$ and 0\% for $e^+$ at CLIC). Linear colliders are expandable to higher energies (ILC to 0.5 and 1 TeV, CLIC to 3 TeV). Both proposals have well established international collaborations, which indicate readiness to start construction soon; their demand for AC wall plug power of 130--170 MW is less than that of the LHC complex ($\sim$200 MW). 

At the same time, one has to pay attention to the following factors: (i) the cost of these facilities equals or somewhat exceeds the LHC cost; (ii) the ILC and CLIC luminosity projections are in general lower than those for rings (see Fig.~\ref{HFlumis} and discussion below), and luminosity upgrades (such as via a two-fold increase of the number of bunches $n_b$ and doubling the repetition rate from 5 Hz to 10 Hz in the ILC) will probably come at additional cost; (iii) operational experience with linear colliders is limited only to SLAC's SLC, whose ten-year commissioning experience hints the possible operational challenges; (iv) the CLIC's two-beam scheme is quite novel (klystrons are therefore a backup RF source option); and (v) AC wall plug power demand may grow beyond 200 MW for the proposed luminosity and energy upgrades. 

\begin{figure}[htbp]
\centering
\includegraphics[width=0.99\linewidth]{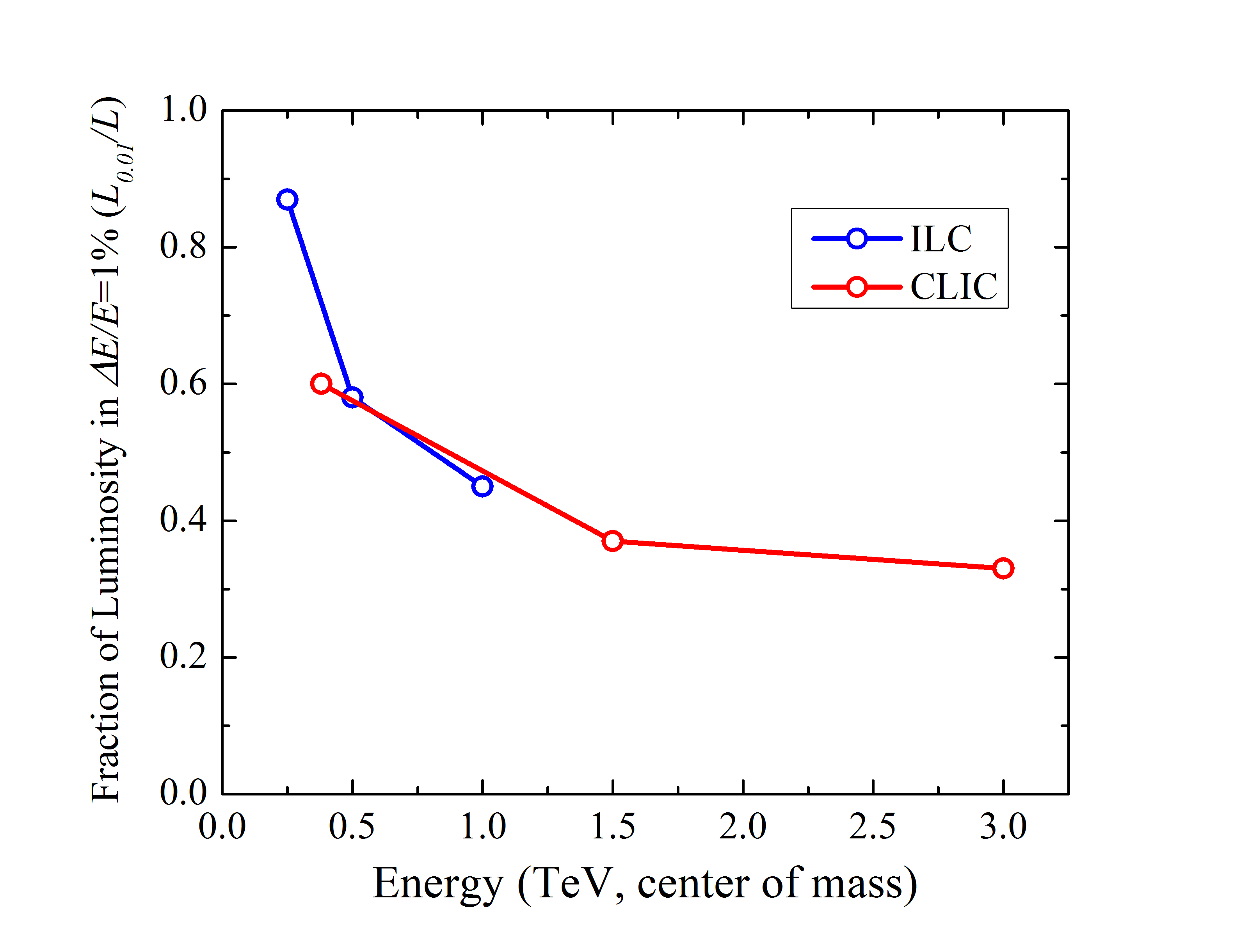}
\caption{Beamstrahlung effects in ILC and CLIC luminosity 
--- fraction of the luminosity within 1\% of c.m.e.~vs.~energy.}
\label{fig:lcbeamstrahlung}
\end{figure}

Linear $e^+e^-$ colliders for TeV and multi-TeV c.m.e.~face even more formidable challenges: their lengths grow to 40--50 km, AC power requirements become 300--600 MW, the beamstrahlung leaves only 30--40\% of the luminosity within 1\% of maximum energy (see Fig.~\ref{fig:lcbeamstrahlung}) and project costs grow to \$17B for 1 TeV ILC (the TDR cost estimate) and 18.3 BCHF for 3 TeV CLIC (CDR). 

\begin{figure}[htbp]
\centering
\includegraphics[width=.95\linewidth]{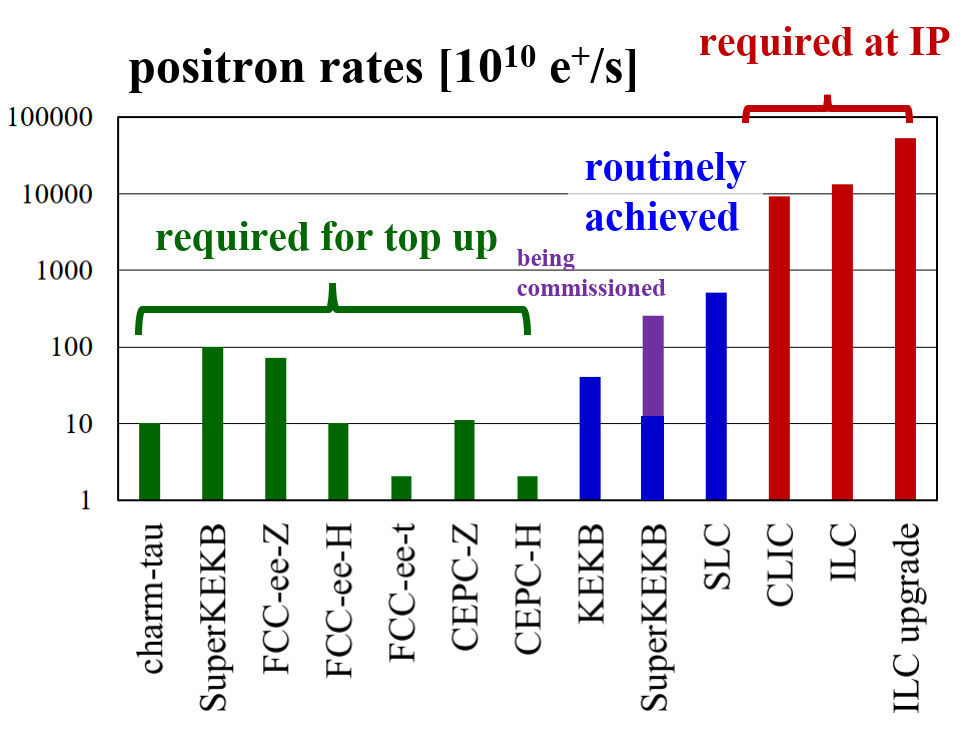}
\caption{\label{possource}
Positron production rates achieved at the SLC, KEKB and SuperKEKB compared with the need for top-up injection at future circular and linear e$^+$e$^-$ colliders (also in Ref.~\protect\cite{fcc-nature}).}
\end{figure}

To reach their design luminosities, both CLIC and ILC require unprecedented rates of positron production. The ILC baseline foresees about 40 times the world record set by the SLC positron source, and the ILC luminosity upgrade calls for improvement by another factor of four. Figure \ref{possource} compares the demonstrated $e^+$ production rates at the SLC, KEKB, and SuperKEKB with the needs for top-up injection at future circular $e^+e^-$ colliders, and at the collision point of future linear $e^+e^-$ colliders.

\subsubsection{Circular $e^+e^-$ colliders for the electroweak sector: FCC-$ee$ and CEPC}
\label{circfutureee}
The 2012 discovery of the Higgs boson at the LHC has stimulated interest in circular Higgs factories \cite{blondel2013report}, and in particular, in the construction of a large circular tunnel which could host a variety of energy-frontier machines, including high-energy electron-positron, proton-proton, and lepton-hadron colliders. Such projects are being developed by the global Future Circular Collider (FCC) collaboration hosted at CERN \cite{fccee} and, in parallel, by the Circular Electron-Positron Collider (CEPC) study group based in China \cite{cepc}, following an earlier proposal for a Very Large Lepton Collider (VLLC) \cite{sen2002vllcvlhc} in the US, which would have been housed in the 230-km long tunnel of the Very Large Hadron Collider (VLHC) \cite{accel:vlhc}.

In order to serve as a Higgs factory, a new circular e$^+$e$^-$ collider needs to achieve a c.m.e.~of at least 240 GeV \cite{benedikt2019futurecircol}. 
The unprecedentedly high target luminosity  
$L$ of FCC-ee and CEPC implies a short beam lifetime, 
\begin{equation}
\tau_{\rm beam} \le \frac{n_b N}{n_{IP} \sigma L}
\; 
\label{radb}
\end{equation}
of order 1 hour, 
due to the unavoidable radiative Bhabha scattering
with a cross section of $\sigma \approx 200$~mbarn \cite{Burkhardt:1994wk}. In Eq.~(\ref{radb}) $n_b$ signifies the number of bunches,
$n_{IP}$ the number of collision points, and $N$ the bunch population.
The short beam lifetime due to radiative Bhabha scattering, which can be further reduced
by beamstrahlung (see below), 
is sustained by quasi-continuous top-up injection. The technique of 
top-up injection was routinely and successfully 
used at both PEP-II and KEKB \cite{Seeman:2015gpa,Satoh:2010zz}, where physics runs with nearly constant beam currents and constant luminosity were only occasionally interrupted (e.g., 
a few times per day) by fast beam aborts due to hardware failures. 
Top-up injection for FCC-ee or CEPC calls for a full-energy fast-ramping booster ring, with the same circumference as the collider,
i.e., installed in the same tunnel.

At high energy, the performance of a circular collider is limited by synchrotron radiation. The maximum beam current is directly proportional to synchrotron radiation power $P_{\rm SR}$, and to the bending radius $\rho$, and it scales with the inverse fourth power of the beam energy $E_{b}$ or the Lorentz factor $\gamma$, that is $I_{b} = P_{\rm SR}/(\Delta E_{SR}) \propto  P_{\rm SR} \rho / \gamma^{4}$ (see Eq.~(\ref{SR})). 
Correspondingly, the luminosity Eq.~(\ref{eq:lumiee}) scales as the product of the ring radius $\rho$, beam-beam parameter $\xi_y$, beta-function at the IP $\beta_y^*$, and the RF power $P$, and with the inverse of $E_b^3$. The beam-beam parameter is limited to about $\xi_y$=0.13 by a new type of beam-beam instability
occurring for collisions with a nonzero crossing angle 
\cite{Ohmi:2017cwi,oide2018bbinstabfccee}. 
Average beam energy loss per turn due to the synchrotron radiation varies between 0.1 and 5\% (from $Z$ energy to 365 GeV), and, at the higher energies, it is significantly larger than the equilibrium energy spread due to beamstrahlung, which can be of order 0.1--0.2\%.
It is worth noting that, in the lower energy running modes
of the circular 
Higgs/electroweak factories ($Z$ and $WW$ runs),
the beamstrahlung
significantly increases the energy spread and bunch length,
by up to a factor of 3--4 over what would be obtained without collisions, i.e.~with the natural bunch length set by the quantum fluctuation in the low-field collider arcs. 
This large relative increase of the energy spread  
due to beamstrahlung is due to the weak radiation 
damping at these lower energies, where each electron, 
or positron, emits multiple beamstrahlung photons 
during one radiation damping time.    
At the higher beam energies, especially for 
${\rm t \bar{t}}$ operation, radiation damping is stronger
and the energy spread increase due to beamstrahlung 
becomes much less pronounced. However, here 
the single emission of hard beamstrahlung photons at the collision point introduces an additional limit on the beam lifetime \cite{telnov-prl}, which is about 20 minutes even in a sophisticated  {\it crab-waist} optics with $\beta^*_y = 0.8$--$1.6$~mm with large energy acceptance \cite{bogomyagkov2014beamstrahlfccee}.


The FCC-ee would be installed in a $\sim$100 km tunnel,
which can afterwards host a 100 TeV hadron collider (FCC-hh, see below). The FCC complex would be connected to the existing CERN infrastructure. 
CEPC is a project under development in China \cite{cepc}, that is similar to FCC-ee.
CEPC would also be followed by a highest-energy hadron collider in the same tunnel, called the Super proton-proton Collider (SppC).

%

FCC-ee operation is staged, starting on the $Z$ pole (91 GeV c.m.e.) with approximately $10^{5}$ the luminosity of the previous LEP collider, then operating at the $WW$ threshold (160 GeV), followed by the Higgs production peak (240 GeV), and finally at the $t\bar{t}$ threshold (365 GeV). 
An optional FCC-ee run at 125 GeV for direct Higgs 
production with monochromatization  \cite{ValdiviaGarcia:2019ezi}  
could access the Higgs-electron Yukawa coupling 
\cite{dEnterria:2017dac,Blondel:2019ykp}.
Possibly this constitutes the only available path to 
address the origin of the electron mass. 
On the $Z$ pole and at the $WW$ threshold, radiative self polarization allows for an extremely precise c.m.e.~energy calibration, at the $10^{-6}$ level, based on resonant depolarization \cite{Blondel:2019jmp}. Even at the highest FCC-ee collision energy, 365 GeV c.m.e., the luminosity, limited by 100 MW of synchrotron radiation power, would still exceed 10$^{34}$ cm$^{-2}$s$^{-1}$ at each of two or four collision points.  
The FCC-ee SRF system is optimally adapted for each mode of operation, i.e., it is optimized for the respective beam current and the RF voltage required.
Specifically, at the $Z$ pole, the FCC-ee deploys single-cell 400 MHz Nb/Cu cavities, while at the $WW$ threshold and the $ZH$ production peak 400 MHz five-cell Nb/Cu cavities will operate at 4.5 K. For $t\bar{t}$ running they will be complemented  by additional 800 MHz bulk Nb cavities at 2 K.
At the $t\bar{t}$ energy, the RF cavities are shared by the two beams, in common RF straights, which saves cost and is possible thanks to the small number of bunches in this mode of operation.


CEPC operation is scheduled to start at the Higgs production peak (240 GeV c.m.e.), continues on the Z pole (91 GeV), and ends with the WW threshold (160 GeV). Operation at the $t\bar{t}$ energy is not foreseen -- see Fig.~\ref{HFlumis}.   CEPC deploys the same 
650 MHz bulk Nb cavities at all beam energies. 
However, the total number of installed RF cavities varies from 240 (240 GeV) over 120 (91 GeV) to 216 (160 GeV), in the three modes of operation.  
At the highest (initial) center-of-mass 
energy of 240 GeV  the 240 installed cavities are shared by the two beams. The impedance and higher-order mode power of the 650 MHz RF cavities limit the projected CEPC luminosity at the $Z$ pole to a value about an order-of-magnitude lower than for FCC-ee (Fig.~\ref{HFlumis}). 

The optics designs of FCC-ee \cite{oideoptics} and CEPC contain several novel features, which boost their performance.   
For the crab waist collision scheme, a large crossing angle is needed and 30 mrad was found to be optimal for FCC-ee;
for CECP a similar value of 33 mrad has been chosen. 
The two colliding beams always approach the IP from the inside,
with bending magnets on the incoming side selected to be so weak that, for the FCC-ee, 
the critical energy of the photons emitted towards the detector stays below 100 keV over the last 450 m from the IP, even at the highest beam energy for $t\bar{t}$ operation
(365 GeV c.m.e.). Similarly, 
for CEPC in $ZH$ production mode (240 GeV), 
the critical energy of the synchrotron-radiation photons 
is less than 120 keV over the last 400 m upstream of the IP. 
 
\begin{figure}[htbp]
\centering
\includegraphics[width=0.9\columnwidth]{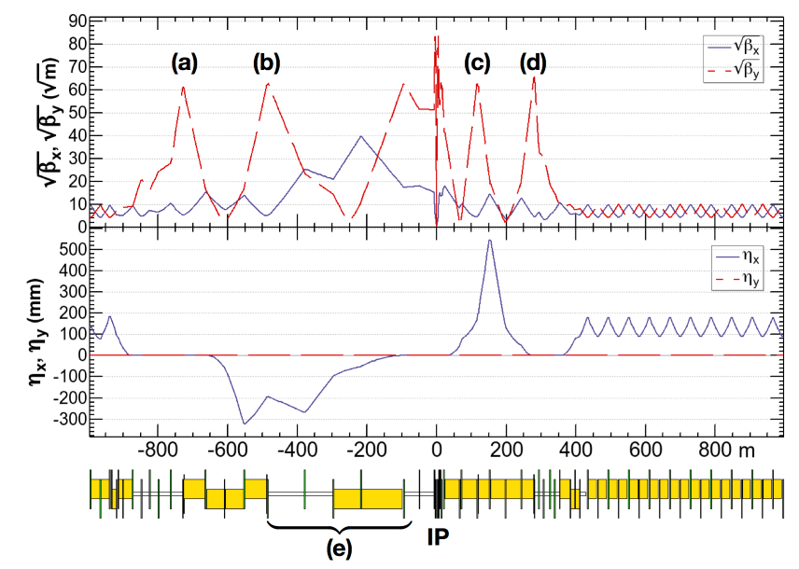}
\caption{Asymmetric final-focus optics of FCC-ee, 
featuring four sextupoles (a--d) 
for local vertical chromaticity correction 
combined with a virtual 
crab waist (see text for details) \protect\cite{oideoptics,fccee}.} 
\label{fcceeff}
\end{figure}

\begin{figure}[htbp]
\centering
\includegraphics[width=0.98\columnwidth]{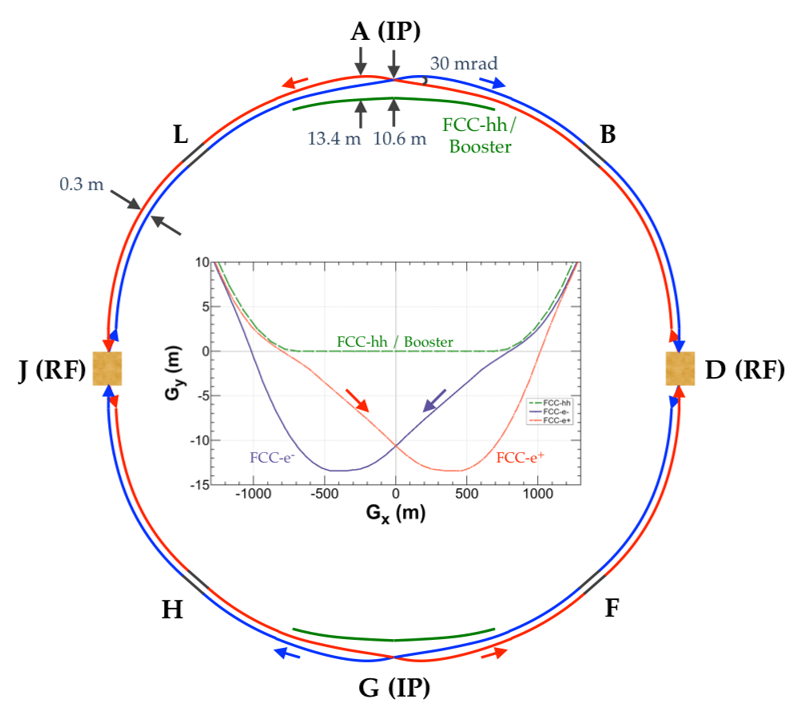}
\caption{Layout of the FCC-ee double ring collider with two long RF straights and two interactions points, sharing a tunnel with the full-energy top-up booster \protect\cite{oideoptics,fccee}.} 
\label{fccee-layout}
\end{figure}

As a result, the FCC-ee and CEPC final focus optics are asymmetric (see Fig.~\ref{fcceeff} for the FCC-ee). 
Figure \ref{fccee-layout} illustrates how asymmetric bending also separates the FCC-ee collider IP from the beam line of the full-energy top-up booster by more than 10 meters, leaving sufficient space for the experimental detector. The CEPC layout is similar. 

Stronger magnets and a shorter final focus system are 
installed on the outgoing side of the IP.  Each final focus accommodates a pair of sextupoles, separated by a {\it minus identity} ($-I$) optics transformer,  to accomplish a local correction of the 
vertical chromaticity. 
Thanks to the $-I$ transformer, geometric aberrations 
can be exactly 
cancelled between the two sextupoles of a pair.
However, by construction, the dominant aberration generated when reducing the strength of the outer sextupole of the pair
generates the desired crab waist at the IP,
while maintaining the chromatic correction. 
Hence, instead of adding one more sextupole as for the crab-waist implementation in other colliders, the FCC-ee utilizes an elegant and novel ``virtual crab waist'' scheme. Here, the total number and strength of nonlinear magnets is even reduced, compared with the case of no crab waist, with a positive effect on the dynamic aperture.

%

Both FCC-ee and CEPC proposals call for very high SRF power transfer to beams (100 MW in FCC-ee and 60 MW in CEPC), leading to total site power of about 300 MW. Cost estimate of the FCC-ee is 10.5 BCHF (plus additional 1.1 BCHF for the option to operate at the higher ${\rm t\bar{t}}$ energy) and \$5B to \$6B for the CEPC (``less than 6 BCHF'' cited in the CEPC CDR). 

The advantages of circular Higgs Factories include a quite mature SRF acceleration technology, with vast operational  experience from numerous other rings, suggesting a lower performance risk, along with a higher luminosity and better luminosity/cost ratio. 
They can also host detectors at several (2 to 4) IPs, which could further strengthen their role as {\it EW (electroweak) factories}. The 100 km long tunnels can both be reused, and are required, by follow-up future $pp$ colliders. 

Transverse polarization occurs naturally at $Z$ and $W$ energies, and can be employed, with the help of pilot bunches and, possibly, polarization wigglers at the $Z$ energy, for precise energy calibration at the 100 keV level. The strong and broad {\it global FCC collaboration} has issued a comprehensive CDR \cite{fccee}, that addresses key design points and indicates a possible start date ca.~2039. The schedule of the CEPC CDR \cite{cepc}  
is more aggressive, and foresees a start of machine operation some 7--9 years sooner. 
Prior to this, the FCC and CEPC R\&D programs are expected to address several important items, such as high efficiency RF sources (e.g.~targeting over 85\% for 400/800 MHz klystrons, up from the current 65\%), high efficiency SRF cavities (to achieve 10--20 MV/m CW gradient and a high cavity quality factor $Q_0$, and to develop new technologies like advanced 
Nb-on-Cu or Nb$_3$Sn cavities), 
the exploration of the crab-waist collision scheme (the SuperKEKB experience will be extremely helpful in this regard), energy storage and release (so that the 
energy stored in cycling magnets can be reused), and the efficient usage of excavated materials (some 10 million cu.m.~will need to be taken out of a 100 km tunnel).

\begin{table*}
\begin{center}
\begin{tabular}{|l|ccc|cc|cc|cc|}
\hline\hline
 & \multicolumn{3}{c|}{FCC-ee \protect\cite{fccee}} &
 \multicolumn{2}{c|}{CEPC \protect\cite{cepc}} &
 \multicolumn{2}{c|}{ILC \protect\cite{accel:ILC3}} &
 \multicolumn{2}{c|}{CLIC \protect\cite{accel:CLICupdate}} 
\\ \hline
Species&
\multicolumn{3}{c|}{$e^+e^-$} &
\multicolumn{2}{c|}{$e^+e^-$} &	
\multicolumn{2}{c|}{$e^+e^-$} &	
\multicolumn{2}{c|}{$e^+e^-$} 
\\ 
Beam energy (GeV)& 
45.6 &  120 &  183 &
45.5 & 120 &
125 & 250 &
190 & 1500
\\ 
Circumf., length (km) &
\multicolumn{3}{c|}{97.75}&
\multicolumn{2}{c|}{100}	
	& 20.5 &  31
        & 11 &  50
\\ 
Interaction regions  &
\multicolumn{3}{c|}{2 (or 3)}&
\multicolumn{2}{c|}{2}&
\multicolumn{2}{c|}{1} & 
\multicolumn{2}{c|}{1}
\\ 
Int.~lum./exp.~(ab$^{-1}$/year)&
	26& 0.9& 0.17&	4& 0.4 &
        0.2& 0.2&
        0.2 & 0.6
\\
Peak lum.~(10$^{34}$ cm$^{-2}$ s$^{-1}$)&
200 & 7& 1.5&
32& 3&
1.4 & 1.8 &
1.5 & 5.9 
\\
Rep.rate (Hz, $f_{rev}$ for rings)&
\multicolumn{3}{c|}{3067}&
\multicolumn{2}{c|}{3000}&
\multicolumn{2}{c|}{5} & 
\multicolumn{2}{c|}{50}
\\
Polarization (\%)&
$\ge$10 &  0 &  0 & 
5--10 &  0 &     
\multicolumn{2}{c|}{
80, 30\% ($e^-,e^+$)} 
& 
\multicolumn{2}{c|}{
80\%, 0\% } 
\\ 
Time betw.~collisions ($\mu s$)&
0.015& 0.75 & 8.5 &
	0.025 & 0.68 & 
0.55 & 0.55 & 
0.0005 & 0.0005
\\
\hline
Energy spread (rms, $10^{-3}$)&
	1.3 &  1.65 &  2.0 & 
	0.4 & 1.0 &
	1.9, 1.5  & 
	1.2, 0.7  &
	3.5 & 3.5 
\\
 &
 &  &  & 
 &  &
 ($e^-$,$e^+$) & 
 ($e^-$,$e^+$) &
 &
\\
Bunch length (rms, mm)&
	12.1 &  5.3 &  3.8 &
	8.5 &  4.4 &
	0.3 & 0.3 & 
	0.07 &  0.044
\\
Norm.~rms emit. (H,V $\mu$m)&
	24,  0.09 & 
	148, 0.3 &  
520, 1.0 &
	16,  0.14 & 
	284, 0.6
&
	5, 0.035 & 
	10, 0.035 & 
	0.9, 0.03
	& 
	0.66, 0.03
\\
$\beta^*$ at IP (H,V cm)&
15, 0.08
&  30, 0.1 & 100 0.16&
	20,  0.1 & 36, 0.15 &
	1.3, 0.041 & 2.2, 0.048 &
	0.8, 0.01 & 0.69,  0.007 
\\
Hor.~IP beam size ($\mu$m)&
6.4 &  14 &  38 &
6.0 & 21
& 0.52 &
0.47 & 
0.15 & 
0.04 
\\ 
Vert.~IP beam size (nm)&
28 &  36 &  68 &
40 & 60
& 8 &
 6 & 
 3 & 
 1 
\\ 
Full crossing angle (mrad)&
\multicolumn{3}{c|}{30} &
\multicolumn{2}{c|}{33} & 
\multicolumn{2}{c|}{14} & 
\multicolumn{2}{c|}{20} 
\\
Crossing scheme&
\multicolumn{3}{c|}{crab waist} &
\multicolumn{2}{c|}{crab waist} &
\multicolumn{2}{c|}{crab crossing }&
\multicolumn{2}{c|}{crab crossing }	
\\ 
Piwinski angle $\Phi$ 
&
28.5 &  5.8 &  1.5 &
23.8 & 2.6 &
\multicolumn{2}{c|}{0} &
\multicolumn{2}{c|}{0} 
\\
Beam-b.~param.~$\xi_{y}$ ($10^{-3}$) &
133 & 118 &  144 &
    72 & 109 &
\multicolumn{2}{c|}{n/a} &
\multicolumn{2}{c|}{n/a} 
\\ 
RF frequency (MHz)&
400 &  400 &  400\&800 &
650 & 650 & 1300 & 1300 & 11994 & 11994 \\

Particles per bunch ($10^{10}$) &
	17 &  15 &  27 &
	8 & 15&
	2 & 2 &
        0.52 &  0.37 
\\
Bunches per beam  &
16640 &  328 &  33 &
12000 & 242 & 1312 & 1312 & 
352 & 312  
\\
Avg.~beam current (mA)  &
1390 &  29 &  5.4 & 19.2 & 19.2 & 
0.021 & 0.021 & 
0.014 & 0.009
\\
\hline
Injection energy (GeV)&
\multicolumn{3}{c|}{on energy (top up)}&
\multicolumn{2}{c|}{on energy (top up)}&
\multicolumn{2}{c|}{5.0 (linac)} & 
\multicolumn{2}{c|}{9.0 (linac)}
\\ 
RF gradient (MV/m)  &
1.3 &  9.8 &  19.8 &
3.6 & 19.7 &
31.5 & 31.5 & 
72 & 100 
\\ 
SR power loss  (MW) &
\multicolumn{3}{c|}{100} &
\multicolumn{2}{c|}{64} &
\multicolumn{2}{c|}{n/a} &
\multicolumn{2}{c|}{n/a} 
\\ 
Beam power/beam (MW) &
\multicolumn{3}{c|}{n/a} &
\multicolumn{2}{c|}{n/a} &
5.3 & 10.5 & 3 & 14 
\\ 
Novel technology required &
\multicolumn{3}{c|}{---} &
\multicolumn{2}{c|}{---} &
\multicolumn{2}{c|}{high grad.~SC RF} &
\multicolumn{2}{c|}{two-beam accel.} \\
\hline\hline
\end{tabular}

\end{center}
\caption{Tentative parameters of selected future $e^+e^−$ high-energy colliders.  
}. 
\label{eefuturetable}
\end{table*}

\subsection{Energy frontier colliders (HE-LHC, FCC-hh, SppC, Muon Colliders)}
\label{frontiercolliders}

Several hadron and lepton colliders have been proposed to extend the energy 
reach beyond the present LHC at CERN. 
The physics program that could be pursued by a next-energy frontier collider with sufficient luminosity would include: understanding the mechanism behind mass generation, the Higgs mechanism and the role of the Higgs boson in the electroweak symmetry breaking; answering the question of whether the Higgs boson is a fundamental or composite particle; searching for, and possibly discovering, supersymmetric or other exotic particles, which could be part of the Universe's dark matter; and hunting for signs of extra spacetime dimensions and quantum gravity \cite{quigg2019, arkani2016physics100tev}. As alluded to above, ambitious plans have been proposed to upgrade the FCC and CEPC to hadron colliders (FCC-hh and SppC, respectively) by means of next- or next-next generation SC magnets installed in the arc sections of the 100 km rings, so as to enable collision energy of the order 100 TeV or above  \cite{benedikt2019futurecircol,cepc}. Among the lepton colliders, there is a growing community interest in cost efficient muon colliders that can possibly provide collision energies ranging from 3 TeV to 14 TeV, significantly beyond the reach of practical  $e^+e^-$ linear colliders.  

\subsubsection{Post-LHC hadron colliders}

Circular hadron colliders are known as discovery machines. Their discovery reach is determined by beam energy, which depends on only two parameters: the dipole magnetic field $B$ and the bending radius $\rho$, $E_{\rm c.m.} \propto \rho  B$ (see Eq. \ref{eq:energy_c}). Historically, new  colliders were always larger and used stronger magnets than their predecessors. For example, the Tevatron near Chicago was the first hadron collider based on SC magnet technology, with a dipole field of 4.2 T, and it was installed in a 6.3 km ring. In comparison, the LHC uses 8.3 T dipoles in a 26.7 km tunnel. 
A proposed high energy ``upgrade'' of the LHC based on 16 T Nb$_3$Sn SC magnets --- the High-Energy LHC (HE-LHC) --- would allow for 27 TeV c.m.e.~hadron collisions re-using the LHC tunnel \cite{helhc}.
A further increase in the collider size by a factor of about 4 compared with the LHC, i.e., to a  circumference of 100 km, yields a c.m.e.~of 100 TeV with similar 16 T dipole magnets (Future Circular Collider, hadron version ``FCC-hh'') \cite{fcchh}. This goal defines the overall infrastructure requirements for the FCC accelerator complex. A proton-electron collision option (FCC-he) calls for a 60 GeV electron beam from an ERL (the same as for the LHeC), which would be collided, at a single interaction point, 
with one of the two 50 TeV proton beams circulating in the FCC-hh; see Chapter \ref{LHeC}.

\begin{figure}
\centering
    \includegraphics*[width=0.99\columnwidth]{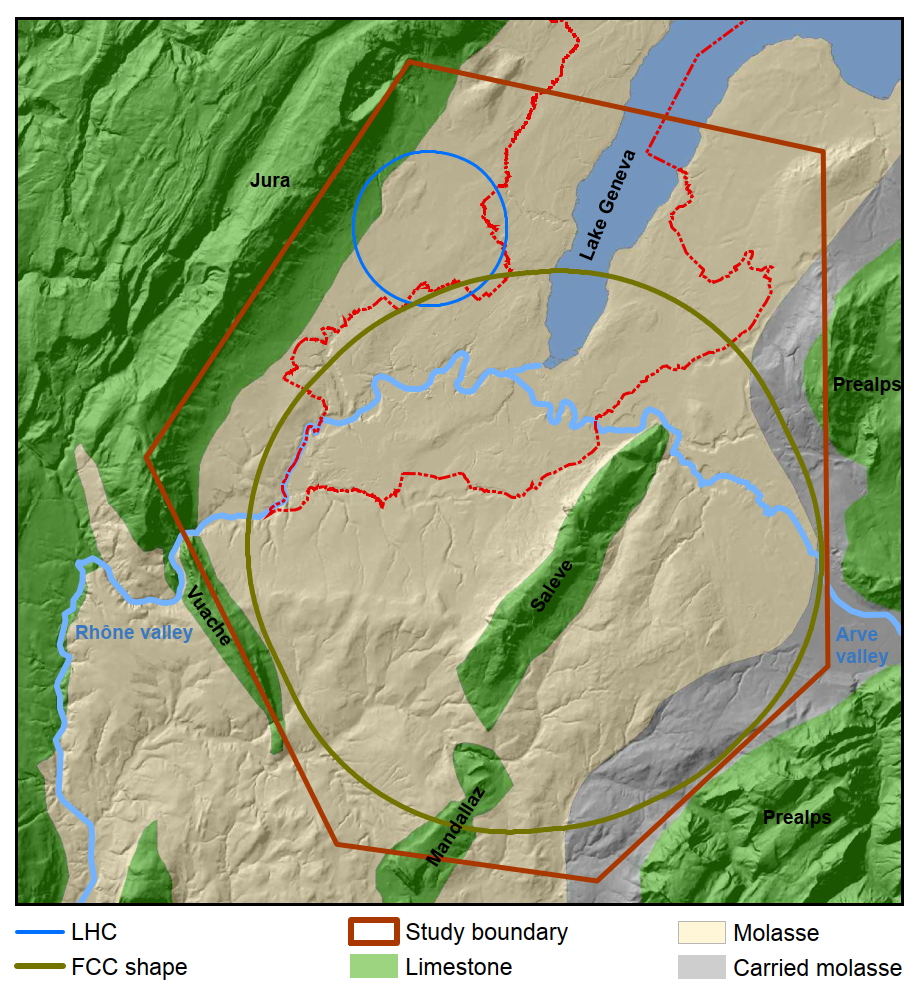}
\caption{Study boundary (red polygon), showing the main topographical and geological structures, LHC (blue line) and FCC tunnel trace (brown line) \protect\cite{fccee,fcchh}.}
\label{fig:schematic_summary}
\end{figure}

 CEPC and SppC are two colliders similar to
 FCC-ee/FCC-hh,  
 which are being studied by another international collaboration, centered at IHEP Beijing \cite{cepc}. These two machines have almost the same circumference as the FCC, of about 100 km. Several possible locations in China are under study. The SppC hadron collider relies on 12 T (later 24 T) iron-based high-temperature superconducting magnets, which could be installed in the same tunnel as the CEPC.  
Table \ref{tb:hadron} compiles key parameters of the HE-LHC, FCC-hh, and SppC.

Figure~\ref{fig:schematic_summary} indicates 
the proposed location of the FCC in the Lake 
Geneva basin,  
connected to the existing CERN/LHC accelerator complex.
The principal structure for the successively installed 
FCC lepton and hadron 
colliders (FCC-ee and FCC-hh) is a quasi-circular 97.75 km long tunnel composed of arc segments interleaved with straight sections. 
Approximately 8 km of bypass tunnels, 18 shafts, 14 large caverns and 
12 new surface sites are also planned. 
The tunnel location and depth were optimized taking into account the local geology.   

Collider luminosity should ideally increase with the square of the energy, since cross sections decrease as the inverse square  of energy. However, due to the nonlinear parton distribution inside the colliding protons, even a lower luminosity can produce exciting physics, with the most important parameter of a hadron collider remaining its energy. 
Nevertheless, at a given energy the discovery reach grows with higher luminosity \cite{salam} --- this is one of 
the motivations for upgrading the LHC to the HL-LHC. 
The LHC design has already dramatically 
increased luminosity compared with previous machines. 
Much higher luminosities still are expected for the proposed High-Energy LHC (HE-LHC) and FCC-hh, and also for the approved HL-LHC, which will lower its peak luminosity by {\it leveling}  in order to make it acceptable for the  physics experiments \cite{mbdsfz}. A high instantaneous luminosity would result in an event pile-up per bunch crossing of order 500 (from up to 50 in LHC), and, for the HL-LHC experiments, this could significantly degrade the quality of the particle detector data collected for the physics analysis. The technique of luminosity leveling allows sustaining the operational luminosity, and the associated event pile-up, at a constant level over a significant length of time via several techniques: (i) a gradual reduction of the beta function at the interaction point $\beta^*$, (ii) crossing angle variation, (iii) changes in the RF voltage of crab cavities or more sophisticated crabbing schemes \cite{fartoukh2014pileup}, (iv) dynamic bunch-length reduction, or (v) controlled variation of the transverse separation between the two colliding beams. Of note is that the luminosity of the highest energy hadron colliders, such as HE-LHC and FCC-hh, will profit from significant radiation damping due to the associated high beam energies and magnetic fields.
This radiation damping will naturally level their  luminosity evolution. 


Hadron-collider luminosity will increase linearly 
with energy due to the shrinking beam sizes $\sigma^*_{x,y}=\sqrt{\beta^*_{x,y} \varepsilon_{n \, x,y} / \gamma}$, when keeping the normalized beam emittances, beam currents, the beta functions at the interaction point (IP), $\beta_{x,y}^{\ast}$, and beam-beam tune shift constant.  Even higher luminosity can be achieved by reducing the IP beta functions. Perhaps surprisingly, until now all hadron colliders, starting from the ISR, have operated with similar beta functions, with minimum values of about 0.3 m (see Table \ref{tab:beta}). With a value of 0.15~m (or even 0.10~m) the HL-LHC will set a new record. An ongoing study aims at pushing the FCC-hh $\beta^{\ast}$ down to 5 cm  \cite{martin2017interaction}. 

For proton-proton colliders with many bunches, such as the HL-LHC and FCC-hh, a crossing angle is required to avoid or mitigate parasitic beam-beam collisions. Unfortunately, this crossing angle needs to be increased as $\beta_{x,y}^{\ast}$ is reduced. Without countermeasures, this would dramatically degrade the geometric overlap of the colliding bunches and all but eliminate any benefit from reducing the IP beam size. To avoid such a degradation, the HL-LHC, HE-LHC, and FCC-hh will all use novel crab cavities \cite{verdu2016crabcavities}. In 2018,
first beam tests of such crab cavities with protons were successfully performed at the CERN SPS \cite{Carver:2696108}.


\begin{table}[htbp]
\caption{Beta function values $\beta^*$ at IPs of hadron colliders (adapted and extended from 
Ref.~\protect\cite{xbeam17}).}
\label{tab:beta}
\begin{center}
\begin{tabular}{|l|c|c|}
\hline\hline
collider & $\beta_{x}^{\ast}$ [m] & $\beta_{y}^{\ast}$ [m] \\ 
\hline 
ISR & 3.0 & 0.3 \\
\hline
S$p\bar{p}$S & 0.6 & 0.15 \\
\hline
HERA-p & 2.45 & 0.18\\
\hline
RHIC & 0.50 & 0.50 \\
\hline
Tevatron & 0.28 & 0.28 \\
\hline
LHC & 0.3 & 0.3 \\
\hline
HL-LHC& 0.15 & 0.15 \\ 
\hline
HE-LHC & 0.25 & 0.25 \\
\hline
FCC-hh & 1.1$\rightarrow$ 0.3 (0.05) & 1.1$\rightarrow$ 0.3 (0.05)\\
\hline
SppC & 0.75 & 0.75  \\
\hline\hline
\end{tabular}
\end{center}
\end{table}

\begin{center}
\begin{table*}[ht]
{\small
\hfill{}
\begin{tabular}{|l|c|c|c|ccc|}
\hline
\hline
&  
HE-LHC & FCC-hh & SppC & \multicolumn{3}{c|}{$\mu\mu$ Collider} \\
\hline  
Species & 
	$pp$ &
	$pp$ &
	$pp$ &
	\multicolumn{3}{c|}{$\mu^{+}\mu^{-}$} 
\\
Beam energy (TeV) 
&  13.5 &  50 & 37.5 & 0.063 & 3 & 7$^*$
        \\
Circumference (km) & 
26.7	
        & 97.75 
        & 100 
        & 0.3  & 6 & 26.7 
        \\
Interaction regions  
& 2 (4) & 4 & 2 & 1 & 2 & 2 
\\
Peak luminosity (10$^{34}$ cm$^{-2}$ s$^{-1}$)
& 15 & 5$-$30 & 10 & 0.008 & 12 & 33 \\
Int.~lum. per exp. (ab$^{-1}$/year)
& 0.5 & 0.2--1.0 & 0.4  & 0.001 & 1.0 & 3 
\\
Time between coll.~($\mu s$) 
& 0.025 & 0.025 & 0.025 & 1 & 20 & 90 
\\
Events per crossing
& 800 & 170$-$1000 & $\sim$300 &  &  &  \\
\hline
Energy spread (rms, $10^{-3}$)&
0.1 & 0.1 & 0.2 & 0.04 & 1 & 1 \\
Bunch length (rms, mm)& 
80 & 80 & 75.5 & 63 & 2 & 1 \\
IP beam size ($\mu$m)&
6.6
& 6.8 (init.) & 6.8 (init.) &
75 & 1.5 & 0.6 
\\
Injection energy (GeV) &
1300 &3300 & 2100 & \multicolumn{3}{c|}{on energy} \\
Transv. emittance (rms norm., $\mu$m)&
2.5 & 2.1 (init.) & 2.4 (init.) & 200 & 25 & 25
\\
$\beta^*$, amplitude function at IP (cm)&
45 & 110$-$30 & 75 & 1.7 & 0.25 & 1 
\\
Beam-beam tune shift / IP ($10^{-3})$ &
5 & 5$-$15 & 7.5 & 20 & 90 & 100 \\
RF frequency (MHz) &
400 & 400 & 400/200 & 805 & 805 & 805
\\
Particles per bunch ($10^{10})$ &
22 & 10 & 15 & 400 & 200 & 200 \\
Bunches per beam  &
2808 & 10600 & 10080 & 1 & 1 & 1\\
Average beam current (mA)  &
1120 & 500 & 730 & 640 & 16 & 4 \\
Length of standard cell (m)  &
137 & 213 & 148 & \multicolumn{3}{c|}{n/a} \\
Phase advance per cell (deg)  &
90 & 90 & 90 & \multicolumn{3}{c|}{n/a} \\
\hline
Peak magnetic field (T)  &
16 & 16 & 12 (24) & 10 & 10 & 16 \\
SR power loss/beam  (MW) &
0.1 & 2.4 & 1.2 & 
0 & 0.07 & 0.5 \\
Long. damping time  (h) &
3.6 & 1.1 & 2.4 & 
 &  &  \\
Initial burn-off time  (h) &
3.0 & 17$-$3.4 & 13 & 
 &  &  \\
Total facility AC power (MW) &
200 & 580 & n/a & 
200 & 270 & 290 \\
Novel technology & 
16 T magnets & 16 T magnets & HTS magnets & \multicolumn{3}{c|}{$\mu$ prod./10-16 T magn.}
\\
\hline\hline
\end{tabular}}
\hfill{}
\caption{Tentative parameters of selected future high-energy 
hadron and muon colliders. Parameters of the $\mu^+\mu^-$ Higgs factory are given for reference only. $^*$ The 14 TeV c.m.e. muon collider design is not completed yet, the numbers are a projection \cite{neuffer2018}.}
\label{tb:hadron}
\end{table*}
\end{center}

Future hadron colliders are characterized by record high stored beam energy, rendering machine protection a paramount concern. A very challenging multi-stage collimation system is needed to avoid local beam loss spikes near cold magnets, which would induce magnet quenches. 
Beam injection and beam extraction are particularly sensitive operations, as injection or dump kickers are among the fastest elements in the machine. The collider design must be robust against the sudden asynchronous firing of a kicker unit. The collimators are likely to be the first element to be hit by the beam in case of any fast failure and must be able to withstand the impact of one or a few bunches. The primary and secondary 
 collimators of the LHC are based on carbon-carbon composite material. For the HL-LHC and future hadron colliders, ever stronger materials are being developed and examined, which also feature higher conductivity and, hence, lower ``impedance''. More advanced options include the use of short 
bent crystals as primary collimators \cite{scandale2016lhccrystal} and the deployment of hollow electron-beam lenses as non-destructible collimators \cite{stancari2011collimation}. Acceptable performance of the collimation system, along with small IP beta function, also requires an excellent optics control.

In view of the substantial ring circumference and the associated low momentum compaction, hadron beam intensity in very large accelerators may be limited by conventional instabilities. In particular the resistive wall instability becomes a concern due the low revolution frequency, and TMCI could appear at injection due to the low synchrotron tune \cite{burov2000beamVLHCstability, shiltsev2015intlJournMPA}. 

For future higher-energy hadron colliders, synchrotron radiation damping becomes significant. In such a situation,  longitudinal emittance needs to be kept constant during the physics store through controlled longitudinal noise excitation, in order to maintain longitudinal Landau damping \cite{zimmermann2001}. At the same time, the transverse emittance shrinks due to strong radiation damping, while proton intensity rapidly decreases as a result of the ``burn-off'' due to the high luminosity. The initial proton burn-off time can be computed as:
\begin{equation}
\tau_{\rm bu} = \frac{N_{p}N_{b}}{\Lumi_0 \sigma_{\rm tot} N_{\rm IP}}\; ,
\end{equation}
where $N_{p}$ denotes the proton bunch population, $\Lumi_{0}$ the initial luminosity, $\sigma_{\rm tot}$ the total proton-proton cross section, $N_{b}$ the number of bunches per beam, and $N_{\rm IP}$ the number of high-luminosity IPs; $N_{\rm IP}=2$ for all three colliders under consideration. The total hadron cross section grows with the c.m.e.~collision energies (see Ch.~46 in \cite{pdg2018}), which implies shorter beam lifetimes for higher energy hadron colliders even at a constant target luminosity and could strain requirements on the machines' injector chains. 

For the FCC-hh, the emittance damping time is shorter than the proton burn-off time. As a result, the total beam-beam tune shift $N_{\rm IP} \xi$ increases during the store.  At some point, the beam-beam limit is reached, and, from this point onward, the transverse emittance must  be controlled by transverse  noise excitation, so as to keep the beam-beam tuneshift at, or below, the empirical limit. This limit determines the  further luminosity evolution during the store and the  optimum run time \protect\cite{mbdsfz}. By contrast, at the HE-LHC, the proton burn off time is slightly shorter than the radiation damping time. This situation is qualitatively different from that of the FCC-hh.  For the HE-LHC, there is almost a natural luminosity leveling, while the beam-beam tune shift naturally decreases during the store.

\begin{figure}[htbp]
\centering
\includegraphics[width=0.9\columnwidth]{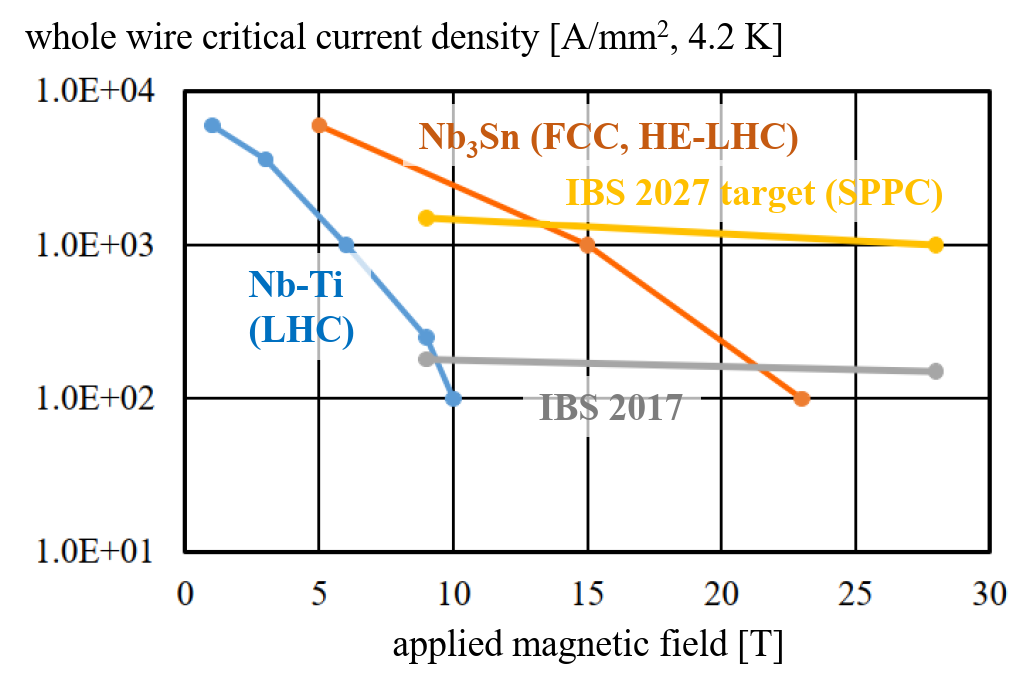}
\caption{Field limits for LHC-type Nb-Ti conductor, Nb$_{3}$Sn conductor as used for 
HL-LHC, FCC-hh and HE-LHC, and iron-based superconductor (present and 
10-year forecast) for SppC 
(after P.J.~Lee \protect\cite{plee}, and private 
communication J.~Gao).}
\label{sc2}
\end{figure}

The primary technology of future hadron colliders is high-field magnets, both dipoles and quadrupoles. Magnets made from  Nb-Ti superconductor were the core technology of the present LHC, Tevatron, RHIC, and HERA. Nb-Ti magnets are limited to maximum fields of about 8 T. The HL-LHC will, for the first time in a collider, use some tens of dipole and quadrupole magnets with a peak field of 11--12 T, based on a new high-field magnet technology using a Nb$_{3}$Sn superconductor. This will prepare the ground for the development of 16 Tesla Nb$_{3}$Sn magnets, and for the later production of about 5000 Nb$_{3}$Sn magnets required by the FCC-hh. The Chinese SppC magnets will utilize cables based on an iron-based high-temperature superconductor, a material discovered at the Tokyo Institute of Technology in the year 2006 \cite{ibssc}. Figure \ref{sc2} sketches the respective current densities and field limits. It is clear that Nb$_{3}$Sn can approximately double the magnetic field reached with Nb-Ti . The R\&D target for SppC looks aggressive. The SppC goal is to increase magnet performance ten times while simultaneously reducing its cost by an order of magnitude. If successful, the iron-based HTS magnet technology could become a game changer for future hadron colliders.

Also important is the minimum field of SC magnets allowing for efficient operation. This minimum field is determined by balancing various considerations such as the injected beam size and magnet aperture, the magnetic field quality at injection, 
machine protection against accidental beam loss due to injection-system failures, beam losses, injection kicker system strength and rise time, severity of beam instabilities, overall cost, etc. Typically, the dynamic range (energy swing) of SC circular accelerators lies in the range 10--20. An increase in c.m.e.~eases the beam dynamics, but implies additional acceleration stages in the injector complex, thereby potentially 
affecting the overall cost, the collider filling time, and the overall machine efficiency. 

Recently, several important milestones were accomplished in the development
of high-field Nb$_{3}$Sn magnets. In the US, an FNAL team has completed a 15 T accelerator dipole short model demonstrator \cite{zlobin-napac19}. In a staged approach, as a first step this magnet was pre-stressed for a maximum field of about 14 T. In successful tests in the fall of 2019 its field indeed reached 14.1 T at 4.5 K \cite{zlobin2020test14T}.  Higher field is facilitated by a higher-quality conductor. Advanced US wires with Artificial Pinning Centers (APCs) produced by two different teams (FNAL, Hyper Tech Research Inc., and Ohio State; and NHMFL, FAMU/FSU) have reached the target critical current density for FCC, of 1500 A/mm$^2$ at 16 T \cite{uswire1, uswire2}, which is 50\% higher than for the HL-LHC wires.   The APCs allow for better performance; they decrease magnetization heat during field ramps, improve the magnet field quality at injection, and reduce the probability of flux jumps \cite{xu2014refinement}. 

\begin{figure}[htbp]
\centering
\includegraphics[width=.95\linewidth]{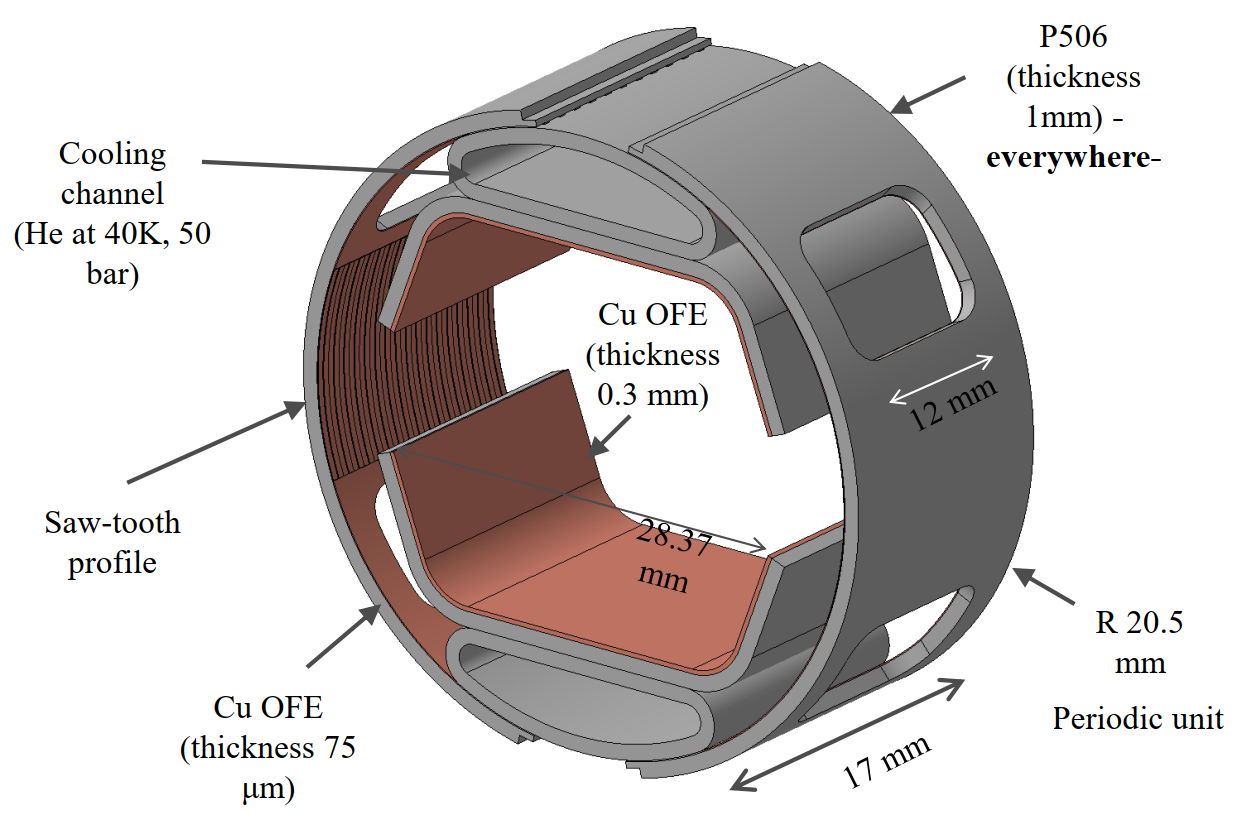}
\caption{\label{fcc-hh-bs}
A periodic unit of the FCC-hh vacuum beamscreen, which will be
mounted inside the magnet cold bore (1.9 K) \protect\cite{perez}. 
This beamscreen will be operated at an elevated
temperature of about 50 K
for efficient removal of the heat from 
synchrotron radiation.}
\end{figure}

Another important technology is the cryogenic beam vacuum system, which has to cope with unusually high levels of synchrotron radiation 
(about 5 MW in total for FCC-hh) in a cold environment. 
The design of the 
beamscreen intercepting the radiation inside the cold bore of the magnets and the 
choice of its operating temperature (50 K -- significantly higher than the 5--20 K chosen for the LHC beamscreen) are key ingredients of the FCC-hh design. The first hardware prototypes for FCC-hh were tested in 2017 at the KIT ANKA/KARA facility at Karlsruhe, with synchrotron radiation from an electron beam, whose spectrum and flux resembled those of the FCC-hh.  
These beam measurements at ANKA/KARA validated the basic design assumptions \cite{gonzalez2019fcchhANKA}.
The latest version of the FCC-hh beamscreen design 
is shown in Fig.~\ref{fcc-hh-bs}.

Further key technologies of energy frontier hadron colliders include the collimators, the kicker and septa required for the extremely high beam energy, and the SC RF systems, e.g., for acceleration and for compensation of synchrotron-radiation 
energy losses, as well as for ever more demanding crab cavities.

\begin{figure*}[htbp]
\centering
\includegraphics[width=0.8\textwidth]{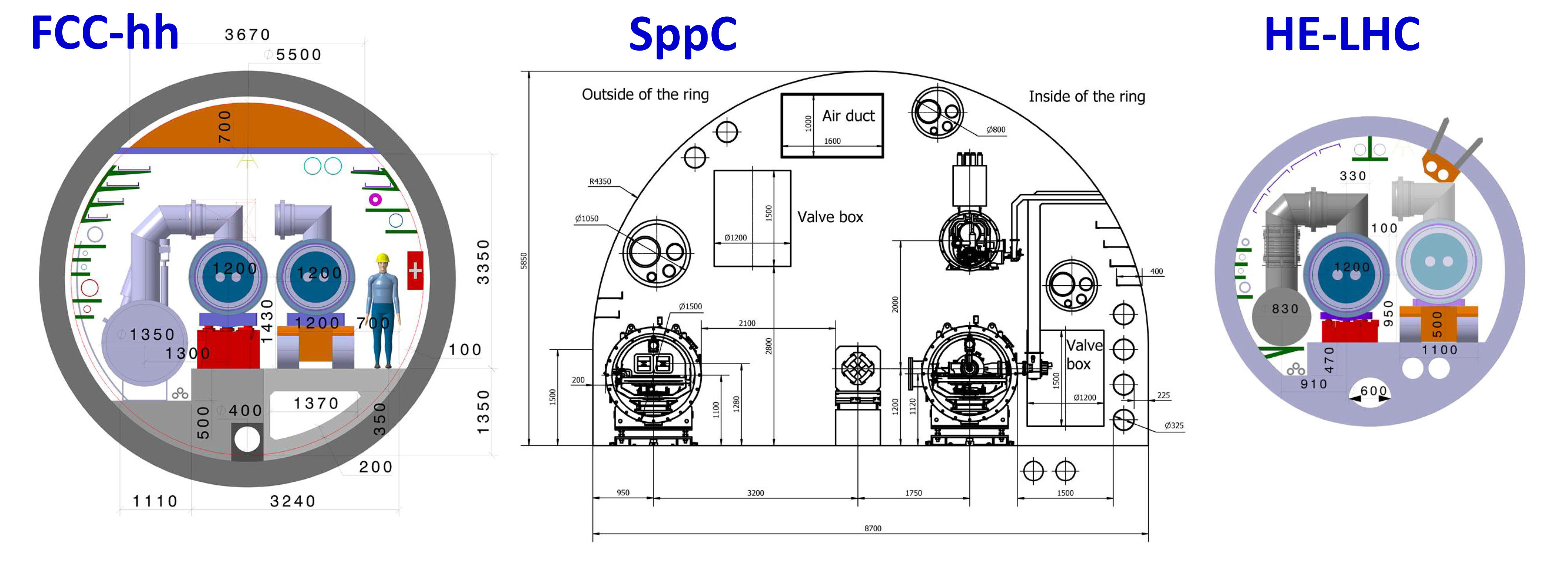}
\caption{Tunnel cross sections for FCC-hh, SppC, and HE-LHC, approximately to  scale  (from Ref.~\cite{Benedikt-NIMA}). }
\label{tunnel}
\end{figure*}

The tunnel is a core element of any new collider. The FCC-hh (FCC-ee) and SppC (CEPC) tunnels are to be constructed differently, by tunnel boring machines and drill/blast 
techniques, respectively. The tunnel shapes and sizes are also rather different, as is illustrated in Fig.~\ref{tunnel}. 
The HE-LHC must fit into the existing LHC tunnel, with a diameter of 3.8 m. The HE-LHC dipole magnets must, therefore, be made as compact as possible, with a maximum outer diameter of 1.2 m.  In addition, half-sector cooling is proposed to reduce the diameter of the cryogenics lines and relax tunnel integration, calling for additional 1.8 K refrigeration units. 
The new round tunnel for the FCC-hh will have a significantly larger diameter of 5.5 m, to host the (possibly larger) 
16 T magnets and enlarged cryogenics lines, plus allow for additional safety features, such as smoke extraction, ventilation, and escape passages. This large a tunnel still 
does not offer enough space to accommodate both a lepton and a hadron machine at the same time. If the FCC-ee is built as a first step, it will need to be disassembled prior to the installation of the FCC-hh hadron collider. The SppC tunnel is even larger, with a transverse width of 8.7 m. It is meant to provide enough space for both lepton and hadron machines, also including a lepton booster ring for top-up injection, which, in principle, could all be operated in parallel.

The HE-LHC cryogenic system reuses the existing LHC helium refrigerators, which will be upgraded by doubling the number of 1.8 K refrigeration units (2 units per sector instead of 1 unit) and by adding specific refrigeration units for temperatures above 40 K to handle, in particular, the beam-induced heat load on the beamscreens, which is dominated by the synchrotron radiation  power  of  about  190 kW.
It should be noted that by 2040, some of the existing LHC cryoplants will be 50 years old, and that the associated 
ageing issues need to be carefully studied. 
In view of their much larger circumference and very high synchrotron radiation power, the FCC-hh and SppC will both still need substantially larger cryogenic facilities. Specifically, the FCC-hh foresees 10 cryoplants, each with 50--100 kW at 4.5 K including 12 kW at 1.8 K, and requiring a helium inventory of 800 tons, about 6 times the helium inventory of the
present LHC.  The FCC-hh  will  use  a  cryogenics  system  based  on  a  neon-helium mixture (nelium), which leads to electrical energy savings of about 10\% with respect to the LHC-type helium cryogenics infrastructure The electrical power consumption of the FCC-hh cryoplants is about 200 MW \cite{laurent2017}.
 
To summarize, the key challenges for the energy frontier $pp$ colliders such as the HE-LHC, FCC-hh, and SppC are associated with the need for long tunnels (27 km, 100 km and 100 km, respectively), high field SC magnets (16 T, 16 T and 12 T, respectively), and total AC site wall plug power ranging from about 200 MW (HE-LHC) to $\sim$500 MW. The cost estimates extend from 7.2 BCHF for the HE-LHC to 17.1 BCHF for the FCC-hh (assuming that the 7 BCHF tunnel is available) --- see Table \ref{tb:hadron}. In all these options, the detectors will need to operate at luminosities of $O(10^{35}$ cm$^{-2}$s$^{-1}$) and the corresponding pile-up of events per crossing will be $O$(500). A 12--18 year long R\&D program is foreseen to address the most critical technical issues, such as: (i) development of accelerator quality 16 T dipole magnets based on Nb$_3$Sn (or 12 T iron-based HTS magnets for the SppC); (ii) effective intercept of synchrotron radiation (5 MW in FCC-hh and 1 MW in SppC); (iii) beam halo collimation with circulating beam power 7 times that of the LHC; (iv) choice of optimal injector (e.g., a new 1.3 TeV SC SPS, or 3.3 TeV ring either in the LHC tunnel or the FCC tunnel, for the FCC-hh), and (v) overall machine design issues (IRs, pile-up, vacuum, etc),  power and cost reduction, etc. It is noteworthy that such machines can additionally be used for ion-ion/ion-proton collisions;  high energy proton beams can also be collided with high intensity $O$(60) GeV electrons from an ERL.

\subsubsection{Muon colliders}
\label{muoncolliders}
\begin{figure}[htbp]
\centering
\includegraphics[width=80mm]{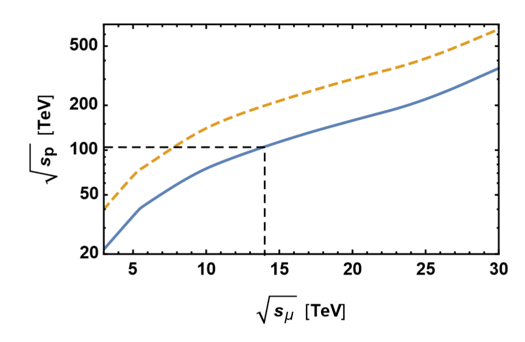}
\caption{Energy reach of muon-muon collisions: the energy at which the proton collider cross-section equals that of a muon collider (from Ref.~\cite{delahaye2019arxiv}). The dashed line assumes comparable Feynman amplitudes for muon and proton production processes.} 
\label{MuonProton}
\end{figure}

The lifetime of the muon, 2.2 $\mu$s in the muon rest frame, is sufficient to allow fast acceleration to high energy before  the muon decays into an electron, a muon-type neutrino, and an electron-type antineutrino ($\mu^- \rightarrow e^- \nu_μ \nu_e$) and storage for some 300$\times B$ turns in a ring with average field $B$ (Tesla). The muon to electron mass ratio of 207 implies that all synchrotron radiation effects are smaller by a factor of about $(m_\mu/m_e)^4 \approx 2\times 10^9$.  
Even a multi-TeV $\mu^+\mu^-$ collider can be highly power efficient, while being circular, 
and, therefore, may have 
quite a compact geometry, that will fit on existing accelerator sites or tunnels. The c.m.e.~spread for 3 to 14 TeV $\mu^+\mu^-$ colliders is $dE/E < 10^{-3}$ (see parameters of such facilities in Table \ref{tb:hadron}), which is an order of magnitude smaller than for an  $e^+e^-$ collider of the same energy. Much like in $e^+e^-$ colliders, the muon collider center of mass energy $\sqrt{s}$ is entirely available to produce short-distance reactions rather than being spread among proton constituents. A 14 TeV muon collider with sufficient luminosity might be very effective as a direct exploration machine, with a physics potential similar to that of a 100 TeV proton-proton collider --- see Fig.~\ref{MuonProton} from Ref.~\cite{delahaye2019arxiv}. 

In general, muon colliders are predicted to be significantly less expensive than other energy frontier hadron or $e^+e^-$ machines \cite{shiltsev2014}. They need lower AC wall plug power \cite{jpdmuon, boscolo2018futuremuon} and, due to compact size, a smaller number of elements requiring high reliability and individual control for effective operation \cite{shiltsev2010muonfeasible}. In addition, a $\mu^+\mu^-$ Higgs factory would have the advantages of a large Higgs production cross-section via $s$-channel production, and of a beam energy equal to about one half of the standard $e^+e^-$ Higgs production mode at 240--250 GeV c.m.e.~(i.e., 2$\times$63 GeV  for $\mu^+\mu^- \rightarrow H_0$). It would, therefore, offer a small footprint, a low energy spread in non-radiating muon beams, $O$(3 MeV), and low total site power of $\sim$200~MW \cite{MC1999, alexahin2018muonlattice}. Finally, a neutrino factory could potentially be realized during the course of its  construction \cite{geer1998neutrino, geer2009, boscolo2018futuremuon}.

\begin{figure}[htbp]
\centering
\includegraphics[width=80mm]{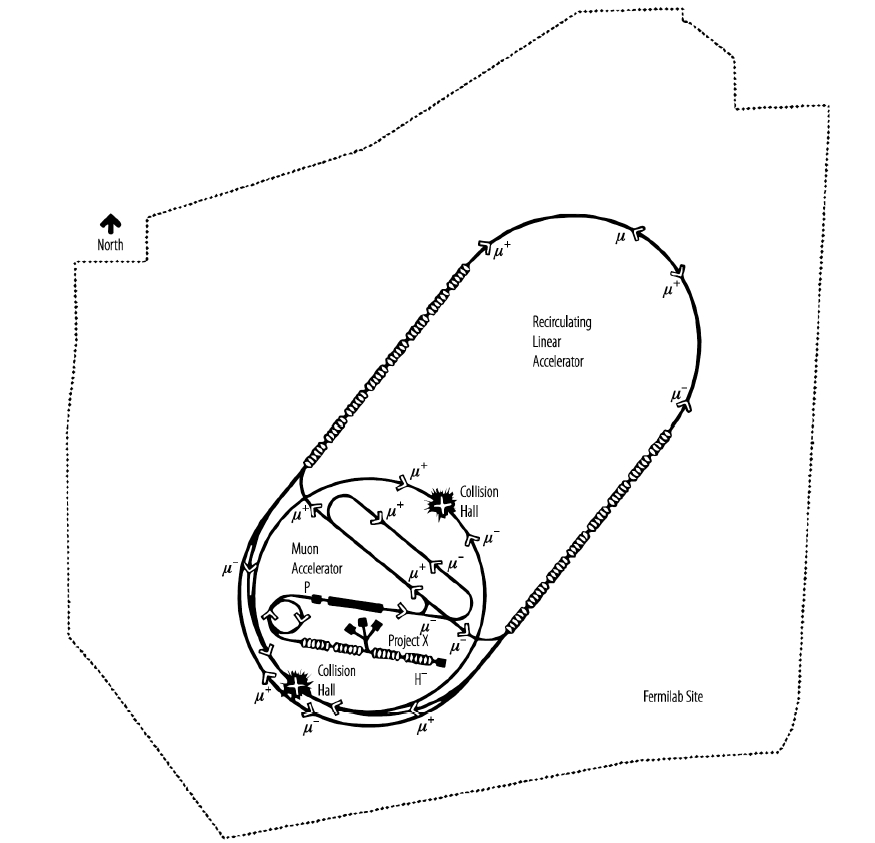}
\caption{Schematic of a 4 TeV Muon Collider on the 6$\times$7 km FNAL site (from Ch.12.2 Ref. \cite{myers2013accelerators}).}  
\label{FNAL_MuColl}
\end{figure}

Muon colliders were proposed by F.~Tikhonin and G.~Budker at the end of the 1960s \cite{tikhonin1968, budker1969, budker1970} and conceptually developed later by a number of authors and collaborations (see a  comprehensive list of references in Refs.~\cite{geer2009, boscolo2018futuremuon}). Figure~\ref{FNAL_MuColl} presents a possible layout of a multi-TeV c.m.e.~high luminosity $O$($10^{34}$~cm$^{-2}$s$^{-1}$) muon collider, consisting of: (i) a high power proton driver (SRF 8 GeV 2--4 MW $H^-$ linac); (ii) pre-target accumulation and compressor rings, in which high-intensity 1--3 ns long proton bunches are formed; (iii) a liquid mercury target for converting the proton beam into a tertiary muon beam with energy of about 200 MeV; (iv) a multi-stage ionization cooling section that reduces the transverse and longitudinal emittances and, thereby, creates a low emittance beam; (v) a multistage acceleration (initial and main) system --- the latter employing recirculating linear accelerators (RLA) to accelerate muons in a modest number of turns up to 2 TeV using superconducting RF technology; and, finally, (vi) a roughly 2 km diameter collider ring located some 100 m underground, where counter-propagating muon beams are stored and collide over the roughly 1000--2000 turns corresponding to the muon lifetime. 

Since muons decay quickly, large numbers of them must be produced to operate a muon collider at high luminosity. Collection of muons from the decay of pions produced in proton-nucleus interactions results in a large initial 6D phase-space volume for the muons, which must be reduced (cooled) by a factor of 10$^6$ for a practical collider.  Without such  cooling, the luminosity reach will not exceed $O$($10^{31}$~cm$^{-2}$s$^{-1}$). The technique of ionization cooling proposed in \cite{ado1971use, skrinsky1981cooling, neuffer1983principles} is very fast and uniquely applicable to muons because of their minimal interaction with matter. It involves passing the muon beam through some material absorber in which the particles lose momentum essentially along the direction of motion via ionization energy loss, commonly referred to as $dE/dx$. Both transverse and longitudinal momentum are reduced via this mechanism, but only longitudinal momentum is then restored by reacceleration, leaving a net loss of transverse momentum (transverse cooling). The process is repeated many times to achieve a large cooling factor. 

The rate of change of the normalized transverse emittance $\varepsilon_{x,y}=\varepsilon_{\perp}$ as the beam passes through an absorber is given approximately by 
\begin{equation}
\frac{d\varepsilon_\perp}{dz} \simeq - \frac{\varepsilon_\perp}{\beta^2 E_\mu} \left|\frac{dE_\mu}{dz} \right| + \frac{\beta_\perp (13.6 {\rm MeV}/c)^2} {2 \beta^3 E_\mu m_\mu X_0}
\label{ionizcooling}
\end{equation} 
where $\beta c$ denotes the muon velocity, $E_\mu$ the muon  energy, $\big| dE_\mu/ dz \big|$ the mean energy loss per unit path length, $X_0$ the radiation length of the absorber, and $\beta_\perp$ the transverse betatron function at the absorber. The first term of this equation describes the cooling effect by ionization energy loss and the second describes the heating caused by multiple Coulomb scattering. Initially the cooling effect dominates over the heating one, leading to a small equilibrium emittance. The energy spread acquired in such a process due to fluctuation of ionization losses (Landau straggling) can be reduced by introducing a transverse variation in the absorber density or thickness (e.g., a wedge) at a location where there is dispersion $D_{x,y}$ (a correlation between transverse position and energy). This method results in a corresponding increase of transverse phase space, represents an exchange of longitudinal and transverse emittances, and allows cooling in all dimensions, thanks  to the fast transverse cooling \cite{palmer2014muonrast}. 

\begin{figure}[htbp]
\centering
\includegraphics[width=80mm]{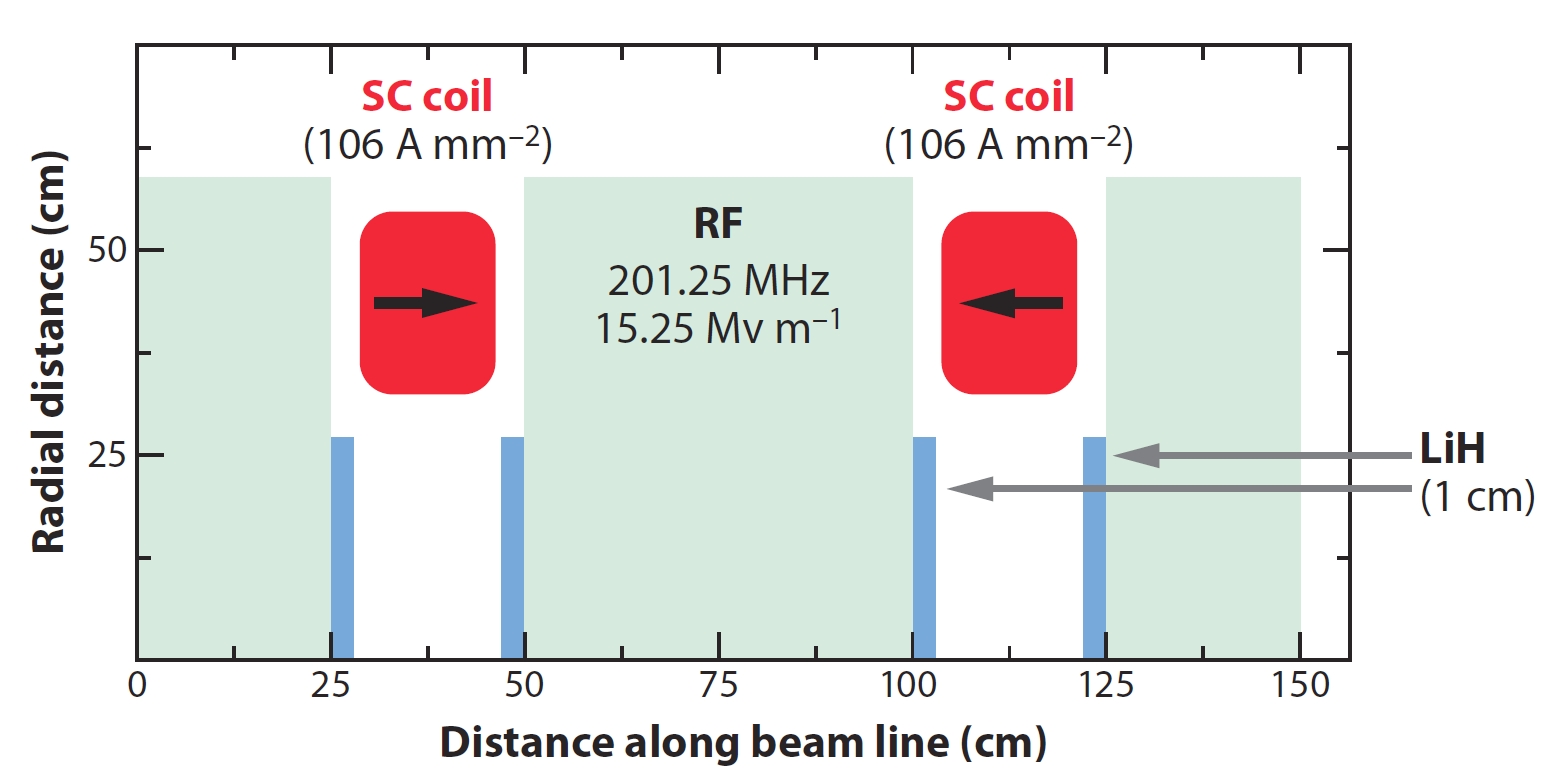}
\caption{Ionization cooling-channel section. 200 MeV muons lose energy in lithium hydrate (LiH) absorbers (blue) that is replaced when the muons are reaccelerated in the longitudinal direction in RF cavities (green). The few-Tesla SC solenoids (red)  confine the beam within the channel and radially focus the beam at the absorbers. Some representative component parameters are also shown (from Ref.\cite{geer2009}).}  
\label{MuCooling}
\end{figure}
 
Theoretical studies  \cite{palmer1996, sessler1998} and numerical simulations \cite{sayed2015mucoolsimulations} have shown that, assuming realistic parameters for cooling hardware, ionization cooling can be expected to reduce the phase space volume occupied by the initial muon beam by a factor of 10$^{5}$ to 10$^{6}$. A complete cooling channel would consist of 20 to 30 cooling stages, each yielding about a factor of 2 in 6D phase space reduction; see Fig.~\ref{MuCooling}. 

The ionization cooling method, though relatively straightforward in principle, faces some practical implementation challenges. These include RF breakdown suppression and attainment of high accelerating gradients in relatively low frequency NC RF cavities immersed in strong magnetic fields. The International Muon Ionization Cooling Experiment (MICE) \cite{sandstrom2008mice, mice2019first} at RAL (UK) has recently demonstrated effective $O$(10\%) reduction of transverse emittance of initially dispersed 140 MeV/c muons passing through an ionization cooling channel cell consisting of a sequence of LiH or liquid hydrogen absorbers within a lattice of up to 3.5 T solenoids that provide the required particle focusing \cite{mice2018a, Mice2020Nature}. 

\begin{figure}[htbp]
\centering
\includegraphics[width=80mm]{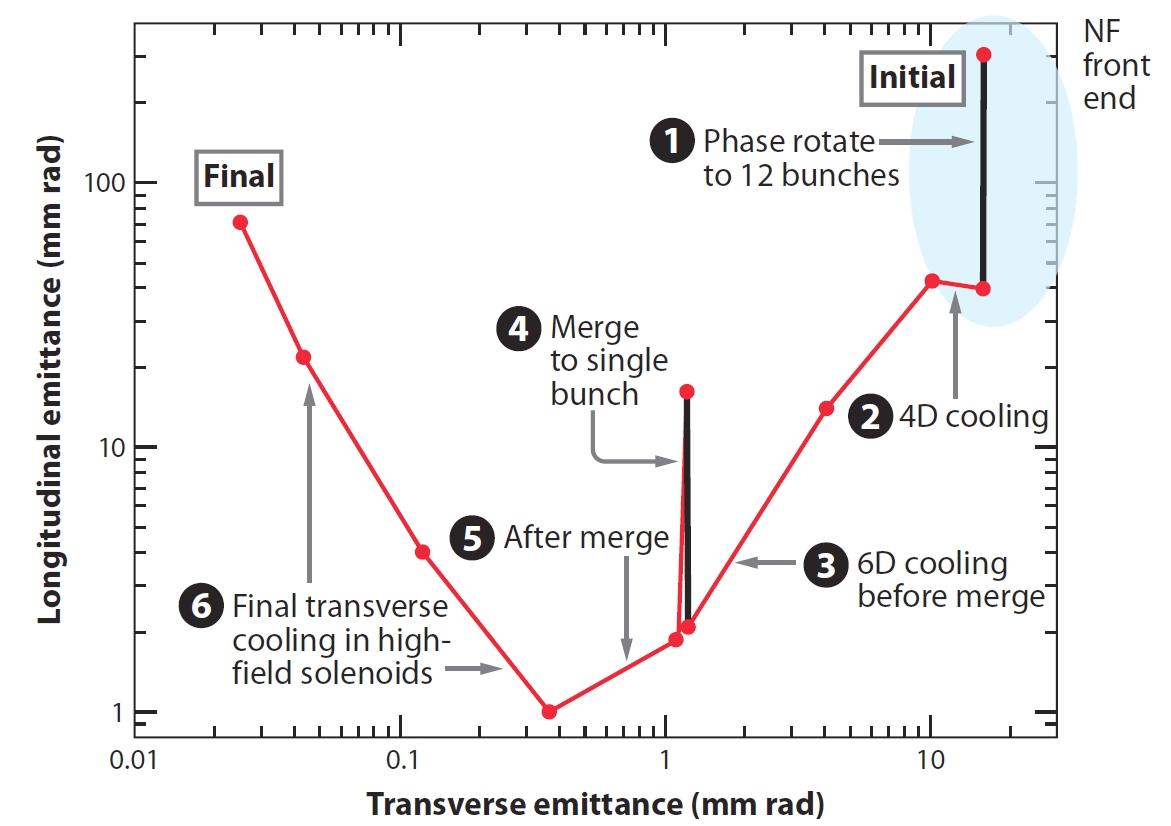}
\caption{Simulated six-dimensional (6D) cooling path \cite{palmer2014muonrast} corresponding to one particular candidate muon collider cooling channel. The first part of the scheme (blue ellipse) is identical to the present baseline neutrino factory front end (from Ref.\cite{geer2009}).}  
\label{MuCool6D}
\end{figure}

Constructing and operating a muon collider with usable luminosity requires surmounting significant technical challenges associated with the production, capture, cooling, acceleration, and storage of muons in large quantities and with appropriate phase-space densities. References~\cite{palmer2014muonrast, boscolo2018futuremuon} provide comprehensive overviews of the  significant progress achieved in developing the concepts and technologies needed for high luminosity energy frontier muon colliders. Beside the pioneering demonstration of ionization cooling by MICE, muon collider R\&D has led to a number of remarkable advances in the past decade. The Mercury Intense Target experiment (MERIT) \cite{mcdonald2009merit} has successfully injected a high intensity proton beam from the CERN proton synchrotron into a liquid mercury jet inside a 15 T solenoid, proving the feasibility of beam power in excess of 4 MW on such targets. Accelerating gradients of 50 MV/m were obtained in vacuum and pressurized gas-filled NC RF immersed in a 3 T magnetic field at Fermilab \cite{bowring2018ncrf, chung2013pressurizedrf}. Also at Fermilab, rapid cycling HTS magnets achieved a record field ramping rate of 12 T/s \cite{piekarz2019recordhts}. The first RF acceleration of muons was demonstrated at the JPARC MUSE RFQ \cite{bae2018muonrfq}.  Some 16--20 T small bore HTS solenoids were built at BNL --- an important step toward the 30--40 T magnets needed for the final muon cooling stage \cite{gupta2013hts}. The US Muon Accelerator Program collaboration and its international partners have successfully carried out complete 6D muon ionization cooling simulations --- see Fig. \ref{MuCool6D} --- as well as overall facility feasibility studies, demonstrating that muon colliders can be built with present-day SC magnet and RF technologies, and developed initial designs for 1.5 TeV, 3 TeV, 6 TeV, and 14 TeV colliders (see Table \ref{tb:hadron}). 

A novel concept is being proposed of using 45 GeV positrons to generate muon pairs through $e^+e^-$ annihilation just above threshold \cite{antonelli2016lemma}, allowing low emittance beams to be obtained directly, without any cooling. This scheme may allow operation of a very high energy muon collider with manageable neutrino radiation on and off the site. Major directions of the R\&D to evaluate the possible luminosity reach of this concept and to address key issues of corresponding technologies have been outlined in \cite{boscolo2018futuremuon}. 

Another novel approach, called the Gamma Factory (GF)  \cite{Krasny:2015ffb,Krasny:2018alc}, could potentially 
help make a muon collider become reality. 
The GF would generate frequent  
bursts of gamma rays by repeatedly colliding a partially stripped heavy-ion beam circulating in the LHC, or in a future higher-energy hadron storage ring like the FCC-hh, 
with a conventional laser pulse, profiting from two Lorentz 
boosts. 
Impinging on a suitable target, 
the GF's intense gamma bursts 
could produce positrons or muons at an unprecedented rate. 
Thereby, the GF could deliver positrons at the 
rate required for the aforementioned 
positron-based muon production, 
or, alternatively, it could directly provide a     
low-emittance muon beam \cite{Zimmermann:2018wfu}. 
In 2018, first beam tests confirmed the predicted long beam lifetime, of more than a day, for a partially-stripped 
heavy-ion beam stored in the LHC at top energy  
\cite{Schaumann:2019evk}. 
The next series of proof-of-principle GF experiments, consisting of laser-beam collisions in the 
SPS, is planned for 2021.

Under active study are concepts for muon collider detectors which must operate in the presence of various backgrounds originating from muon decay and effective measures to control neutrino radiation \cite{bartosik2019mcneutrinoradiation}. Any straight section within the collider ring produces a beam of muon-decay neutrinos in the direction of the straight section. These neutrinos exit the Earth at some point, perhaps a few tens of kilometers away if the ring is deep. At the exit point, neutrino interactions within the rock create radiation at the surface. The radiation level increases rapidly with the stored muon energy. Besides the straightforward approach of placing the collider ring tunnel at sufficient depth, there are several mitigation ideas on how to keep neutrino radiation below the commonly accepted limit of 1 mSv/yr. E.g., the radiation density can be reduced by about an order of magnitude by adding a vertical collider orbit variation of a few mm.

\section{Advanced collider concepts}
\label{advancedcolliders}

\subsection{Acceleration in plasma and plasma-based collider proposals}
\label{accelerationinplasma}

Since about the mid-1950s, it has been understood that collective plasma-based accelerators offer the great promise of extremely large accelerating gradients \cite{veksler1957principle}. Ionized plasmas can sustain electron plasma density waves with electric fields in excess of $E_0=cm_e \omega_p /e$ or 
\begin{equation}
E_0\approx 96\, {\rm [V/m]}\, \sqrt{n_0{\rm [cm^{-3}]}},
\label{plasmagradient}
\end{equation} 
(so-called {\it cold nonrelativistic wave-breaking field} \cite{dawson1959nonlinear}) where $n_0$ denotes the ambient electron number density, $\omega_p=\sqrt{e^2 n_0/(m_e \varepsilon_{0})}$  the electron plasma frequency, $m_e$ and $e$ electron rest mass and charge, respectively, $c$ the speed of light in vacuum, and $\varepsilon_{0}$ the electrical  permittivity of free space. 
For example, a plasma density of about 10$^{18}$~cm$^{-3}$ yields $E_0\sim$100 GV/m, approximately three orders of magnitude greater than $\sim$100 MV/m obtained in conventional breakdown limited RF structures.  

\begin{figure}[htbp]
\centering
\includegraphics[width=80mm]{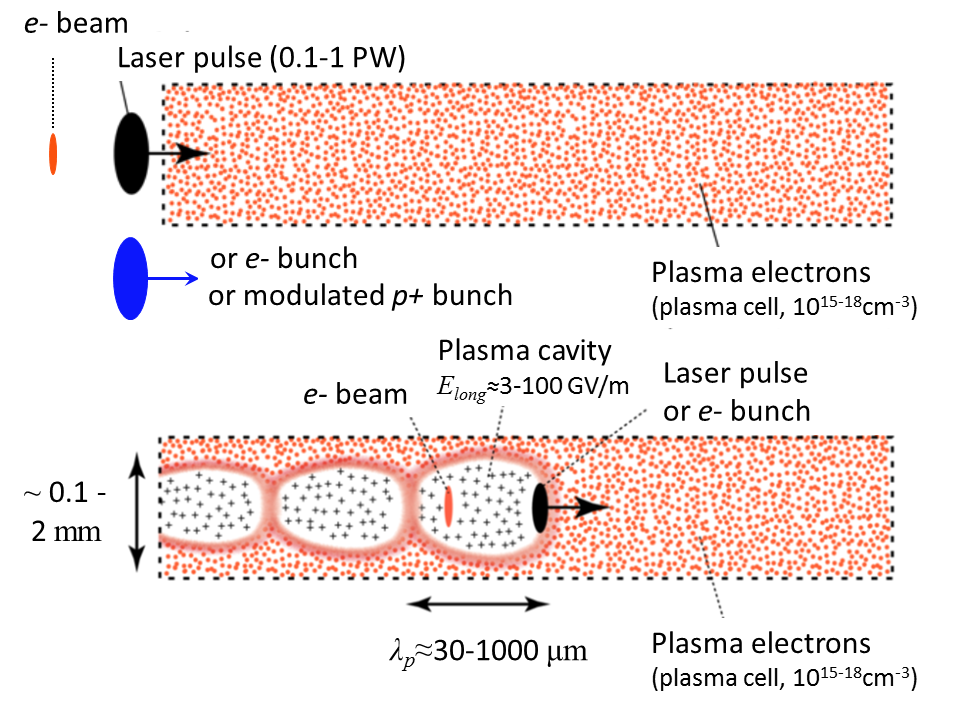}
\caption{Concept of the plasma wakefield acceleration driven either by a short laser pulse (LWFA), or by a short electron bunch or by long(er) modulated proton bunch (PWFA, adapted from \cite{assmann2014plasmaquality}). } 
\label{PWAconcept}
\end{figure}

Such gradients can be effectively excited by either powerful external pulses of laser light or by electron bunches if they are shorter than the plasma wavelength $\lambda_p=c/\omega_p \approx1$~mm$\times \sqrt{10^{15}\, {\rm cm^{-3}}/n_0 }$, or by longer beams of charged particles if their charge density is modulated with the period of $\lambda_p$.  
Figure~\ref{PWAconcept} illustrates the  concept  of  plasma  acceleration. For example, the plasma response to a short laser pulse is as follows \cite{tajima1979laser}: (i) the  laser  pulse  enters  the  plasma  and  transversely accelerates plasma electrons (ponderomotive force as transverse  driving  force), (ii) the  plasma  ions  move  a  negligible amount and a positively charged ion channel is formed along the laser path, (iii) after the passage of the laser pulse, the plasma electrons rush back in, attracted by the transverse  restoring force of the positively charged ion channel, pass the center of the ion channel, rush back out and are attracted back by the ion channel, and (vi) a space charge driven oscillation is formed, leaving  alternating   regions   of negative  and positive   net   charge  with strong induced longitudinal  fields behind the  laser  pulse ({\it plasma wakefields}). If a  short test bunch of charged particles, e.g., electrons, is   placed  behind  the  laser  pulse  at  a  proper    distance,    then  it will be accelerated  with  high  gradient. The  process could be  limited  by  depletion  of  laser pulse power,   dephasing  between  the  relativistic  test  bunch  and the wakefield,  and  the  Rayleigh length of the laser beam (unless  counteracted by self-guiding or external guiding of the laser in a plasma channel). Similar concepts have been 
proposed  for  plasma  wakefields  driven  by  short  
electron bunches  \cite{chen1985beamplasma} and by self-modulated high energy proton bunches with 
an rms bunch length of order 10 cm 
\cite{caldwell2009proton}. 

\begin{figure}[htbp]
\centering
\includegraphics[width=0.99\linewidth]{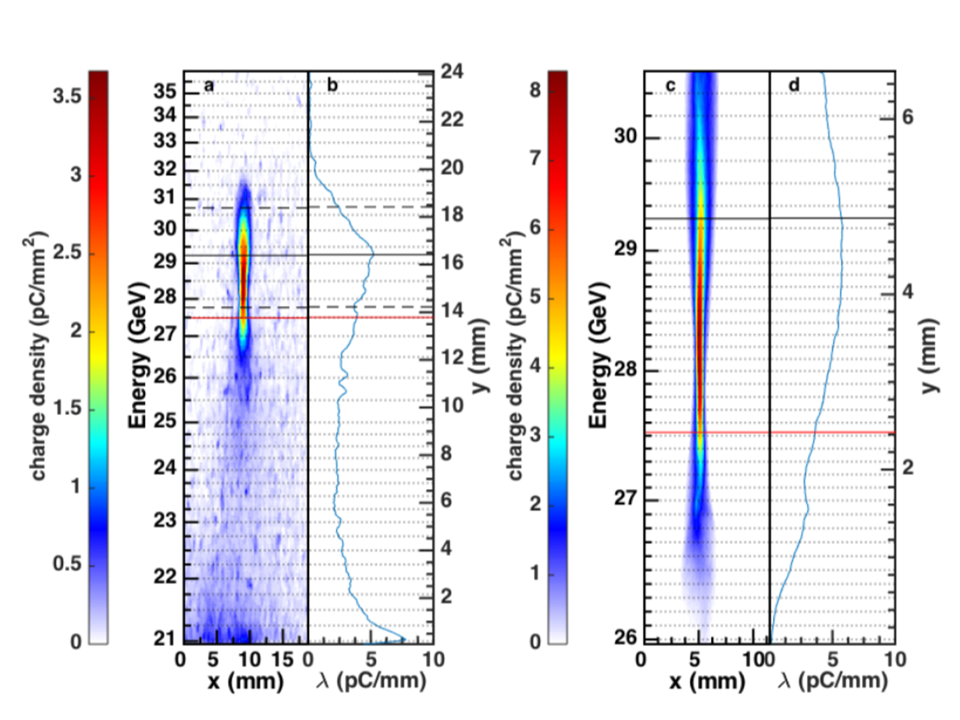}
\caption{A 0.1 nC  electron  bunch  gained  a  maximum  energy  of  9  GeV in a 1.3 m-long electron plasma wakefield accelerator driven by a 20.35 GeV $e^-$ beam at the FACET facility at SLAC: (a)-(d) show the energetically dispersed transverse charge density profile spectra and the horizontally integrated spectral charge density profiles as observed on the wide-field of view  Cherenkov screen and on the order of magnitude more sensitive Lanex screen, respectively (from Ref.~\cite{litos20169gev}). }  
\label{fig:facet9gev}
\end{figure}

The three plasma driver technologies have been explored theoretically and experimentally, and corresponding reviews and references can be found in \cite{esarey2009rmp, hogan2016bpwa, adlimuggli2016proton}. In the past decade, we have seen impressive progress of the plasma wakefield acceleration of high quality beams. Laser driven electron energy gain of about 8 GeV over 20 cm of plasma with density 3$\times$10$^{17}$cm$^{-3}$ has been demonstrated at the BELLA facility at the Lawrence Berkeley National Laboratory (LBNL) \cite{gonsalves2019petawatt8Gev}. Short electron bunches were used to boost the energy of externally injected electron bunches by 9 GeV over 1.3~m of $\sim$10$^{17}$cm$^{-3}$ plasma at the FACET facility in SLAC \cite{litos20169gev} --- see Fig. \ref{fig:facet9gev}. The AWAKE experiment at CERN used 
self-modulating long 450 GeV  
proton bunches to accelerate electrons to 2 GeV over 10 m of 10$^{15}$~cm$^{-3}$ plasma \cite{adli2018awakenature}. 

In principle, the plasma wakefield acceleration scheme has the 
potential  to  be  staged,  e.g.,  several  plasma cells  of  
the  same  kind  can  be  placed  in  series,  resulting  in  
potentially higher beam energy. That makes possible attainment of very high energies and designs of TeV or multi-TeV  $e^+e^-$ colliders, such as proposed in Refs. \cite{leemans2009laser, schroeder2010lpwacollider, adli2013bpwacollider}. 
The primary advantage that a plasma wakefield accelerator could present is a considerably greater compactness and, hence, a much lower `real-estate' investment for the collider.  
There are a number of critical issues that need to be resolved along that path \cite{assmann2014plasmaquality, lebedev2016luminositylimits, schulte2016applicationadv} including acceleration of positrons (which are defocused by the positively charged ion cavity when accelerated in a plasma --- see Fig.~\ref{PWAconcept}), instabilities in accelerated beams and beam emittance control in scattering media,  final focusing of $e^+$ and $e^-$ bunches with significant energy spread acquired during acceleration, and efficiency of staging (beam transfer and matching from one $O$(1m) long plasma cell to another). Indeed, strong  transverse  focusing  gradients  $O$(10MT/m) are  generated  inside  the  ion  channel of plasma  accelerators.  Such focusing is equivalent to  very small beta functions 
$\beta_{x,y}$ in the range of a couple of cm to a few  mm  for high energy beams accelerated in the $n_0=10^{14-17}$~cm$^{-3}$ plasma. Matching electrons and positrons in  and out of these multiple plasma cells is  difficult and compared to ``low-$\beta$'' insertions of traditional colliders, transverse injection error tolerances $O$(1$\mu$m) become highly demanding. 

Comparative analysis of initial strawman designs of high luminosity 3 TeV to 10 TeV to 30 TeV laser-driven and beam-driven $e^+e^-$ colliders \cite{plasmacolliders2019granada, adli2013bpwacollider} with that of CLIC does not show significant AC-to-beam-power efficiency advantage of the advanced schemes; AC wall plug power needs are $\sim$0.5 GW for 10 TeV c.m.e.~and over 1 GW for 30 TeV c.m.e.~machines. The total facility length will still be considerable (6 to 8 km for $\sqrt{s}$=3 TeV, 10--18 km for 10 TeV) and the beamstrahlung effect will ultimately be severe --- the expected rms energy spread at the IP is about 30\% for 10 TeV machines and 80\% for 30 TeV colliders. 

Reference~\cite{caldwell2016vheep} proposes an LHC upgrade for a very high energy 9 TeV c.m.e.~electron--proton collider using a 3 TeV electron beam accelerated by the proton driven plasma wakefields.  In this scenario, one of the two 7 TeV LHC proton beams is used as the proton driver to create plasma wakefields accelerating electrons to 3 TeV, which then collide with the other 7 TeV LHC proton beam.  Assumed are 3000 LHC bunches per fill with a 30 min machine cycle time, 10$^{11}$ electrons and $4\times 10^{11}$ protons per bunch  (about twice the value foreseen for the LHC luminosity upgrade), as well as a transverse RMS beam size of 4 $\mu$m, to reach a relatively low luminosity of $4 \times 10^{28}$~cm$^{-2}$s$^{-1}$.

Plasma wakefield acceleration concepts have not yet achieved the level of a reliable conceptual design for an affordable high luminosity multi-TeV $e^+e^-$ collider. The ILC and CLIC studies have emphasized that the performance reach of a linear collider is essentially proportional to the beam power and at present the plasma wake-field acceleration technology is far from the wall-plug efficiency of NC and SRF linacs. Correspondingly, 
the focus of the current R\&D activities carried out by several groups and collaborations \cite{plasmacolliders2019granada}, including EuPRAXIA (European Plasma Research Accelerator with eXcellence In Applications) \cite{assmann2017eupraxia} and ALEGRO (the Advanced LinEar collider study GROup) \cite{muggli2018alegro}, is less on breaking the accelerating gradient records and more on mundane, but critical issues such as energy transfer efficiency, production of high quality high repetition rate beams with the various driver technologies, positron acceleration, staging, and exploration of the possibilities offered by recent advances in high peak power laser technologies (similar to how {\it chirp pulse amplification} boosted the laser plasma acceleration technique \cite{mourou2019nobel}, awarded with the 2018 Nobel Prize in Physics). A number of beam test facilities addressing these major scientific challenges are either operating, coming on-line, or in the planning phase. In the US, roadmaps of  advanced accelerator R\&D have been developed with  the primary long-term goal of a technical design report (TDR) of a multi-TeV collider in the 2035--40 time period and a
secondary, nearer-term goal of the completion of a TDR for potential early application of these acceleration techniques in the 2025--30 time period \cite{colby2016roadmap}.


\subsection{Other advanced approaches for colliding beam schemes}
\label{other advanced}

In addition to the designs and concepts presented above, many ideas and approaches have been proposed to extend the energy reach of future particle colliders, reduce their cost, and improve their luminosity and energy efficiency. Below we present some which have shown promise and have been considered in at least some detail for applications in future nuclear physics or particle physics colliders. 

{\it Economical magnets for very large hadron colliders.}
The potential benefits of using modest or relatively low field magnets to reach ultra-high proton beam energies in extremely large circular colliders were first discussed by E.~Fermi who in the mid-1950s thought of an Earth-encircling
``Globaltron'' with circumference of $C$=40,000 km and energy reach of 5000 TeV (5 PeV) \cite{cronin2004fermi}. Attempts to figure out a cost-feasible variation of such a concept include the ``Collider in the Sea'' ($C$=1,900 km, underwater in the Gulf of Mexico, $\sqrt{s}$=500 TeV with economical 3.2 T SC magnets)  \cite{mcintyre2017sea500Tev}, the 300 km circumference 300 TeV 
``Eloisatron'' with 10 T magnets \cite{zichichi1990eloisatron, barletta1996eloisatron}, and the 233 km long Very Large Hadron Collider (VLHC) \cite{accel:vlhc}. In the VLHC design, the Stage 1 machine was to accelerate 20 TeV proton beams in a 2 T double-aperture superferric SC magnet synchrotron ring and collide them at $\sqrt{s}$=40 TeV. 
Afterwards, the Stage 1 complex would act as an injector accelerator to a 200 TeV c.m.e.~collider in the same tunnel based on 12 Tesla Nb$_3$Sn magnets. A 1.5 m long single turn 100 kA SC transmission line twin-aperture combined function dipole magnet prototype for  the VLHC Stage 1 has been built at Fermilab and demonstrated  good field quality at the design 2 Tesla field \cite{piekarz2006vlhctest}. Opportunities to reduce the cost of the 100 km FCC-hh collider by using 6 T  to  8 T economical NbTi SC magnets --- resulting in $\sqrt{s}$=37.5--50 TeV --- are also being discussed. As mentioned above, in China, prospects of having inexpensive 12--24 T iron-based HTS superconductors have initiated machine design studies of the Super proton-proton Collider (SppC) in the 100 km CEPC tunnel \cite{tang2017update, Benedikt-NIMA}. 

\begin{figure}[htbp]
\centering
\includegraphics[width=0.9\columnwidth]{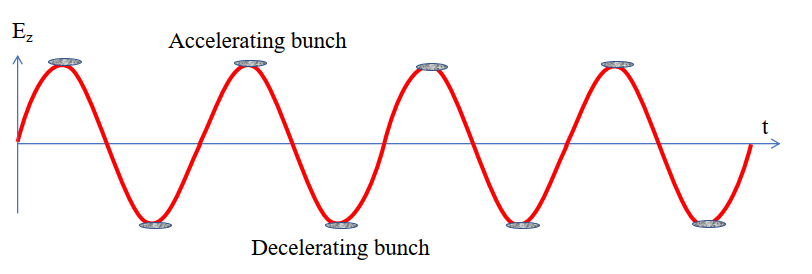}
\caption{Basic ERL principle: accelerating bunches take energy 
from SRF linac, while decelerating bunches return energy back \cite{litvinenko2019fccerl}.  } 
\label{fig:erl}
\end{figure}

{\it Energy recovery linacs (ERLs).} Another promising and actively developing technology is that of recirculating linear accelerators (RLAs) and energy recovery linacs (ERLs). RLAs are accelerators in which the accelerating structure of an RF linac is used multiple (a few to dozens of) times to accelerate the same beam. Return beamlines that are needed to take the beam out of the linac and to reinject it back at proper phase tend to be much cheaper to build than additional RF linac length, thus offering a cost--effective option to achieve the highest possible energy from a given RF installation. Such a hybrid arrangement of linac and ring also allows superior electron beam quality compared to a storage ring. Indeed, the beam dwells a short time in the accelerator and avoids many storage ring processes leading to emittance growth (due to, e.g., synchrotron radiation) or depolarization. With proper care for beam dynamics, electron beam brightness is then determined by the electron source and can thus be high \cite{merminga2003erl, benzvi2016srferl}, \cite[Ch.39]{bruning2016challenges}. In  instances where high average current is required, as for high luminosity colliders, the RLA concept can be augmented with a reverse process of energy recovery: the energy invested in accelerating a beam is returned to the device  powering the acceleration by decelerating the beam after it has been put to some use. 
The basic principle of the energy recovery process is illustrated in
Fig.~\ref{fig:erl}. 

The original idea of an SRF ERL is due to Tigner \cite{tigner1965}, but only in the past two decades has the SRF technology matured enough to render the full potential of ERLs accessible \cite{liepe2006srferl}.  Superconducting RF cavities allow efficient operation in either CW (continuous wave) or in a long pulse regime (due to very high quality factor $O$)). 
Thanks to lower frequency and high gradients, which can potentially exceed 50 MV/m \cite{grassellino49MVm}, they offer larger stored energy compared to NC RF structures and lower beam impedances, which could excite detrimental instabilities at high beam current. Envisioned SRF ERL applications include accelerators for the production of synchrotron radiation and free electron lasers \cite{gruner2002erllightsource}, high-energy electron cooling devices, and $e-p/e-$ion 
colliders \cite{benzvi2016srferl}. 

ERL applications for the JLEIC electron cooling system and in  the LHeC and FCC-eh electron-proton colliders are presented above in Sec. \ref{iieic}. An alternative option of an ERL-based eRHIC design was studied in sufficient detail and summarized in the CDR document \cite{litvinenko2014erhic}. The 10 mA polarizied $e^{-}$ ERL in the RHIC tunnel needs 12 passes through a 1.32 GeV SRF linac to produce 15.9 GeV polarized electrons which are then collided with 250 GeV protons with high $e-p$ luminosity of order $10^{33-34}$~cm$^{-2}$s$^{-1}$. The major challenges of such an approach include: a) the need to suppress excitation of high order modes (HOMs) by the beam passing the ERLs SRF cavities to avoid current-limiting beam break up instability; b) generation of high average current of polarized electron beam out of an RF gun; c) halo and beam loss control in the ERL to avoid undue heating and potential damage; d) collective effects due to {\it coherent synchrotron radiation} and space-charge effects; e) precise magnetic field quality control in numerous return beamlines of the ERL and f) eRHIC design specific coherent electron cooling scheme for hadron beams \cite{litvinenko2009coherentecool}. In 2019, a demonstrator facility, the Cornell-BNL ERL Test Accelerator (CBETA), has accelerated electrons from initial 6 MeV to 42, 78, 114, and 150 MeV in four passes through the SRF cavities and subsequently decelerated them during four additional passes through the same cavities back to their original 6 MeV energy
\cite{CBETA2020}. CBETA was also the first accelerator to use a single beamline with fixed magnetic fields to transport seven different accelerating and decelerating energy beams \cite{cbeta2019ipac}.

{\it ERL-based Higgs factories and $\gamma \gamma$ colliders.}
A similar concept was also proposed as an option for the FCC-ee ring collider in a 100 km tunnel in which two 33.7 GeV linacs would accelerate $e^+$ and $e^-$ beams in four passes to $\sqrt{s}$=250 GeV needed for Higgs boson physics research \cite{litvinenko2019fccerl}. Flat electron and positron beams with emittances two orders of magnitude smaller than those in the ring-ring FCC-ee  design (see Ch.~\ref{leptoncolliders} and Table \ref{eefuturetable}) would be generated in 2 GeV cooling rings with top-up injection, then extracted out of the rings with the frequency required by the collider and accelerated to collision energy in a 4 to 6 pass ERL bypassing the interaction regions. Each path requires an individual 100 km arc made of either permanent magnets or low-cost, very low power consumption 0.04 T electromagnets.
As the top energy beams collide at the IPs, their phases are changed to deceleration and they return up to 81\% of the energy back into the SRF cavities. Some 14 GeV  of beam energy will be lost to synchrotron radiation in the arcs, but given that the total required beam current is very small, total SR power losses will be an order of magnitude lower than in the FCC-ee design, i.e., only $\sim$10 MW for a design luminosity of few $10^{34}$~cm$^{-2}$s$^{-1}$. Low average current would render the ERL relatively free of HOMs and coherent instability concerns, but preservation of the ultra-small beam emittances over hundreds of km of beamlines might be as challenging as for linear $e^+e^-$ colliders. 

\begin{figure}[htbp]
\centering
\includegraphics[width=0.99\linewidth]{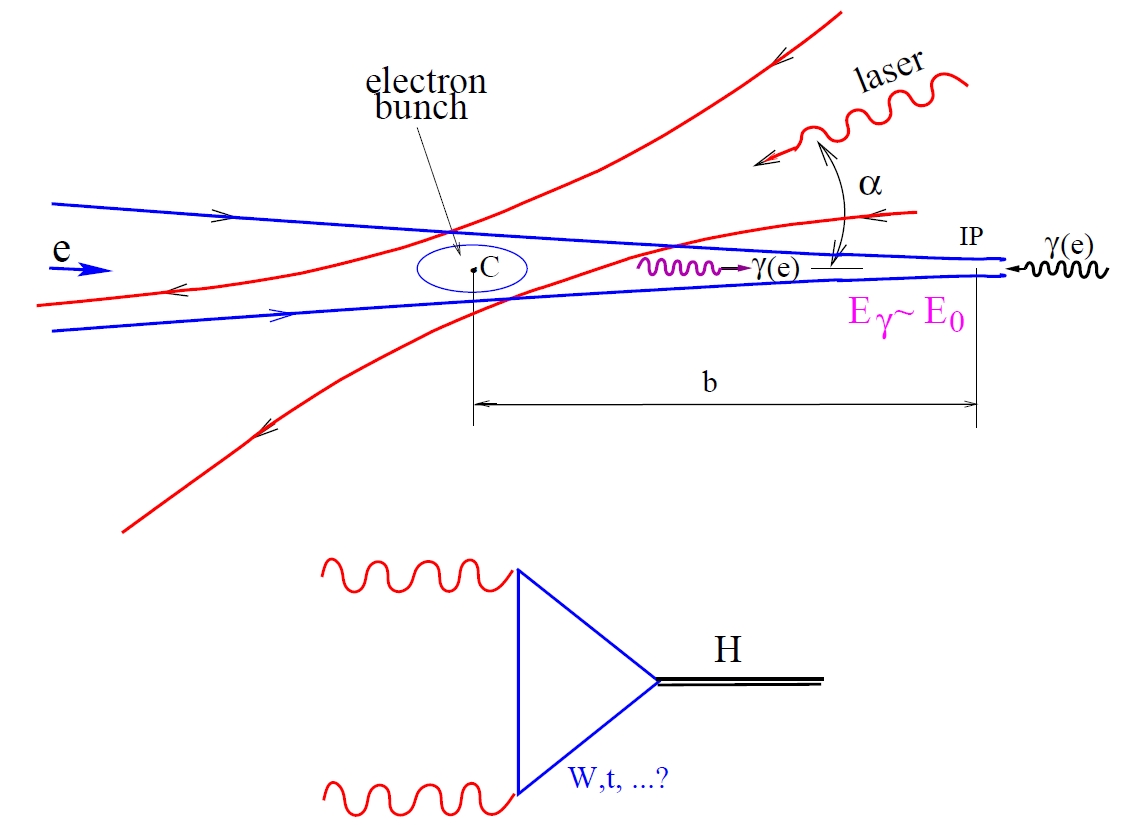}
\caption{(top) Scheme of $\gamma \gamma$ , $\gamma e$ collider; (bottom) Higgs production diagram in $\gamma \gamma$  collisions (from Ref.\cite{telnov2014photonhiggs}).}  
\label{fig:gammagamma}
\end{figure}

The idea of a photon--photon collider through near-IP conversion of high energy electron beams into intense $\gamma$ beams by backward Compton scattering of a high power laser was put forward in the early 1980s \cite{ginzburg1981colliding}. The spectrum of resulting $\gamma '$s will be close to the incident electron energy, so with a proper laser system, such a scheme (see Fig.~\ref{fig:gammagamma}), can obtain $\gamma\gamma$ and $\gamma e$ collisions with energy and luminosity comparable to electron-position luminosity, will be free of the beamstrahlung effect, and would not need a positron beam production complex \cite{telnov1995gammagamma}. An additional advantage for Higgs physics studies is that the energy of photons has to be only half of the $m_H$ for the $s$-channel production reaction $\gamma \gamma \rightarrow H$ , i.e., about 62.5 GeV, requiring lower initial electron beam energy of only $\sim$80 GeV vs $E_e$=125 GeV in the $e^+e^-$ collider Higgs factory designs. 
Besides elimination of the positron production system and reduced electron energy, ERL-based concepts for facilities that could reach the Higgs mass in $\gamma\gamma$ collisions offer additional opportunities to minimize accelerator costs by  using a flat beam electron gun instead of electron damping rings and minimizing the total required accelerating voltage of RF sections \cite{gronberg2014rast,bogacz2012sapphire}. In order to lower beam energy losses due to synchrotron radiation in the ERL arcs, such Higgs factories should be placed in longer circumference tunnels. Two examples are the HFiTT proposal to employ a total of 10 GeV of SRF accelerators in the existing 6.3 km circumference Tevatron tunnel at Fermilab \cite{chou2013hfitt} and the SAPPHiRE proposal with 22 GeV of SRF linacs in the 9 km long racetrack tunnel  \cite{bogacz2012sapphire}. Of concern for such machines is the problem of emittance dilution due to synchrotron radiation and other effects in their very long arcs \cite{telnov2014photonhiggs}. The pulse structure of the ERL based $\gamma\gamma$ Higgs factories with short distance between collisions is very well suited for fiber lasers, and breakthroughs in coherent amplification of short pulses in such lasers \cite{mourou2013fiberfuture} may eventually spark  serious interest in the $\gamma\gamma$ colliders \cite{takahashi2019rastgammagamma}. 

{\it ``Cold'' normal--conducting RF.}
The concept of a TeV-class linear $e^+e^-$ collider based on  NC copper accelerating cavities operating at liquid nitrogen temperature offers promise of significantly lower linac cost and power per GeV than in the ILC (SRF cavities at 2 K) and CLIC (room temperature RF structures) \cite{tantawi2018coldncrf}. The linac design is based on two features: a 5.7 GHz accelerator structure with a separate feed to each cavity permitting the iris to be optimized for high gradient 117 MV/m and lower breakdown rate, and linac RF operation at 77 K, causing Cu (or Cu alloy) conductivity to increase and reducing RF power requirements by about a factor of 2.5. Preliminary design studies indicate some 342 MW of total AC power would be needed for a 2 TeV c.m.e.~collider with luminosity $5 \times 10^{34}$~cm$^{-2}$s$^{-1}$.


{\it Dielectric Wakefield Accelerators (DWFA).} Substantial research efforts have been carried out to extend the two-beam acceleration scheme (similar to that of CLIC) in which resonant dielectric accelerating structures are fed by ultra-short RF pulses of wakefields driven by either collinear or preceding high charge electron bunches \cite{gai1988experimental, gai2009dwa, jing2016dwarast}. In the latter case,  electromagnetic power is radiated by an ultrashort,  intense  ``driving'' electron  bunch  propagating in a high impedance  environment, and then used to accelerate another ``witness'' bunch. Better breakdown properties of some dielectric materials (quartz, ceramics, diamond) and improvement of the BDR with shorter RF pulse length $\tau_{RF}$  (see Eq.~(\ref{eq:bdr})) allow gradients in excess of 1 GV/m for ps exposure times, as demonstrated with simple $O$(0.1 mm) diameter  hollow   dielectric tubes driven by short, narrow and intense 28.5 GeV SLAC linac bunches \cite{thompson2008dlabreakdown}. For collider applications, much longer pulses are needed to drive many bunches and attain high average currents. For example, in the 3 TeV c.m.e.~$e^+e^-$ Argonne Flexible Linear Collider proposal \cite{gai2012shortpulsecollider}, some 20 ns long 26 GHz RF pulses  (12 times shorter than in CLIC) are generated by 32 50 nC drive beam bunches out of 0.86 GeV 1.3 GHz RF linacs  passing through decelerating structures. This scenario should allow 270 MV/m operational accelerating gradients for the main beams. To date the beam accelerating gradient achieved in 26 GHz alumina structures is about 30 MV/m (1.8 MeV acceleration over 6.5 cm) and some 70 MV/m in 11.7 GHz structures (4.9 MeV over 7 cm) \cite{shao2018dwarecent}. Application of this concept to colliding beams faces many challenges, such as fabrication of efficient dielectric high gradient RF structures, drive beam production with bunch charge an order of magnitude greater than typically achieved in the most common efficient RF guns, and wakefield damping to assure main beam stability and attainment of overall AC power to beam efficiency comparable or exceeding that of CLIC \cite{schulte2016applicationadv, jing2016dwarast}. Design, construction and testing of a smaller module for free electron laser (FEL) applications \cite{zholents2016preliminary} may greatly help to advance DWFA technology. 

\begin{figure}[htbp]
\centering
\includegraphics[width=0.9\columnwidth]{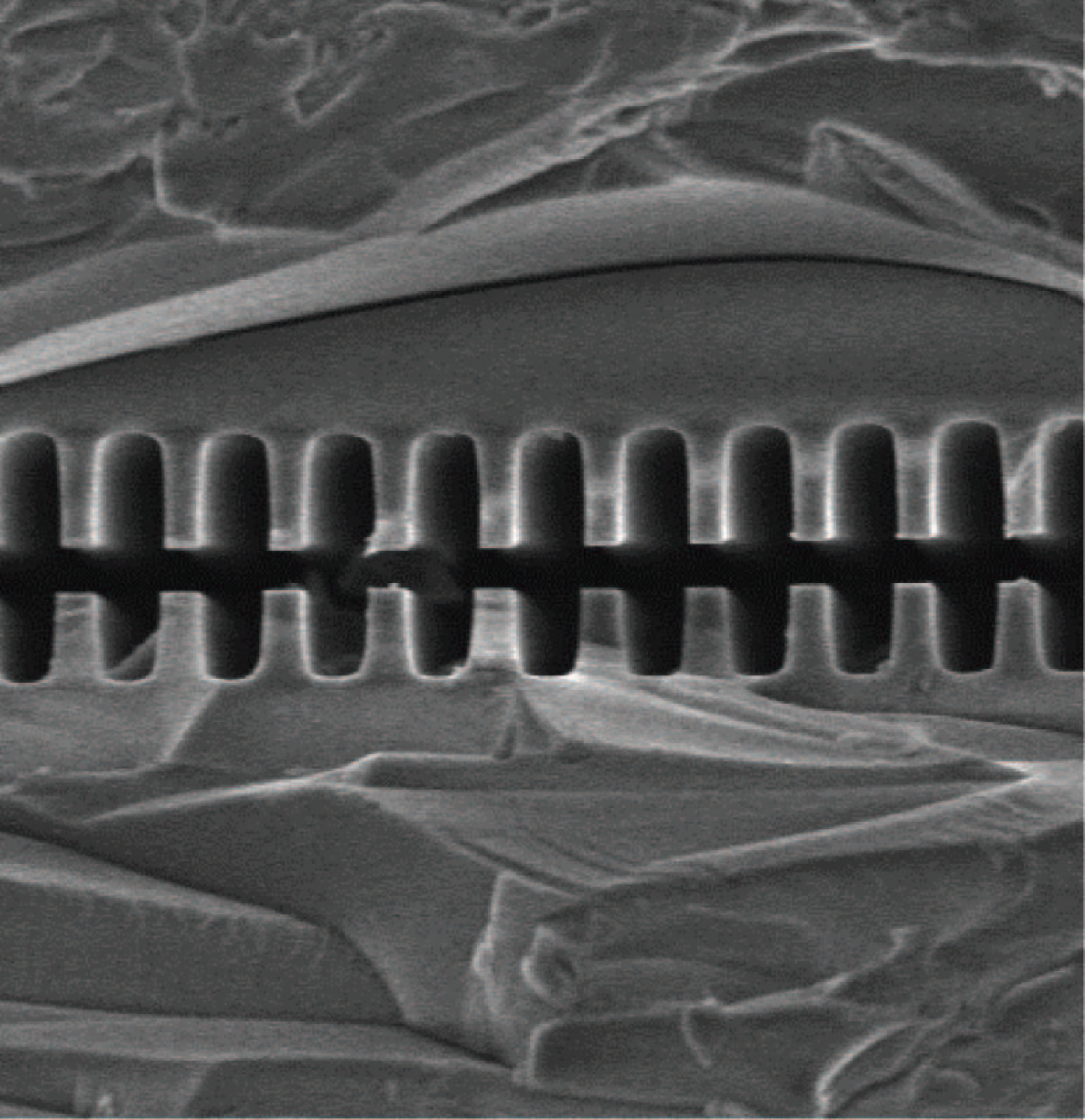}
\caption{ Scanning electron microscope image of the longitudinal cross-section of a dielectric laser acceleration structure
with 400 nm gap \cite{peralta2013dlademonstration}.} 
\label{fig:dla}
\end{figure}

{\it Dielectric Laser Accelerators (DLA).} 
Micron-size dielectric accelerating structures can be driven not by conventional RF, but rather by a laser \cite{peralta2013dlademonstration}, and they can support accelerating fields an order of magnitude higher than can RF cavity-based accelerators. For example, some 35 keV electron energy gain over only 50 microns (700 MV/m gradient) was achieved in a fused silica DLA structure with 800 nm grating period driven by a 90 fs 800nm Ti:sapphire laser pulse \cite{wootton2016demodla}. Despite relatively modest accelerating gradients as compared with plasmas, 
the prospects of using commercial lasers as a power source, which are smaller and less expensive than the RF klystrons powering present day accelerators, and low-cost fabrication lithographic techniques for mass production of optical structures, like the one depicted in Fig.~\ref{fig:dla}, have generated interest in DLA-based linear electron-positron colliders \cite{england2014rmpdla}. 
Strawman parameter tables for 3 TeV and 30 TeV DLA machines presented in Ref.~\cite{plasmacolliders2019granada} indicate that a path exists to high luminosities of $10^{34-36}$~cm$^{-2}$s$^{-1}$ via a combination of very high repetition rate ($f_{\rm r}=$20 MHz) operation of 2 micron  wavelength fiber lasers, small bunch population of some 30,000 electrons and positrons per bunch, and very small emittances, nanometer bunch length and spot sizes at the IP, etc. In such a scheme, beamstrahlung-induced energy spread is minimal while the luminosity enhancement factor $H_D$ of about 10 is due to the pinching effect from the beam-beam interaction at the IP. The required AC wall plug power is 360 MW for 3 TeV and 30 GW for 30 TeV machines and scales with luminosity. In addition to critical issues of production of ultra-small emittance beams, which will be particularly challenging for positrons, and preservation of these very small emittances over many kilometers of optical linacs, the overall power efficiency of the DLA-based collider scheme will require extensive research and development on laser power generation and distribution to achieve a level comparable to or better than that of CLIC or ILC. 

{\it Linear muon crystal colliders.} 
Wakefield acceleration of muons (instead of electrons or hadrons) channeling between the planes in crystals \cite{tajima1979laser} or inside carbon nanotubes (CNT) \cite{shin2013x} with charge carrier density $\sim$10$^{20-22}$ cm$^{-3}$ holds the promise of the maximum theoretical accelerating gradients of 1--10 TeV/m \cite{tajima1979laser} --- see Eq.~(\ref{plasmagradient}) --- allowing envisioning of a compact 1 PeV linear crystal muon collider \cite{shiltsev2012}. The choice of muons is beneficial because of small scattering on solid media electrons, absence of beamstrahlung effects at the IP, and continuous focusing while channeling in crystals, i.e., acceleration to final energy can be done in a single stage. Muon decay becomes practically irrelevant in such very fast acceleration gradients as muon lifetime quickly grows with energy as 2.2[$\mu$s]$\times \gamma$.  Initial luminosity analysis of such machines assumes  a small number of muons per bunch $\sim$10$^3$, a small number of bunches $\sim$100, high repetition rate $\sim$1 MHz and ultimately small sizes and overlap of the colliding beams $\sim 1$~\AA. Excitation of plasma wakefields in crystals or nanostructures can be possible by either short sub-$\mu$m high density bunches of charged particles or X-ray laser pulses \cite{zhang2016cntxray, tajima2019rastsinglecyclelasers}, by heavy high-$Z$ ions, or by pre-modulated or self-modulated very high current bunches \cite{shiltsev2019ultimate}. The concept of acceleration in the crystal or CNT plasma requires a proof-of-principle demonstration \cite{shin2015tev}, extensive theoretical analysis, modeling and simulations \cite{Shiltsev2019xtalworkshop}.


\section{Conclusions}
\label{conclusions}

High energy particle colliders are unique facilities in many ways. They are the pinnacle of almost a century of developments in accelerator and beam physics. Most advances in physics and in technologies of single beam accelerators for various branches of contemporary science have been utilized in  colliders over their half a century long history. The opposite is true, too: many breakthroughs in the development of the collider method are widely used in modern accelerators for industry, medicine, and scientific research in biology, chemistry, and solid state and nuclear physics. Numerous technological advances from other fields of science and technology --- most notably from solid state physics, lasers, plasma, high-energy physics, computers and information technology, cryogenic devices and RF generation, radiation control and ultra-high vacuum, among others --- have effectively been applied to construct better and more powerful colliders. The center of mass energy of colliding-beam facilities has grown by five orders of magnitude and their luminosity by about seven. Collisions of high energy particles offer unique opportunities to answer the most fundamental questions of modern science regarding the composition and evolution of the Universe, and there is a growing aspiration for colliders with order(s) of magnitude higher energies and luminosities.

\begin{table*}[t]
\begin{center}
\caption{Main parameters of proposed colliders for high-energy particle physics research: center of mass energy, number of detectors in simultaneous operation $N_{\rm det}$, total integrated luminosity in these detectors, expected collider operation time, average AC wall plug power, cost estimate, the cost per ab$^{-1}$ of integrated luminosity and integrated luminosity per TWh of electricity consumption. Most of the parameters are taken from the input documents submitted to the European Particle Physics Strategy Update \cite{eppsu2020granada} and cost estimates are given with some 20-30\% accuracy. Note that the cost accounting is not uniform across the projects, as well as the currency. E.g., the ILC cost is given in ``ILC Units'', 1 ILCU is defined as 1 US dollar (USD) in January, 2012. $^*$ Estimates for LHeC and muon collider are pro-rated from the costs other projects, see Refs. \cite{bruning2018lheccost} and \cite{neuffer2018}, respectively. }

\begin{tabular}{|l|c|c|c|cc|c|c||c|c|}
\hline 
\hline
\textbf{Project} & \textbf{Type} & \textbf{Energy} & \textbf{$N_{\rm det}$} & \textbf{$\Lumi_{\rm int}$} &
\textbf{Time} & \textbf{Power} & \textbf{Cost} & \textbf{Cost/$\Lumi_{\rm int}$} & \textbf{$\Lumi_{\rm int}$/Power} \\
 & & (TeV, c.m.e.) & & (ab$^{-1}$) & (years) & (MW) & & (BCHF/ab$^{-1}$) & (ab$^{-1}$/TWh) \\
\textbf{ILC} & $e^+e^-$ & 0.25 & 1 & 2 & 11 & 129 & 4.8-5.3BILCU & 2.7 & 0.24 \\
 &  & 0.5 & 1 & 4 & 10 & 163(204) & 8.0 BILCU & 1.3 & 0.4 \\
 &  & 1 & 1 & &  & 300 & +(n/a) &  &  \\
\textbf{CLIC} & $e^+e^-$ & 0.38 & 1 & 1 & 8 & 168 & 5.9 BCHF & 5.9 & 0.12 \\
 & & 1.5 & 1 & 2.5 & 7 & 370 & + 5.1 BCHF & 3.1 & 0.16 \\
 & & 3 & 1 & 5 & 8 & 590 & +7.3 BCHF & 2.0 & 0.18 \\
\textbf{CEPC} & $e^+e^-$ & 0.091\&0.16 & 2 & 16+2.6 & 2+1 & 149 & 5 B USD & 0.27 & 7.0 \\
 &  & 0.24 & 2 & 5.6 & 7 & 266 & +(n/a) & 0.21 & 0.5 \\
\textbf{FCC-ee} & $e^+e^-$ & 0.091\&0.16 & 2 & 150+10 & 4+1 & 259 & 10.5 BCHF & 0.065 & 20.5 \\
 & & 0.24 & 2 & 5 & 3 & 282 &  & 0.064 & 0.9 \\
 & & 0.365 \& 0.35 & 2 & 1.5+0.2 & 4+1 & 340 & +1.1 BCHF & 0.07 & 0.15 \\
\hline
\textbf{LHeC} & $ep$ & 1.3 & 1 & 1 & 12 & (+100) & 1.75$^*$ BCHF & 1.75 & 0.14 \\
\textbf{HE-LHC} & $pp$ & 27 & 2 & 20 & 20 & 220 & 7.2 BCHF & 0.36 & 0.75 \\
\textbf{FCC-hh} & $pp$ & 100 & 2 & 30 & 25 & 580 & 17(+7) BCHF & 0.8
& 0.35 \\
\textbf{FCC-eh} & $ep$ & 3.5 & 1 & 2 & 25 & (+100) & 1.75 BCHF & 0.9 & 0.13 \\
\textbf{Muon Collider} & $\mu\mu$ & 14 & 2 & 50 & 15 & 290 & 10.7$^*$ BCHF & 0.21 & 1.9 \\
\hline
\hline
\end{tabular}
\label{allfuture}
\end{center}
\end{table*}

The physics community of the seven currently operational colliders is very wide and includes the majority of the world population of some 33,000 high--energy physicists and a large fraction of 24,000 nuclear physicists \cite{battiston2019physstat}. Several colliding beam facilities are either under construction or entering the construction project phase (NICA in Russia, eRHIC in the US, etc.). It is easy to see that these mostly aim at serving nuclear physics research needs. One of the main reasons (besides scientific) for such projects to proceed is their relatively modest energy reach (several to 100s of GeV of the center of mass energy $\sqrt{s}$) and as a result, affordable cost in the sub-billion dollar to \$1--2B range.  

The situation differs for the next generation of HEP colliders. At present, aspirations of the HEP community are focused on two opportunities offering exciting physics prospects, namely future Higgs (or electroweak) factories and energy-frontier (EF) colliders. There are four feasible widely-discussed concepts: linear $e^+e^-$ colliders, circular $e^+e^-$ colliders, $pp/ep$ colliders and muon colliders. These all have limitations in energy, luminosity, efficiency, and cost. The most critical requirement for a Higgs factory is high luminosity, and four proposals generally satisfy it: the ILC at 250 GeV c.m.e., CLIC at 380 GeV, CEPC and FCC-ee. The next level criteria include (in order): facility cost, required AC wall plug power, and technical readiness. The construction cost, if calibrated to performance (i.e., in units of GCHF per ab$^{-1}$ of the integrated luminosity) is the lowest for the FCC-ee, followed by the CepC (by a factor of $\times$4), then the ILC (another $\times$10), then CLIC (another $\times$2) --- see Table \ref{allfuture}. The expected AC site power consumption, if calibrated to performance (i.e., in the units of ab$^{-1}$/TWh) also is the lowest for the FCC-ee, followed by the CEPC ($\times$2), then the ILC (another $\times$2), then CLIC (another $\times$2). As for readiness to start construction, the ILC is somewhat ahead of other proposals (it has TDR vs CDRs for CLIC, CEPC, and FCC-ee) and is technologically quite mature, with well understood plans for industrial participation. 
On the other hand, the FCC-ee and CEPC proposals are based on concepts and beam dynamics parameters that have already been proven at many past and presently operating circular colliders. 

\begin{figure*}[htbp]
\centering
\includegraphics[width=0.99\linewidth]{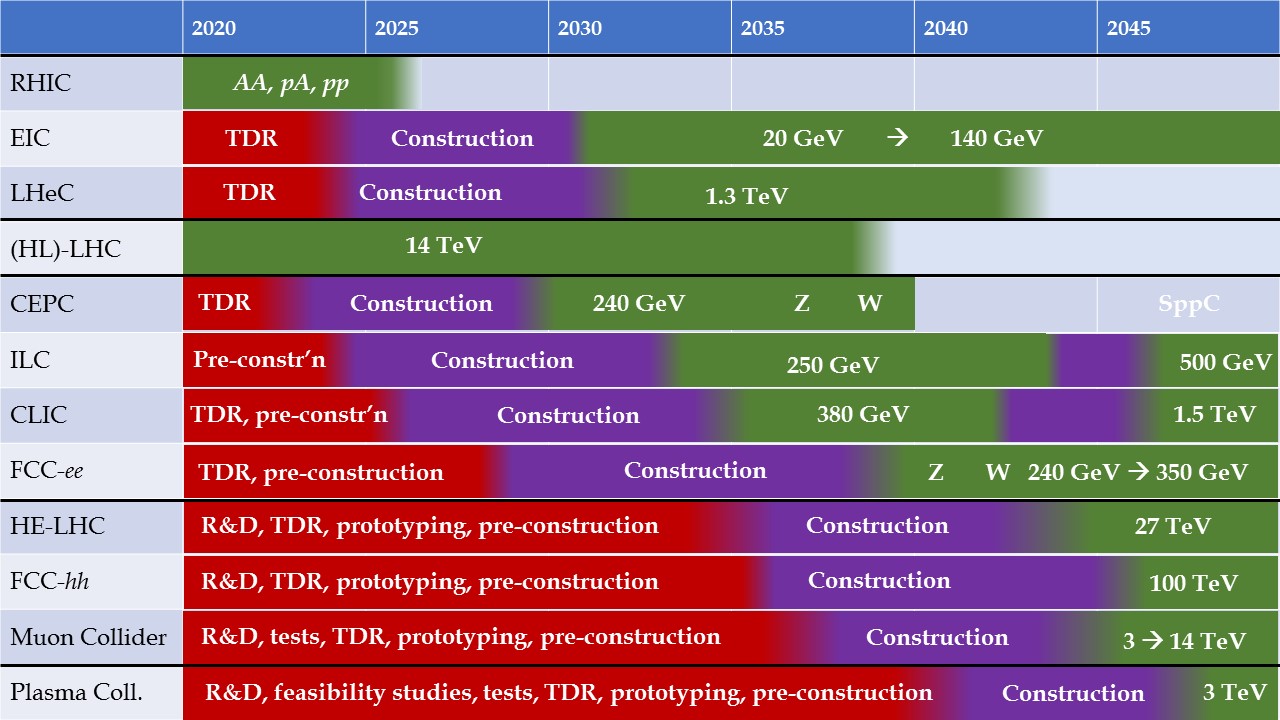}
\caption{Approximate technically limited timelines of future large colliding beam facilities.}  
\label{fig:timelines}
\end{figure*}

The most critical requirement for the EF colliders is the center-of-mass  energy reach. There are four proposals which generally satisfy it (in order of  energy reach): 3 TeV CLIC, HE-LHC, 6 to 14 TeV Muon Collider, and FCC-hh/SppC. The next level criteria for EF machines are (in order):
cost, facility's AC wall plug power, 
machine efficiency and attainable annual integrated luminosity, 
total annual running cost (including manpower),  
and the level of R\&D effort needed to bring the concept to the level of construction readiness (level of comprehensive TDR). 
The construction cost is lowest for the HE-LHC and Muon Collider, followed by the 3 TeV CLIC ($\times$2) and  FCC-hh (another $\times$1.5). The estimated AC site power requirement is  lowest for the HE-LHC and the Muon Collider, followed by CLIC ($\times$2), then by FCC-hh (another $\times$1.5). As for required duration and scale of R\&D efforts to reach the TDR level of readiness, the 3 TeV CLIC project is ahead of other proposals as it requires $\sim$10 years of R\&D vs about twice that for the HE-LHC, FCC-hh/SppC, and Muon Collider (the latter being at present the only concept without a comprehensive CDR). 

Another important factor for any large scale facility is operating cost. Design efforts need to be made, starting from the early concept stage, to enable a sustainable operational efficiency. The history of large-scale collider infrastructures such as at Fermilab and CERN reveals a trend of steadily decreasing normalised operating costs and number of personnel. For example, at the peak of LEP operation CERN had 3,300 staff members, while in the era of LHC, which together with its injectors is a much more complex machine, the laboratory staff has shrunk to 2,300 employees. Such manifestation of progress in technology, operation and maintenance concepts will need to continue for the energy frontier machines. Their designs should place an emphasis on conceiving the individual systems and subsystems such that they can be monitored, maintained and repaired by service suppliers as much
as reasonably possible, on investing early-on in a modular architecture of basic components and equipment to enable streamlined operation, service and repair, and on in-kind, collaborative operation. 

Arguably the biggest technical challenge for EF hadron and muon colliders is the development of bending magnets with a maximum field up to 16 T. There are fundamental challenges in getting the required current density in SC cables and in dealing with the ultimate magnetic pressures and mechanical stresses in the superconductor and associated components. Some experts estimate that at least 15--20 years might be needed for new approaches to mature and for developing the technology required to overcome the aforementioned limits through continuous R\&D efforts. 
Lowering the maximum field requirement to 12--14 T or even to 6--9 T could greatly reduce the time needed for short-model R\&D, prototyping and pre-series work with industry. To realize even higher fields --- beyond 16 T, if needed --- HTS technology will inevitably be required. At present, the most critical constraint for HTS is its much higher cost, even compared with the Nb$_3$Sn superconductor. 

Impressive advances of exploratory plasma wakefield acceleration R\&D over the past decade make it important to find out whether a feasible ``far future'' lepton collider option for particle physics can be based on that technology. One should note that laser- or beam-driven plasma wakefield accelerators (PWFAs) have a significant potential for non-HEP applications and have drawn significant interest and support from the broader community, most notably, because of their possible use in medicine and for generation of X-rays \cite{uesaka2016advanced, albert2016applications}. Several research and test facilities are already built and operated, and many more are being planned \cite{eupraxia2019ipac}. It will be important for HEP accelerator designers to learn from experience, understand the applicability of PWFA advances for particle colliders, and encourage further technological development of the method. 
The push for more effective and cost-efficient methods of particle acceleration continues in several directions, ranging from the use of exotic particles, like muons, over more advanced magnets and RF cavities, to compact high-gradient acceleration in dielectric structures or solid media plasmas. 

Figure~\ref{fig:timelines} illustrates approximate technically limited timelines of future large colliding beam facilities for the next three decades based on the presentations by their proponents given and briefly discussed at the European Particle Physics Strategy Update Symposium (May 13--16, 2019, Granada, Spain) \cite{eppsu2020granada} and Ref.~\cite{colby2016roadmap}. 
In Fig.~\ref{fig:timelines}, 
each of the proposed colliders is considered individually, without any possible interference or interconnection between them, such as a sequential scenario of FCC-hh construction following the completion of FCC-ee operation, as foreseen 
in the FCC integrated project plan \cite{fccee}. 
Several factors are expected to play a role in the actual development: (i) a decisive move --- e.g., the approval of any of the four Higgs factory projects will have an impact on the 
others; (ii) a better understanding of performance, timeline and cost feasibility for the energy-frontier collider proposals after 
further R\&D and more detailed project cost evaluation; and (iii) new discoveries at the LHC or other related particle physics experiments, which might provide clear guidance and preferences for the next generation of accelerator-based HEP programs. 

Under circumstances where projects under consideration in the field are becoming so large and costly that no single country or a group of countries can carry them out in isolation,   coordination of efforts on regional and global levels becomes ever more critical. Discussion forums on  the  future  of  high-energy accelerators such as the Snowmass workshops  \cite{snowmass2013energyfrontier} and the Particle Physics Project Prioritization Panel (P5) in the US \cite{p5report2014}, the European Particle Physics Strategy updates \cite{eppsu2020granada}, the European and Asian Committees for Future Accelerators (ECFA and ACFA) \cite{ecfa,acfa}, the Nuclear Physics European Collaboration Committee (NuPECC \cite{nupecc}) and a number of European-Union co-funded accelerator development and coordination projects 
(e.g.~TIARA \cite{tiara}, ARIES \cite{aries}, E-JADE \cite{ejade} and EuPRAXIA \cite{eupraxia}) 
transcend national or regional  boundaries. 
Even more globally, the International Committee for Future Accelerators (ICFA) \cite{bhat2019icfa}, created, in 1976, 
by the International Union of Pure and Applied Physics 
(IUPAP) \cite{iupap}, 
plays an important role as a facilitator of international collaborations, such as, for example, on the LHC, the ILC and CLIC in the recent past, and promotes international efforts in all phases of construction and exploitation of future global  accelerator facilities for particle physics.

In this review, we have presented only the most promising options for particle colliders; there are many more ideas and avenues that remain to be explored. Collider beam physics, an astonishingly fertile and dynamic research field, is still breaking new ground.  
We are certain that some two decades from now, accelerator and beam physicists will have achieved remarkable accomplishments resulting in better, more effective and more economical colliding-beam facilities, as they have done again and again over the past sixty years.

\section*{Acknowledgements} 
We would like to acknowledge fruitful discussions on particle colliders with many colleagues whom we have collaborated over the years. Special thanks go to our colleagues who provided valuable input and shared their views on various topics of this review, including G.~Arduini, R.~Assmann, M.~Benedikt, P.~Bhat, C.~Biscari, A.~Blondel, A.~Bogacz, J.~Brau, O.~Br\"{u}ning, A.~Canepa, W.~Chou, A.~Grasselino, J.P.~Delahaye, D.~Denisov, V.~Dolgashev, J.~Gao, 
A.~Grasselino, E.~Gschwendtner, 
M.~Klein, W.~Krasny, W.~Leemans, E.~Levichev, B.~List, V.~Litvinenko, E.~M\'{e}tral, H.~Montgomery, P.~Muggli, D.~Neuffer, K.~Ohmi, K.~Oide, H.~Padamsee, M.~Palmer, R.~Palmer, N.~Pastrone, Q.~Qin,  T.~Raubenheimer, L.~Rivkin, A.~Romanenko, M.~Ross, L.~Rossi, G.~Rumolo,
D.~Schulte, M.~Seidel, T.~Sen, A.~Seryi, D.~Shatilov, S.~Stapnes, M.~Syphers, Y.~Tikhonov, F.~Willeke, V.~Yakovlev, A.~Yamamoto, K.~Yokoya,
A.~Zlobin, and M.~Zobov. We are grateful to P.~Derwent and V.~Higgins for carefully reading through the manuscript and helpful feedback.

Some of this material has appeared in other forms written by the authors. Many of the ideas and proposals outlined above were presented and discussed at the European Particle Physics Strategy Update Symposium (May 13--16, 2019, Granada, Spain) \cite{eppsu2020granada}. 

V.S.~was supported by Fermi National Accelerator Laboratory, which is operated by the Fermi Research Alliance, LLC under Contract No. DE-AC02-07CH11359 with the
United States Department of Energy.
F.Z.~was supported by the European Organization for Nuclear Research, and  
by the European Commission under 
the HORIZON 2020 project 
ARIES, grant agreement no.~730871.

\bibliography{RMP_colliders_v1}
\end{document}